\documentclass[aps,prx,twocolumn,showpacs,eqsecnum,superscriptaddress,floatfix,10pt,longbibliography]{revtex4-1}
\usepackage{amsmath,amssymb,amsfonts,stmaryrd,wasysym,graphicx,multirow,color,textcomp,subfigure,xcolor}
\usepackage{url}\usepackage{diagrams}
\usepackage[colorlinks=true,citecolor=blue,urlcolor=blue]{hyperref}
\usepackage[normalem]{ulem}
\newcommand{\CFT}{\hyperlink{CFT}{CFT} }
\newcommand{\TR}{\hyperlink{TR}{TR} }
\newcommand{\AFTR}{\hyperlink{AFTR}{AFTR} }
\newcommand{\BZ}{\hyperlink{BZ}{BZ} }
\newcommand{\TRIM}{\hyperlink{TRIM}{TRIM} }
\newcommand{\WTI}{\hyperlink{WTI}{WTI} }
\newcommand{\AFTI}{\hyperlink{AFTI}{AFTI} }
\newcommand{\GEV}{\hyperlink{GEV}{GEV} }
\newcommand{\FQH}{\hyperlink{FQH}{FQH} }
\newcommand{\QP}{\hyperlink{QP}{QP} }
\newcommand{\PHS}{\hyperlink{PHS}{PHS} }
\newcommand{\ETCR}{\hyperlink{ETCR}{ETCR} }
\newcommand{\ABJ}{\hyperlink{ABJ}{ABJ} }
\newcommand{\SG}{\hyperlink{SG}{SG} }
\newcommand{\ARPES}{\hyperlink{ARPES}{ARPES} }

\begin{document}

%\title{Symmetry preserving gapping of Weyl and Dirac Semimetals}
\title{\texorpdfstring{From Dirac semimetals to topological phases in three dimensions: \\ a coupled wire construction}{From Dirac semimetals to topological phases in three dimensions: a coupled wire construction}}
\author{Syed Raza}
\author{Alexander Sirota}
\author{Jeffrey C. Y. Teo}\email{jteo@virginia.edu}
\affiliation{Department of Physics, University of Virginia, Charlottesville VA 22904, USA}
%\date{\today}

\begin{abstract}
Weyl and Dirac (semi)metals in three dimensions have robust gapless electronic band structures. Their massless single-body energy spectra are protected by symmetries such as lattice translation, (screw) rotation and time reversal. In this manuscript, we discuss many-body  interactions in these systems. We focus on strong interactions that preserve symmetries and are outside the single-body mean-field regime. By mapping a Dirac (semi)metal to a model based on a three dimensional array of coupled Dirac wires, we show (1) the Dirac (semi)metal can acquire a many-body excitation energy gap without breaking the relevant symmetries, and (2) interaction can enable an anomalous Weyl (semi)metallic phase that is otherwise forbidden by symmetries in the single-body setting and can only be present holographically on the boundary of a four dimensional weak topological insulator. Both of these topological states support fractional gapped (gapless) bulk (resp.~boundary) quasiparticle excitations.
\end{abstract}

\maketitle

\section{Introduction}\label{sec:introduction}
Dirac and Weyl semimetals are nodal electronic phases of matter in three spatial dimensions. Their low-energy emergent quasiparticle excitations are electronic Dirac~\cite{Dirac28} and Weyl~\cite{Weyl29} fermions. (Contemporary reviews in condensed electronic matter can be found in Ref.~\onlinecite{Ashvin_Weyl_review,TurnerVishwanath13,HasanXuBian15,RMP,Burkov16,JiaXuHasan16,ArmitageMeleVishwanath16,YanFelser17}.) They are three dimensional generalizations of the Dirac fermions that appear in two dimensional graphene~\cite{NetoGuineaPeresNovoselovGeim09} and the surface boundary of a topological insulator~\cite{HasanKane10,QiZhangreview11,HasanMoore11,RMP}. They follow massless quasi-relativistic linear dispersions near nodal points in the energy-momentum space close to the Fermi level. Contrary to accidental degeneracies which can be lifted by generic perturbations, these nodal points are protected by topologies or symmetries. 

A Weyl fermion is {\em chiral} and has a non-trivial winding of a pseudo-spin texture near the singular nodal point in energy-momentum space. This would associate to a non-conservative charge current under a parallel electric and magnetic field and is known as the Adler-Bell-Jackiw (\hypertarget{ABJ}{ABJ}) anomaly~\cite{Adler69,BellJackiw69}. Thus, in a true three dimensional lattice system, Weyl fermions must come in pairs~\cite{Nielsen_Ninomiya_1981,NielsenNinomiyaPLB1981,NielsenNinomiya83} so that the net chirality, and consequently the anomaly, cancels. Or otherwise, a three dimensional system of a single Weyl fermion must be holographically supported as the boundary of a topological insulator in four dimensions~\cite{ZhangHu01,BernevigChernHuToumbasZhang02,QiHughesZhang08}. On the other hand, a Dirac fermion in three dimensions consists of a pair of Weyl fermions with opposite chiralities. Without symmetries, it is not stable and can turn massive upon inter-Weyl-species coupling. With symmetries, a band crossing can be protected by the distinct symmetry quantum numbers the bands carry along a high symmetry axis. In this article, we focus on the fourfold degenerate Dirac nodal point protected by time-reversal (\hypertarget{TR}{TR}) and (screw) rotation symmetry.

In electronic systems, massless Dirac and Weyl fermions appear in gap-closing phase transitions between spin-orbit coupled topological insulators and normal insulators~\cite{Murakami2007}. When inversion or time-reversal symmetry is broken, nodal Weyl points can be separated in energy-momentum space. Such gapless electronic phases are contemporarily referred to as Weyl (semi)metals~\cite{WanVishwanathSavrasovPRB11,YangLuRan11,burkovBalenstPRL11,BurkovBalentsPRB11}. Their boundary surfaces support open Fermi arcs~\cite{WanVishwanathSavrasovPRB11} that connect surface-projected Weyl nodes. Weyl (semi)metals also exhibit exotic transport properties, such as negative magneto-resistance, non-local transport, chiral magnetic effect, and chiral vortical effect~\cite{Burkov_Weyl_electromagnetic_2012,Hosur_Weyl_develop,Lu_anomaly_Weyl_2013,SonSpivak13,Sid_anomaly_Weyl,Marcel_Weyl_response}. There have been numerous first principle calculations~\cite{WengXiZhong16} on proposed materials such as the non-centrosymmetric (La/Lu)Bi$_{1-x}$Sb$_x$Te$_3$~\cite{LiuVanderbilt14}, the TlBiSe$_2$ family~\cite{SinghSharmaLinHasanPrasadBansil12}, the TaAs family~\cite{WengBernevigDai2015,HuangXuZahidTaAs2015}, trigonal Se/Te~\cite{HirayamaOkugawaIshibashiMurakamiMiyake15} and the HgTe class~\cite{RuanXing16}, as well as the time-reversal breaking pyrochlore iridates \cite{WanVishwanathSavrasovPRB11,witczak_kim_weyl_2012,chen_hermele_weyl}, magnetically doped topological and trivial insulator multilayers \cite{burkovBalenstPRL11}, HgCr$_2$Se$_4$~\cite{XuWengWangDaiFang11} and Hg$_{1-x-y}$Cd$_x$Mn$_y$Te~\cite{BulmashLiuQi14}. At the same time, there have also been abundant experimental observations in bulk and surface energy spectra~\cite{HasanXuBelopolskiHuang17} as well as transport~\cite{WangLinWangYuLiao17}. Angle-resolved photoemission spectroscopy (\hypertarget{ARPES}{ARPES}) showed bulk Weyl spectra and surface Fermi arcs in TaAs~\cite{Xu_Weyl_2015_first,Weyl_discovery_TaAs,YangLiuChenTaAs2015,TaAs_Weyl_obeservationDing,BelopolskiZahid16} as well as similar materials such as NbAs, NbP and TaP~\cite{XuNbAs15,LiuChen16}. Other materials such as Ag$_3$BO$_3$, TlTe$_2$O$_6$ and Ag$_2$Se~\cite{ChangHasan16} were observed to host pinned Weyl nodes at high symmetry points. %theory pinned along screw axis {TsirkinSouzaVanderbilt17}
Negative magneto-resistance was reported in TaAs~\cite{Huang_Weyl_2015,Zhang_anomaly_Weyl_2015} as a suggestive signature of the \ABJ anomaly. Similar properties were also observed in TaP~\cite{HuMaoTaP17}, NbP and NbAs~\cite{CorinnaNiemannFelserNbP17,LiXuNbAsNbP17,GoothNielschNbP17}, although not without controversies~\cite{SudeshPatnaikNbP17}. 

Weyl points with opposite chiralities cannot be separated in energy-momentum space when both inversion and time reversal symmetries are present. Massless Dirac fermions appear between gap-closing phase transitions between topological and trivial (crystalline) insulators, such as Bi$_{1-x}$Sb$_x$~\cite{TeoFuKane08} and Pb$_{1-x}$Sn$_x$Te~\cite{Hsieh:2012fk}. Critical Dirac (semi)metals were investigated for example in the tunable TlBiSe$_{2-x}$S$_x$~\cite{SatoTakahashi11,SoumaAndo12,XuCavaHasan11}, Bi$_{2−x}$In$_x$Se$_3$~\cite{BrahlekSeongshik12,WuArmitage13} and Hg$_{1-x}$Cd$_x$Te~\cite{OrlitaPotemski14}, as well as the charge balanced BaAgBi~\cite{DuWanXYBi15}, PtBi$_2$, SrSn$_2$As$_2$~\cite{GibsonCava15} and ZrTe$_5$~\cite{LiVallaZrTe516} whose natural states are believed to be close to a topological critical transition. A Dirac (semi)metallic phase can be stabilized when the Dirac band crossing is secured along a high symmetry axis and the two crossing bands carry distinct irreducible representations. Theoretical studies include the diamond-structured $\beta$-crystobalite BiO$_2$ family~\cite{BiO3_Dirac_semimetal} (space group (\hypertarget{SG}{SG}) No.~227, Fd3m), the orthorhombic body-centered BiZnSiO$_4$ family~\cite{SteinbergYoungZaheerKaneMeleRappe14} (\SG No.~74, Imma), the tetragonal Cd$_3$As$_2$~\cite{wangCd3As2PRB13} (\SG No.~142, I4$_1$/acd), the hexagonal Na$_3$Bi family~\cite{Dai_predition_Na3Bi}, as well as the filling-enforced non-symmorphic Dirac semimetals~\cite{KonigMermin97,ParameswaranTurnerArovasVishwanath13,WatanabePoVishwanathZaletel15,ChenKimKee16,WatanabePoZaletelVishwanath16,BradlynBernevig17} such as the hexagonal TlMo$_3$Te$_3$ family~\cite{GibsonCava15} (\SG No.~176, P6$_3$/m), the monoclinic Ca$_2$Pt$_2$Ga (\SG No.~15, C2/c), AgF$_2$, Ca$_2$InOsO$_6$ (\SG No.~14, P2$_1$/n), and the orthorhombic CsHg$_2$ (\SG No.~74, Imma)~\cite{ChenPoNeatonVishwanath16}. At the same time, there are numerous experimental confirmations. They include \ARPES observations on Cd$_2$As$_3$~\cite{Cd3As2Chen2014,neupaneDiracHasan,borisenkoPRLCd3As2}, Na$_3$Bi~\cite{Liu21022014,Xu18122014} and ZrTe$_5$~\cite{LiVallaZrTe516}; scanning tunneling microscopy in Cd$_2$As$_3$~\cite{Yazdani_CdAs}; magneto-transport in Bi$_{1-x}$Sb$_x$~\cite{KimLiBiSb13}, Cd$_2$As$_3$~\cite{liangOngTransportCd3As2,HeLi14,XiangChen15,FengLuCd3As215,LiYuCd3As215,LiWangCd3As216,GuoLeeCd2As316,ZhangXiuCd3As217}, Na$_3$Bi~\cite{Xu18122014,XiongOng15}, ZrTe$_5$~\cite{ZhengMingliangZrTe514,LiVallaZrTe516,LiangOngHallZrTe516,YuanXiuZrTe516}, HfTe$_5$~\cite{WangWangHfTe516} and PtBi$_2$~\cite{GaoTianPtBi217}; magneto-optics~\cite{AkrapOrlitaCd2As316} and anomalous Nernst effect~\cite{LiangOngNernstCd3As217} in Cd$_2$As$_3$, and many more. However, there are also contradicting pieces of evidence, especially in ZrTe$_5$ and HfTe$_5$ that suggest a bulk band gap~\cite{WengDaiFangZrTe514,LiXingZrTe516,WuPanZrTe516,MoreschiniGrioniZrTe516,ManzoniCrepaldiZrTe516,ManzoniParmigianiZrTe517,FanZhouZrTe517}.
%Cd3As2
%SdH\cite{HeLi14,XiangChen15,GuoLeeCd2As316} negative magnetoresistance\cite{liangOngTransportCd3As2} anomalous Nernst effect\cite{LiangOngNernstCd3As217} magneto-optics\cite{AkrapOrlitaCd2As316}
%Na3Bi
%negative magnetoresistance\cite{Xu18122014,XiongOng15}
%ZrTe5
%negative magnetoresistance\cite{ZhengMingliangZrTe514,LiVallaZrTe516} anomalous Hall\cite{LiangOngHallZrTe516}
%HfTe5
%negative magnetoresistance\cite{WangWangHfTe516}
%PtBi2
%magnetoresistance\cite{GaoTianPtBi217}

\begin{figure}[htbp]
\includegraphics[width=0.45\textwidth]{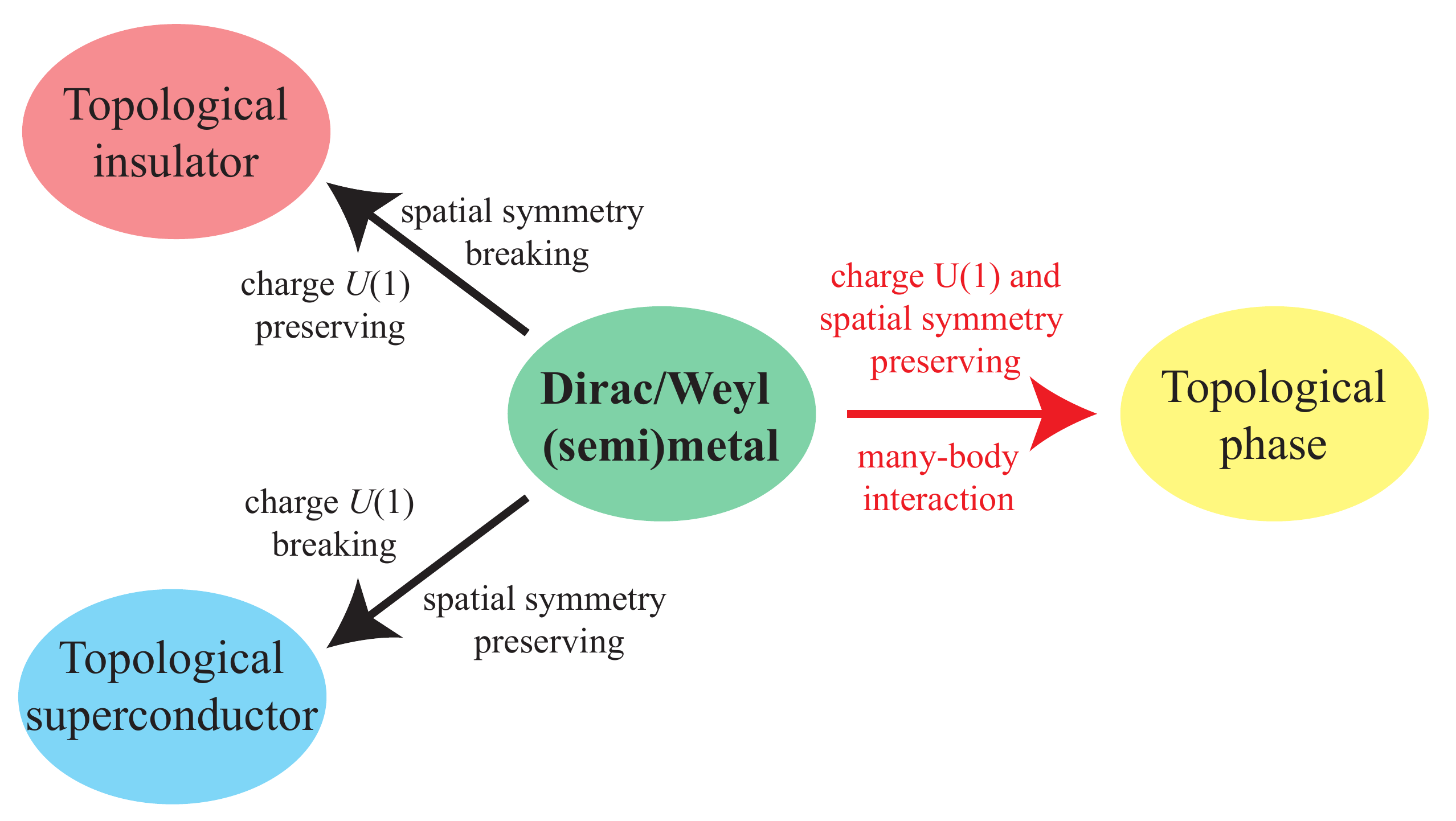}
\caption{Symmetry breaking single-body gapping versus symmetry preserving many-body gapping of a Dirac/Weyl (semi)metal.}\label{fig:intro}
\end{figure}

Dirac/Weyl (semi)metals are the origins of a wide variety of topological phases in three dimensions (see figure~\ref{fig:intro}). By introducing a spatial or charge $U(1)$ symmetry-breaking single-body mass, they can be turned into a topological insulator or superconductor. The focus of this manuscript is on symmetry-preserving many-body gapping interactions. The resulting insulating topological phase can carry long-range entanglement and a non-trivial topological order. Similar phenomena were theoretically studied on the Dirac surface state of a topological insulator~\cite{WangPotterSenthilgapTI13,MetlitskiKaneFisher13b,ChenFidkowskiVishwanath14,BondersonNayakQi13} and the Majorana surface state of a topological superconductor~\cite{LukaszChenVishwanath,MetlitskiFidkowskiChenVishwanath14}, where symmetry-preserving many-body gapping interactions are possible and lead to non-trivial surface topological orders that support anyonic quasiparticle excitations.

Symmetry-preserving gapping interactions cannot be studied using a single-body mean-field theory. This is because the Dirac/Weyl (semi)metallic phase is protected by symmetries in the single-body setting and any mean-field model with an excitation energy gap must therefore break the symmetry either explicitly or spontaneously. The coupled wire construction can serve as a powerful tool in building an exactly-solvable interacting model and understanding many-body topological phases of this sort. The construction involves a highly anisotropic approximation where the electronic degrees of freedom are confined along an array of continuous one-dimensional wires. Inspired by sliding Luttinger liquids~\cite{OHernLubenskyToner99,EmeryFradkinKivelsonLubensky00,VishwanathCarpentier01,SondhiYang01,MukhopadhyayKaneLubensky01}, the coupled wire construction was pioneered by Kane, Mukhopadhyay and Lubensky~\cite{KaneMukhopadhyayLubensky02} in the study of Laughlin~\cite{Laughlin83} and Haldane-Halperin hierarchy~\cite{Haldane83,Halperin84} fractional quantum Hall states. Later, this theoretical technique was applied in more general fractional quantum Hall states~\cite{TeoKaneCouplewires,KlinovajaLoss14,MengStanoKlinovajaLoss14,SagiOregSternHalperin15,KaneSternHalperin17}, anyon models~\cite{OregSelaStern14,StoudenmireClarkeMongAlicea15}, spin liquids~\cite{MengNeupertGreiterThomale15,GorohovskyPereiraSela15}, (fractional) topological insulators~\cite{NeupertChamonMudryThomale14,KlinovajaTserkovnyak14,SagiOreg14,SagiOreg15,SantosHuangGefenGutman15} and superconductors~\cite{mongg2,SeroussiBergOreg14}, as well as the exploration of symmetries and dualities~\cite{MrossAliceaMotrunich16,MrossAliceaMotrunich17}. Moreover, coupled wire construction has already been used to investigate three dimensional fractional topological phases~\cite{Meng15,IadecolaNeupertChamonMudry16,IadecolaNeupertChamonMudry17} and Weyl (semi)metal~\cite{Vazifeh13} even in the strongly-correlated fractional setting~\cite{MengGrushinShtengelBardarson16}. 

The microscopic symmetry-preserving many-body interactions in the Dirac surface state on a topological insulator was discussed by Mross, Essin and Alicea in Ref.\onlinecite{MrossEssinAlicea15}. They mimicked the surface Dirac modes using a coupled wire model and proposed explicit symmetric many-body interactions that lead to a variation of gapped and gapless surface states. Motivated by this and also using a coupled wire construction, the microscopic symmetry-preserving many-body gapping of the Majorana topological superconducting surface state was studied by one of us in Ref.\onlinecite{SahooZhangTeo15}. 

In this article, we focus on (i) a coupled wire realization of a Dirac/Weyl (semi)metallic phase protected by antiferromagnetic time-reversal (\hypertarget{AFTR}{AFTR}) and screw twofold rotation symmetries, (ii) a set of exactly-solvable inter-wire many-body interactions that introduces a finite excitation energy gap while preserving the symmetries, and (iii) an interaction-enabled (semi)metallic electronic phase which is otherwise forbidden by symmetries in the single-body setting.

\subsection{Summary of results}\label{sec:introsummary}

\begin{figure}[htbp]
	\centering\includegraphics[width=0.45\textwidth]{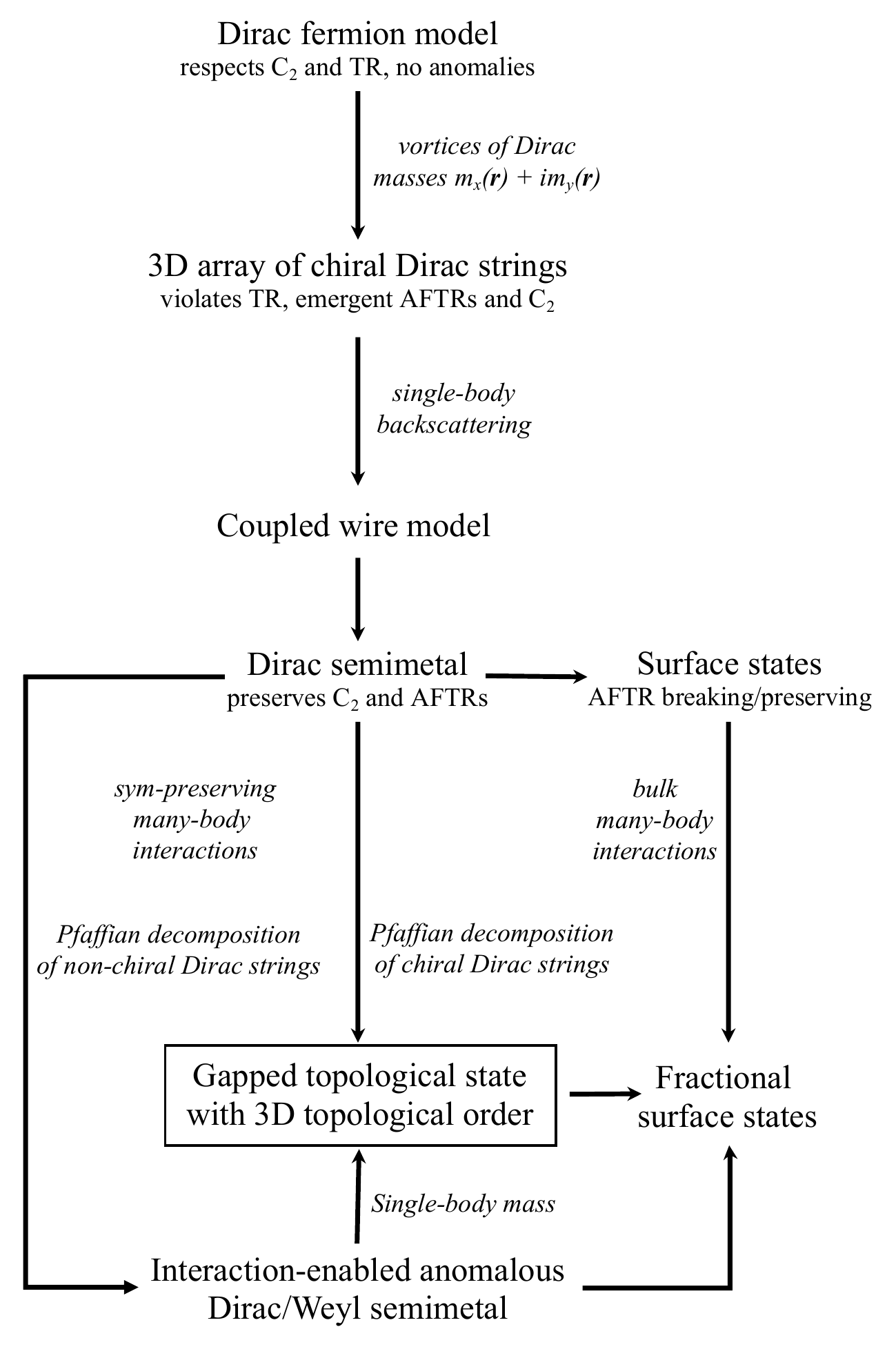}
	\caption{Logical outline of the paper.  It shows the procedure of going from a Dirac fermion model to an interaction-enabled gapped state with three-dimensional topological order. Here, $\mathcal{C}_2$ is the two-fold (screw) rotation symmetry, TR is time-reversal symmetry and AFTR is the antiferromagnetic time-reversal symmetry}\label{fig:Flowchart}
\end{figure}

We now highlight our results. By mapping a Dirac (semi)metal to a model based on a three-dimensional array of wires, we show that the Dirac semimetal can acquire a many-body excitation energy gap without breaking any relevant  symmetries and leads to a three-dimensional topological order. We also construct a new interaction-enabled Dirac semimetallic state which only has a single pair of Weyl nodes and preserves time-reversal symmetry. Such a state is forbidden in a single-body setting. A general outline of the construction of both these states is given in figure~\ref{fig:Flowchart} and also summarized below. 

The starting point of our model is a minimal Dirac fermion model ~(\ref{DiracHam0} and figure~\ref{fig:Diracbands}) equipped with time-reversal and (screw) $\mathcal{C}_2$ rotation symmetries. The model is anomaly-free and so can be realized in a 3D lattice model. The first part of this article addresses a mapping between the isotropic massless Dirac fermion in the continuum limit and an anisotropic coupled wire model where the effective low-energy degrees of freedom are confined along a discrete array of 1D continuous wires. The mapping to a coupled wire model is achieved by first introducing vortices (adding mass terms) that break the symmetries microscopically (\ref{DiracHam}). These vortices are topological line defects that involve spatial winding of symmetry-breaking Dirac mass parameters. Consequently, these vortices host chiral Dirac electronic channels, each of which corresponds to a gapless quasi-1D system where electronic quasiparticles can only propagate in a single direction along the channel and are localized along the perpendiculars (\ref{lowenergy}).	

When assembled together onto a vortex lattice, the system recovers the screw $\mathcal{C}_2$ rotation symmetry as well as a set of emergent antiferromagnetic symmetries, which are combinations of the broken time-reversal and half-translations (figure~\ref{fig:vortexlattice}). Upon nearest-wire single-body electron backscatterings, the electronic band structure in low-energies disperses linearly and mirrors that of the continuous isotropic Dirac parent state. A symmetry-protected massless Dirac fermion (equivalently a pair of Weyl fermions with opposite chiralities) emerges and captures the low-energy long length scale electronic properties (figures~\ref{fig:WeylTB} and \ref{fig:Weylspectrum}). The coupled wire Dirac model and its massless energy spectrum are anomalous with respect to the \AFTR and $\mathcal{C}_2$ symmetry. The three possible resolutions of the anomaly are disussed in section~\ref{sec:anomaly}. The model with an enlarged unit cell which leads to the two momentum-separated Weyl points to collapse to a single Dirac point is also discussed (figure~\ref{fig:DiracTB}). The corresponding Fermi arc surface states are discussed in section~\ref{sec:fermiarc1} and shown in figures~\ref{fig:fermiarc1} and~\ref{fig:fermiarc2}.

The second part of this article addresses non-trivial symmetry-preserving many-body interacting effects beyond the single-body mean-field paradigm. We begin with the anisotropic array of chiral Dirac wires that constitutes a Dirac (semi)metal protected by antiferromagnetic time-reversal (\hyperlink{AFTR}{AFTR}) and (screw) $\mathcal{C}_2$ rotation symmetries (figure~\ref{fig:WeylTB}). We consider an exactly-solvable model of symmetry-preserving inter-wire many-body backscattering interactions. This model is inspired by and can be regarded as a layered version of the symmetric massive interacting surface state of a topological insulator. It is based on a {\em fractionalization} scheme that divides a single chiral Dirac channel into a decoupled pair of identical chiral ``Pfaffian" channels (figure~\ref{fig:glueingsplitting}). Each of these fractional channels carries half of the degrees of freedom of the original Dirac wire. For instance, the fractionalization splits the electric and thermal currents exactly in half. %The bipartition is stabilized by many-body interactions and cannot be realized in any single-body mean-field description. 
It leads to the appearance of fractional quasiparticle excitations. For example, a chiral Pfaffian channel also runs along the 1D edge of the particle-hole symmetric Pfaffian fractional quantum Hall state~\cite{Son15,BarkeshliMulliganFisher15,WangSenthil16}, and supports charge $e/4$ Ising and $e/2$ semionic primary fields.

We consider an explicit combination of many-body interwire backscattering interactions that stabilize the fractionalization. Similar coupled wire constructions were applied in the literature to describe topological insulator's surface state~\cite{MrossEssinAlicea15} and $\nu=1/2$ fractional quantum Hall states~\cite{TeoKaneCouplewires,KaneSternHalperin17}. They are higher dimensional analogues of the Affleck-Kennedy-Lieb-Tasaki (AKLT) spin chain model~\cite{AKLT1,AKLT2}. The pair of chiral Pfaffian channels along each wire is backscattered in opposite directions to neighboring wires by the interaction (figure~\ref{fig:gappinginteraction}). As a result of this dimerization of fractional degrees of freedom, the model acquires a finite excitation energy gap and at the same time preserves the relevant symmetries.

We speculate that such a symmetry preserving-gapping by many-body interactions leads to a three-dimensional topological order supporting exotic point-like and line-like quasiparticle excitations with fractional charge and statistics. The complete characterization of the topological order will be part of a future work~\cite{SirotaRazaTeoappearsoon}. Although there have been numerous field-theoretic discussions on possible properties of topologically ordered phases in 3D, this is the first work with a microscopic model that could possibly lead to its material realization.

In the single-body regime, an (antiferromagnetic) time-reversal symmetric Weyl (semi)metal realizable on a three dimensional lattice has a minimum of four momentum-space-separated Weyl nodes. For a single pair of Weyl nodes with opposite chirality, time-reversal symmetry must be broken. However, a key result of this paper is the realization of a single pair of momentum-space-separated Weyl nodes in the presence of \AFTR symmetry as enabled by many-body interactions. The coupled wire construction suggests a new {\em interaction-enabled topological (semi)metal} in which these Weyl nodes can be realized (figure~\ref{fig:intenable}). 

The many-body interacting coupled-wire model can be turned into a gapless system, where 1) all low-energy degrees of freedom are electronic and freely described in the single-body non-interacting setting by two and only two separated Weyl nodes, 2) the high-energy gapped sector supports fractionalization. Although the model is antiferromagnetic, we conjecture that similar anomalous Weyl (semi)metal can be enabled by interaction while preserving local time-reversal.

The paper is organized as follows. In section~\ref{sec:DiracSemimetal}, we construct a single-body coupled wire model of a Dirac/Weyl (semi)metal equipped with two emergent antiferromagnetic time-reversal (\AFTR) axes and a (screw) $\mathcal{C}_2$ rotation symmetry. In section~\ref{sec:anomaly}, we establish the equivalence between the isotropic continuum limit and the anisotropic coupled wire limit by a coarse-graining mapping. We also discuss the anomalous aspects of the pair of Weyl fermions and different resolutions to the anomaly. In section~\ref{sec:fermiarc1}, we describe the gapless surface states of the coupled wire model. \AFTR breaking and preserving surfaces are considered separately in section~\ref{sec:fermiarcAFTRbreaking} and \ref{sec:fermiarcAFTRpreserving} respectively.

In section~\ref{sec:interaction}, we move on to the effect of symmetry-preserving many-body interactions. The fractionalization of a chiral Dirac channel is explained in section~\ref{sec:gluing}, where we establish the Pfaffian decomposition through bosonization techniques. The splitting of a Dirac channel is summarized in figure~\ref{fig:fractionalization}. In section~\ref{sec:interactionmodels}, we explicitly construct an exactly-solvable interacting coupled wire model that introduces a finite excitation energy gap to the Dirac system while preserving relevant symmetries. The many-body interwire backscattering interactions are summarized in figure~\ref{fig:gappinginteraction}. In section~\ref{sec:AFMstabilization}, we discuss a plausible stabilization mechanism of the desired interactions through an antiferromagnetic order. 

In section~\ref{sec:intenable}, we discuss the other key result of the paper, a variation of the model that enables an anomalous topological (semi)metal. We show how a single pair of Weyl nodes in the presence of time-reversal symmetry can be realized through many-body interactions. Such a state is forbidden in the single-body setting. In section~\ref{sec:fracsurface}, we elaborate on the gapless surface states of both new interacting phases discussed in sections~\ref{sec:interactionmodels} and~\ref{sec:intenable}.

In section~\ref{sec:conclusion}, we conclude with discussions regarding the potential theoretical and high-energy impact of this work. We also discuss the experimental realization and verification of the proposed states and some possible future directions.

\section{Coupled wire construction of a Dirac semimetal}\label{sec:DiracSemimetal}
We begin with a Dirac semimetal in three dimensions. It consists of a pair of massless Weyl fermions with opposite chiralities. In this article we do not distinguish between a Dirac and a Weyl semimetal. This is because the fermion doubling theorem~\cite{Nielsen_Ninomiya_1981,NielsenNinomiyaPLB1981,NielsenNinomiya83} and the absence of the Adler-Bell-Jackiw anomaly~\cite{Adler69,BellJackiw69} require Weyl fermions to always come in pairs in a three dimensional lattice system. A Weyl semimetal therefore carries the same low energy degrees of freedom as a Dirac semimetal. We refer to the case when the pair of Weyl fermions are separated in momentum space as a translation symmetry protected Dirac semimetal. Here, we assume the simplest case where the two Weyl fermions overlap in energy-momentum space. Its low-energy band Hamiltonian takes the spin-orbit coupled form \begin{align}H^0_{\mathrm{Dirac}}({\bf k})=\hbar v{\bf k}\cdot\vec{s}\mu_z\label{DiracHam0}\end{align} where $\vec{s}=(s_x,s_y,s_z)$ are the spin-$1/2$ Pauli matrices, and $\mu_z=\pm1$ indexes the two Weyl fermions.

\begin{figure}[htbp]
\centering\includegraphics[width=0.25\textwidth]{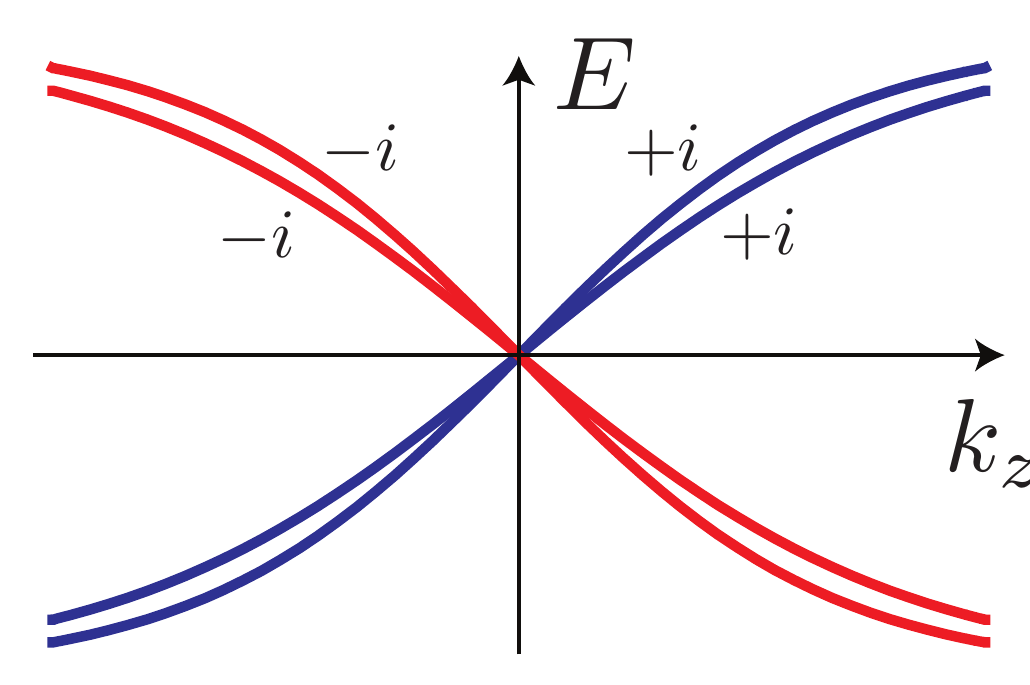}
\caption{The two pairs of counter-propagating Dirac bands along the $k_z$-axis distinguished by eigenvalues of $C_2=\pm i$.}\label{fig:Diracbands}
\end{figure}

Normally the masslessness of the Dirac system is protected by a set of symmetries. Here, we assume the time reversal (TR) $\mathcal{T}$, which is represented in the single-body picture by the spinful operator $\hat{T}=is_y\mathcal{K}$ where $\mathcal{K}$ is the complex conjugation operator, and a twofold rotation $C_2$ about the $z$-axis. In the case when $\mu_z$ has a non-local origin such as sublattice or orbital, it can enter the rotation operator. We assume $\mathcal{C}_2$ is represented in the single-body picture by $\hat{C}_2=is_z\mu_z$. It squares to minus one in agreement with the fermionic statistics, and commutes with the local \TR operator. In momentum space, $\mathcal{T}$ flips ${\bf k}\to-{\bf k}$ while $C_2$ rotates $(k_x,k_y,k_z)\to(-k_x,-k_y,k_z)$. The band Hamiltonian \eqref{DiracHam0} shares simultaneous eigenstates with $C_2$ along the $k_z$-axis. The two forward moving bands have $C_2$ eigenvalues $+i$ while the two backward moving ones have $C_2$ eigenvalues $-i$ (see figure~\ref{fig:Diracbands}). Therefore the band crossing is $C_2$-protected while the fourfold degeneracy is pinned at ${\bf k}=0$ because of \TR symmetry. Noticing that each of the $C_2=\pm i$ sector along the $k_z$-axis is chiral (i.e.~consisting of a single propagating direction), it violates the fermion doubling theorem~\cite{Nielsen_Ninomiya_1981,NielsenNinomiyaPLB1981} and is anomalous. This can be resolved by assuming the $C_2$ symmetry is actually a non-symmorphic screw rotation in the microscopic lattice limit and squares to a primitive lattice translation in $z$. $k_z$ is now periodically defined (up to $2\pi/a$) and the two $C_2$ eigen-sectors wraps onto each other after each period. Focusing on the continuum limit where $k_z$ is small (when compared with $2\pi/a$), $C_2^2=-e^{ik_za}\approx-1$ and the $C_2$ symmetry behaves asymptotically as a proper rotation.

The primary focus of this article is to explore symmetry preserving/enabled interacting topological states that originate from the massless Dirac system. Contrary to its robustness in the single-body non-interacting picture, we show that the 3D Dirac fermion can acquire a many-body mass gap without violating the set of symmetries. To illustrate this, we first make use of the fact that the Dirac system can be turned massive by breaking symmetries. Symmetry breaking inter-valley scatterings introduce two coexisting mass terms \begin{align}H_{\mathrm{Dirac}}({\bf k},{\bf r})=H_{\mathrm{Dirac}}^0({\bf k})+m_x({\bf r})\mu_x+m_y({\bf r})\mu_y\label{DiracHam}\end{align} where $m_x$ (or $m_y$) preserves (resp.~breaks) \TR, and both of them violate $C_2$. We allow slow spatial modulation of the mass parameters, which can be grouped into a single complex parameter $m({\bf r})=m_x({\bf r})+im_y({\bf r})$, and to be precise, momentum ${\bf k}$ should be taken as a differential operator $-i\nabla_{\bf r}$ when translation symmetry is broken. Non-trivial spatial windings of the symmetry breaking mass parameters give rise to topological line defects or vortices that host protected low-energy electronic degrees of freedom. Proliferation of interacting vortices then provides a theoretical path to multiple massive/massless topological phases while restoring and modifying the original symmetries as they emerge in the low-energy long-length scale effective theory.

\begin{figure}[htbp]
\centering\includegraphics[width=0.4\textwidth]{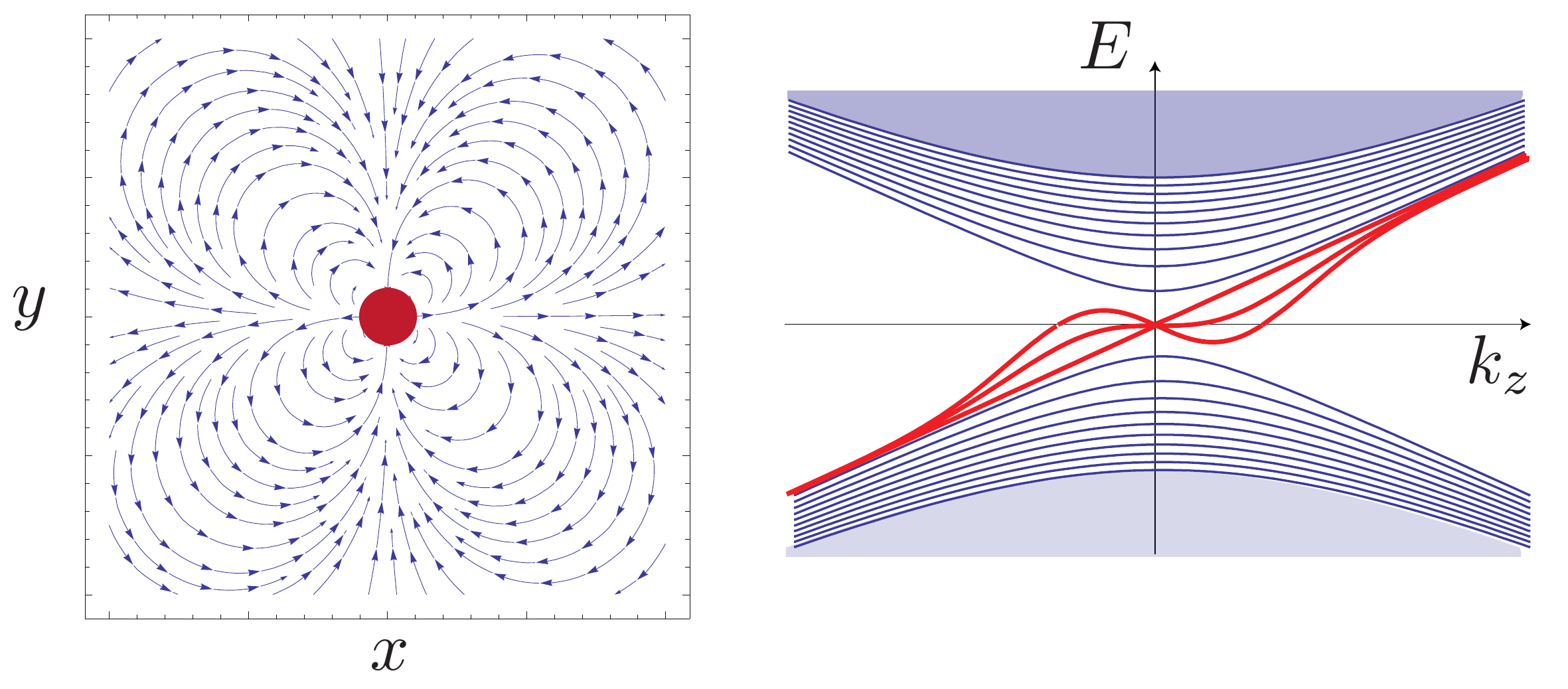}
\caption{Dirac string. (Left) Spatial winding of mass parameters around a Dirac string going out of the paper represented by the center red dot. Stream lines represent the vector field ${\bf m}({\bf r})=(m_x({\bf r}),m_y({\bf r}))$. (Right) Energy spectrum of chiral Dirac fermions. Blue bands represent bulk continuum. Red bands correspond to chiral Dirac fermions localized along the string.}\label{fig:Diracstring}
\end{figure}

A topological line defect is a vortex string of the mass parameter in three dimensions where the complex phase of $m({\bf r})=|m({\bf r})|e^{i\varphi({\bf r})}$ winds non-trivially around the string. The left diagram in figure~\ref{fig:Diracstring} shows the spatial modulation of $\varphi({\bf r})$ along the $xy$ cross-sectional plane normal to a topological line defect, which runs along the $z$ axis. In this example, the complex phase $\varphi({\bf r})$ winds by $6\pi$ around the line defect (represented by the red dot at the origin). The winding number of the complex phase in general can be evaluated by the line integral \begin{align}c=\frac{1}{2\pi}\oint_\mathcal{C}d\varphi({\bf r})=\frac{1}{2\pi i}\oint_\mathcal{C}\frac{\nabla_{\bf r}m({\bf r})}{m({\bf r})}\cdot d{\bf r}\label{winding}\end{align} where $\mathcal{C}$ is a (righthanded) closed path that runs once around the (oriented) line defect. Eq.\eqref{winding} is always an integer given that the mass parameter $m({\bf r})$ is non-vanishing along $\mathcal{C}$.

Massless chiral Dirac fermions run along these topological line defects~\cite{TeoKane}. When focusing at $k_z=0$, the differential operator \eqref{DiracHam} with a vortex along the $z$-axis is identical to the 2D Jackiw-Rossi model~\cite{JackiwRossi81} with chiral symmetry $\gamma_5=s_z\mu_z$. Each zero energy mode corresponds to a massless chiral Dirac fermion with positive or negative group velocity in $z$ depending on the sign of its $\gamma_5$ eigenvalue. (For a concrete example, see appendix~\ref{sec:chiralmodesapp}) These quasi-one dimensional low-energy electronic modes are similar to those that run along the edge of 2D Landau levels and Chern insulators, except they are now embedded in three dimensions. Their wave functions extend along the defect string direction but are localized and exponentially decay away from the defect line. Moreover, such an electronic channel is chiral in the sense that there is only a single propagating direction. The energy spectrum of the topological line defect (for the example with winding number $c=3$) is shown in the right diagram of figure~\ref{fig:Diracstring}, in which, there are three chiral bands (red curves) inside the bulk energy gap representing the 3 chiral Dirac electrons. As a consequence of the chirality, the transport of charge and energy must also be uni-directional. The chiral electric and energy-thermal responses are respectively captured by the two conductances \begin{align}\sigma=\frac{\delta I_{\mathrm{electric}}}{\delta V}=\nu\frac{e^2}{h},\quad\kappa=\frac{\delta I_{\mathrm{energy}}}{\delta T}=c\frac{\pi^2k_B^2}{3h}T\label{conductance}\end{align} where $\nu$ is the filling fraction if the chiral channel is supported by a 2D insulating bulk, and $c$ is called the chiral central charge. For the Dirac case, $c=\nu$ is the number of chiral Dirac channels. Here $c$ can be negative when the Dirac fermions oppose the preferred orientation of the topological line defect. In a more general situation, $c=c_R-c_L$ counts the difference between the number of forward propagating and backward propagating Dirac fermions. There is a mathematical index theorem~\cite{TeoKane,AtiyahSinger63,Nakaharabook} that identifies the topological winding number in \eqref{winding} and the analytic number of chiral Dirac fermions in \eqref{conductance}. Hence, there is no need to distinguish the two $c$'s. 

The massless chiral Dirac channels, described by the low-energy effective theory \begin{align}\mathcal{L}_{\mathrm{Dirac}}=i\sum_{a=1}^{c_R}\psi^\dagger_a(\partial_t+\tilde{v}\partial_x)\psi_a+i\sum_{b=c_R+1}^{c_R+c_L}\psi^\dagger_b(\partial_t-\tilde{v}\partial_x)\psi_b,\label{lowenergy}\end{align} have an emergent conformal symmetry and the index $c=c_R-c_L$ is also the chiral central charge of the effective conformal field theory (\hypertarget{CFT}{CFT}). We refer to the primitive topological line defect with $c=\pm1$ that hosts one and only chiral Dirac fermion $\psi$ as a {\em Dirac string}. (It should not be confused with the Dirac magnetic flux string that connects monopoles.)

\begin{figure}[htbp]
\centering\includegraphics[width=0.45\textwidth]{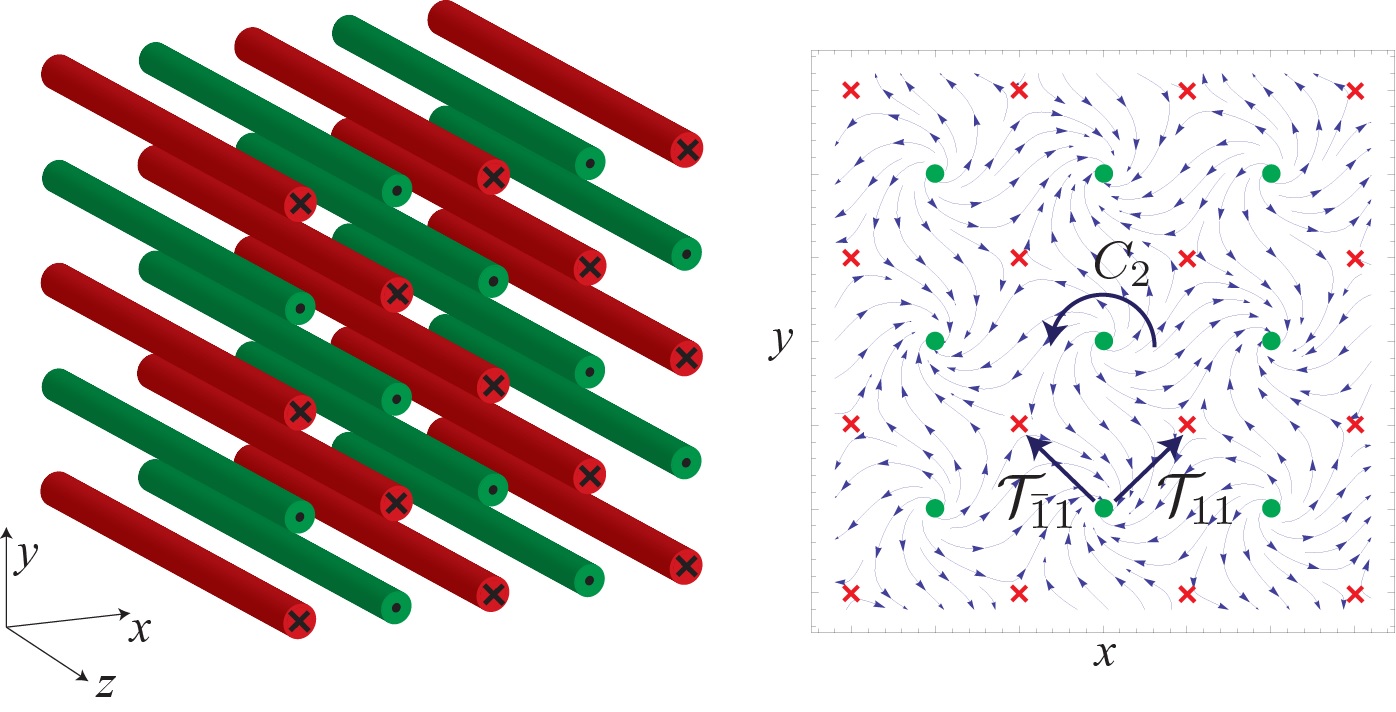}
\caption{(Left) A 3D array of Dirac strings. (Right) Cross section of the array. {\color{red}$\boldsymbol\times$} associates into-the-plane Dirac channel, {\color{green}$\bullet$} represents out-of-plane ones. Stream lines represent the configuration of the mass parameter vector field ${\bf m}({\bf r})=(m_x({\bf r}),m_y({\bf r}))$ of the vortex lattice.}\label{fig:vortexlattice}
\centering\includegraphics[width=0.25\textwidth]{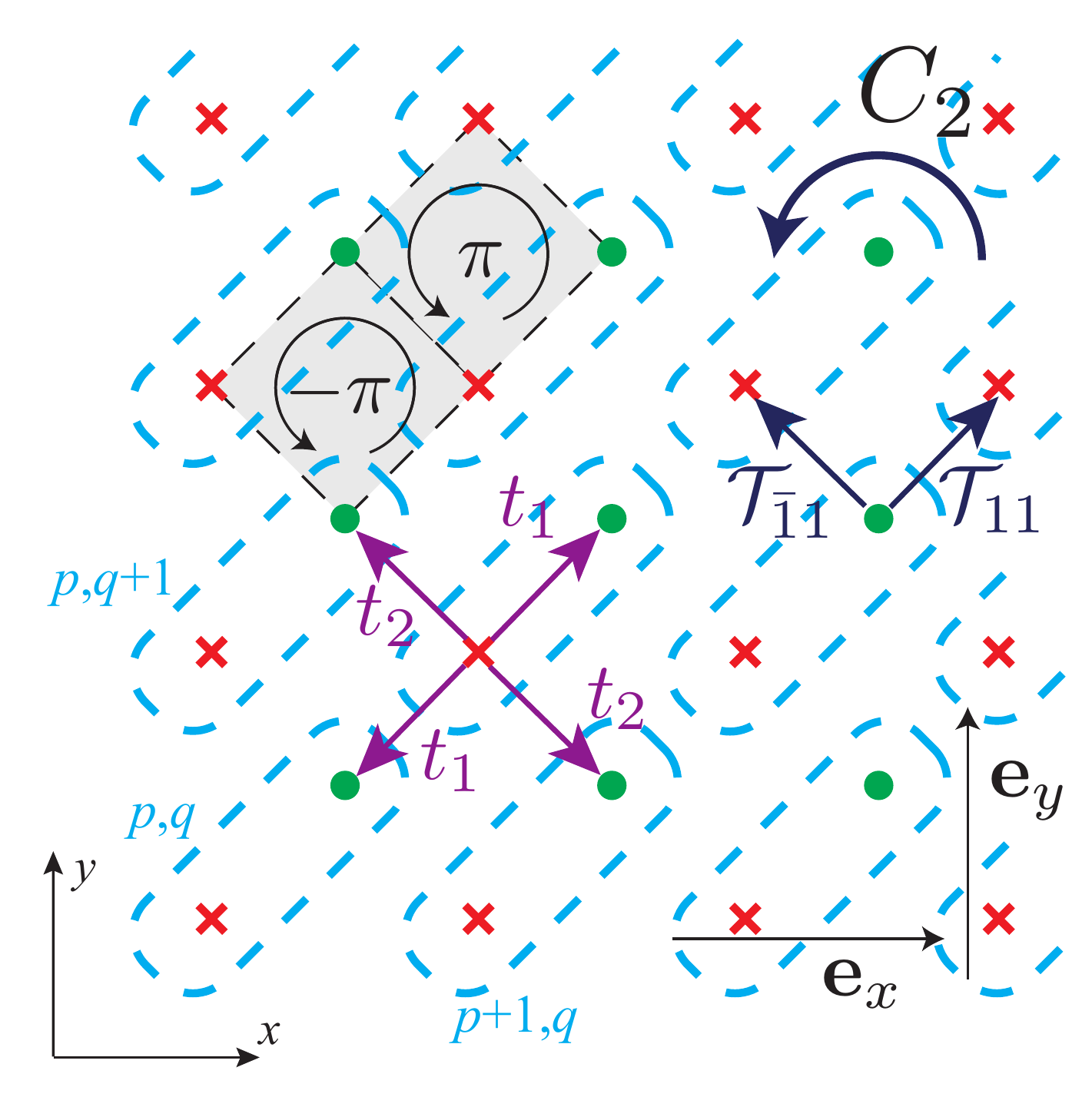}
\caption{Coupled Dirac wire model with tunneling amplitudes $t_1,t_2$. Each unit cell (dashed box) consists a pair of counter-propagating Dirac strings, {\color{red}$\boldsymbol\times$} and {\color{green}$\bullet$}. $\mathcal{T}_{11},\mathcal{T}_{\bar{1}1}$ are the two anti-ferromagnetic directions.}\label{fig:WeylTB}
\end{figure}

A three-dimensional array of Dirac strings (wires) can be realized as a vortex lattice of the mass parameter $m=m_x+im_y$ in a Dirac semimetal. For example, figure~\ref{fig:vortexlattice} shows a vortex lattice generated by the spatially-varying Dirac mass \begin{align}m({\bf r})=m_0\frac{\mathrm{sd}(x+iy)}{|\mathrm{sd}(x+iy)|},\label{Jacobielliptic}\end{align} where $\mathrm{sd}$ is the (rescaled) Jacobian elliptic function~\cite{ReinhardtWalker10} with simple zeros at $p+iq$ and poles at $(p+1/2)+i(q+1/2)$ for $p,q$ integers. It consists of vortices with alternating winding number $c=\pm1$ at the zeros and poles in a checkered board lattice configuration. On the cross section plot on the right side of figure~\ref{fig:vortexlattice}, there is a Dirac string with positive (or negative) winding at each {\color{green}$\bullet$} (resp. {\color{red}$\boldsymbol\times$}). Each vortex string has a chiral Dirac fermion running through it. Figure~\ref{fig:WeylTB} shows the same two-dimensional slice of the array, except suppressing the mass parameters which correspond to irrelevant microscopic high-energy degrees of freedom. We choose a unit cell labeled by $(p,q)$, its $x,y$ coordinates. Each has both a forward moving Dirac fermion $\psi_{p,q}^\odot$ (shown as {\color{green}$\bullet$}) and a backward moving one $\psi_{p,q}^\otimes$ (shown as {\color{red}$\boldsymbol\times$}). 

This array configuration breaks \TR as the symmetry would have reversed the chirality (i.e.~propagating direction) of each Dirac fermion. Instead, it has an emergent {\em anti-ferromagnetic time reversal} (AFTR) symmetry, which is generated by the operators $\mathcal{T}_{11}$ and $\mathcal{T}_{\bar{1}1}$ in the diagonal and off-diagonal directions. Each is composed of a time reversal operation and a half-translation by $({\bf e}_x+{\bf e}_y)/2$ or $(-{\bf e}_x+{\bf e}_y)/2$. \begin{gather}\mathcal{T}_{11}\psi_{p,q}^\otimes\mathcal{T}_{11}^{-1}=\psi_{p,q}^\odot,\quad\mathcal{T}_{11}\psi_{p,q}^\odot\mathcal{T}_{11}^{-1}=-\psi_{p+1,q+1}^\otimes\nonumber\\\mathcal{T}_{\bar{1}1}\psi_{p,q}^\otimes\mathcal{T}_{\bar{1}1}^{-1}=\psi_{p-1,q}^\odot,\quad\mathcal{T}_{\bar{1}1}\psi_{p,q}^\odot\mathcal{T}_{\bar{1}1}^{-1}=-\psi_{p,q+1}^\otimes\label{AFTR}\end{gather} These \AFTR operators are non-local as they come with lattice translation parts. They are anti-unitary in the sense that $\mathcal{T}\alpha\psi\mathcal{T}^{-1}=\alpha^\ast\mathcal{T}\psi\mathcal{T}^{-1}$ and $\langle\mathcal{T}u|\mathcal{T}v\rangle=\langle u|v\rangle^\ast$ because the local time reversal symmetry is anti-unitary. %Normally the local \TR operation for a spinful fermion squares to minus one. However, the non-local nature of the \AFTR symmetry allows us to absorb the sign by a non-local gauge transformation (for example $\psi^{\otimes/\odot}_{p,q}\to(-1)^q\psi^{\otimes/\odot}_{p,q}$) so that no signs appear in \eqref{AFTR}. 
Similar to a spatial non-symmorphic symmetry, the \AFTR symmetries square to the primitive translation operators \begin{align}\mathcal{T}_{11}\mathcal{T}_{\bar{1}1}&=(-1)^F\mbox{translation}({\bf e}_y),\nonumber\\\mathcal{T}_{11}\mathcal{T}_{\bar{1}1}^{-1}&=\mbox{translation}({\bf e}_x),\label{AFTRalg}\end{align} where $(-1)^F$ is the fermion parity operator. Moreover they mutually commute $[\mathcal{T}_{11},\mathcal{T}_{\bar{1}1}]=0$. We notice in passing that the \AFTR symmetry is only an emergent symmetry in the low-energy effective theory. It is not preserved in the microscopic Dirac model \eqref{DiracHam} and is broken by the mass parameter, $m({\bf r})\neq m({\bf r}+({\bf e}_x\pm{\bf e}_y)/2)^\ast$. For instance, the Jacobian elliptic Dirac mass function \eqref{Jacobielliptic} actually has a periodic unit cell twice the size of that of the effective wire model in figure~\ref{fig:WeylTB}. On the other hand, the Dirac mass \eqref{Jacobielliptic} is odd under $C_2$, $m(C_2{\bf r})=-m({\bf r})$. This sign is canceled by the $C_2$ rotations of the Dirac matrices, $\hat{C}_2\mu_{x,y}\hat{C}_2^{-1}=-\mu_{x,y}$, that couple with the Dirac mass in the Hamiltonian \eqref{DiracHam}. Therefore the Dirac wire model in figure~\ref{fig:WeylTB} has a twofold axis along one of the Dirac string, say $\psi^\odot_{0,0}$. The Dirac channel fermions transform unitarily according to \begin{align}\mathcal{C}_2\psi^\odot_{p,q}\mathcal{C}_2^{-1}=i\psi^\odot_{-p,-q},\quad\mathcal{C}_2\psi^\otimes_{p,q}\mathcal{C}_2^{-1}=-i\psi^\otimes_{-p+1,-q+1},\label{C2}\end{align} where the factor of $i$ ensures the fermionic $-1$ twist phase for a $2\pi$ rotation, and the second eqaulity in \eqref{C2} is determined by the first one together with \eqref{AFTR} and the symmetry relations \begin{gather}\mathcal{C}_2\mathcal{T}_{11}=(-1)^F\mathcal{T}_{11}^{-1}\mathcal{C}_2,\quad\mathcal{C}_2\mathcal{T}_{\bar{1}1}=(-1)^F\mathcal{T}_{\bar{1}1}^{-1}\mathcal{C}_2.\label{C2Trelation}\end{gather} Again, in order for the rotation symmetric wire model to be free of anomalies, $C_2$ should really be a screw rotation with respect to some microscopic lattice that has become irrelevant in the low-energy continuum picture. \begin{align}\mathcal{C}_2^2=(-1)^F\mathrm{translation}(a{\bf e}_z)\approx(-1)^F.\label{C2square}\end{align}

When adjacent vortex strings are near each other, their Dirac fermion wave functions overlap and there are finite amplitudes of electron tunneling. We construct a coupled Dirac wire model of nearest-wire single-body backscattering processes with $\pm\pi$ fluxes across each diamond square (figure~\ref{fig:WeylTB}), where the tunneling amplitude $t_1$ (or $t_2$) in the $(11)$ (resp.$(\bar{1}1)$) direction is imaginary (resp.~real). \begin{align}\mathcal{H}=&\sum_{p,q}\hbar\tilde{v}\left({\psi_{p,q}^\odot}^\dagger k_z\psi_{p,q}^\odot-{\psi_{p,q}^\otimes}^\dagger k_z\psi_{p,q}^\otimes\right)\nonumber\\&+it_1\left({\psi_{p,q}^\odot}^\dagger\psi_{p,q}^\otimes-{\psi_{p-1,q-1}^\odot}^\dagger\psi_{p,q}^\otimes\right)+h.c.\label{WeylTBHam}\\&+t_2\left({\psi_{p-1,q}^\odot}^\dagger\psi_{p,q}^\otimes-{\psi_{p,q-1}^\odot}^\dagger\psi_{p,q}^\otimes\right)+h.c.\nonumber\end{align} where the first line is the kinetic Hamiltonian of individual Dirac channels under the Fourier transformation $-i\partial_z\leftrightarrow k_z$ along the wire direction. This tight-binding Hamiltonian preserves the \AFTR symmetry \eqref{AFTR}, $\mathcal{T}\mathcal{H}\mathcal{T}^{-1}=\mathcal{H}$. Fourier transformation of the square lattice $\vec\psi_{p,q}=\int\frac{dk_xdk_y}{(2\pi)^2}e^{-i(k_xp+k_yq)}\vec\psi_{\bf k}$, $\vec\psi=(\psi^\odot,\psi^\otimes)$ turns \eqref{WeylTBHam} into $\mathcal{H}=\int\frac{dk_xdk_y}{(2\pi)^2}\vec\psi_{\bf k}^\dagger H(k)\vec\psi_{\bf k}$, where \begin{align}H({\bf k})=\left(\begin{array}{*{20}c}\hbar\tilde{v}k_z&g(k_x,k_y)\\g^\ast(k_x,k_y)&-\hbar\tilde{v}k_z\end{array}\right)\label{BlochHam}\end{align} is the Bloch band Hamiltonian, for $g(k_x,k_y)=it_1(1-e^{-i(k_y+k_x)})+t_2(e^{-ik_x}-e^{-ik_y})$. Here momentum ${\bf k}$ lives in the ``liquid crystal" Brillouin zone (\hypertarget{BZ}{BZ}) where $-\pi\leq k_x,k_y\leq\pi$ and $-\infty<k_z<\infty$ (in the continuum limit $a\to0$ and $\pi/a\to\infty$). 

\begin{figure}[htbp]
\centering\includegraphics[width=0.45\textwidth]{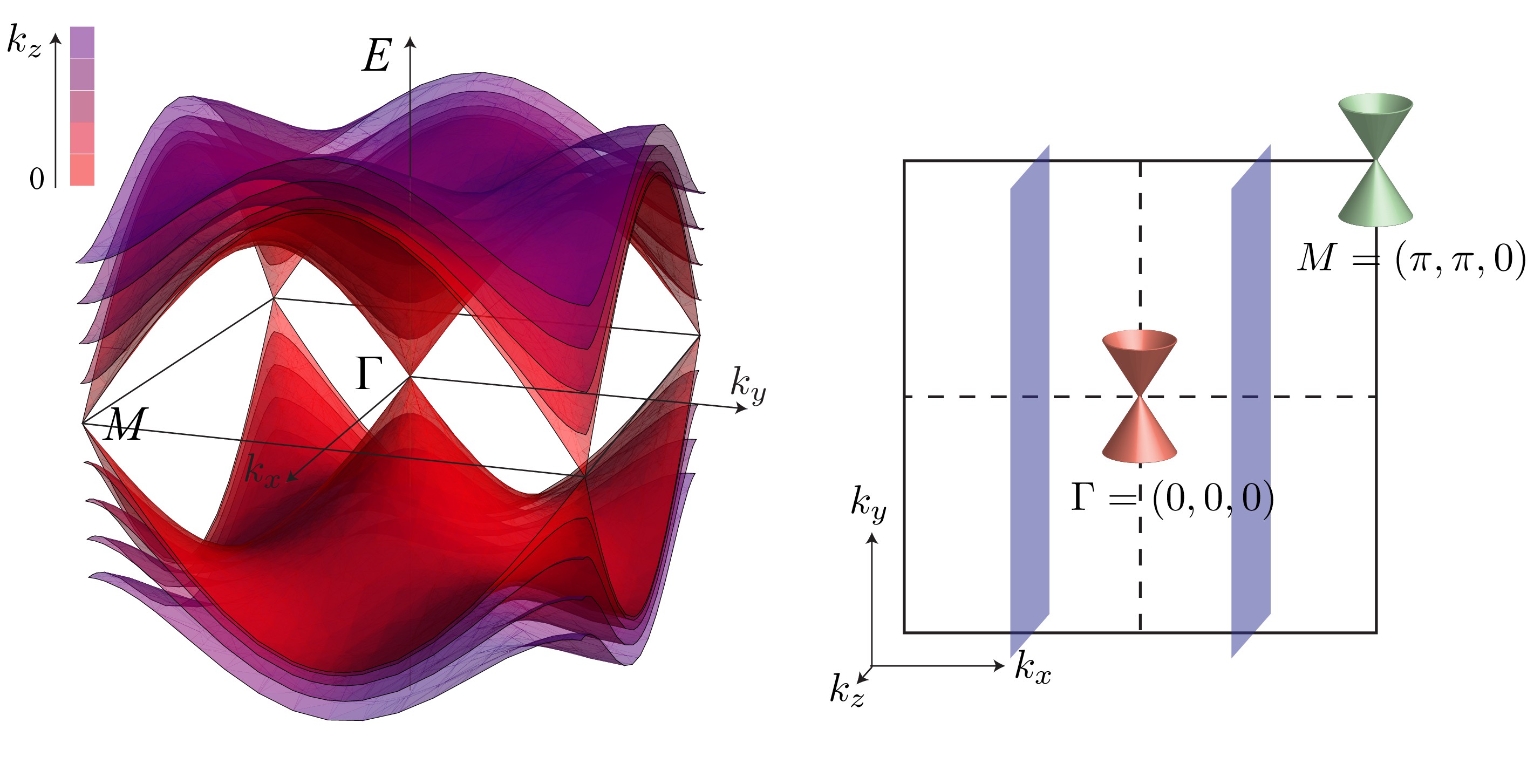}
\caption{Energy spectrum of the coupled Dirac wire model \eqref{WeylTBHam}.}\label{fig:Weylspectrum}
\end{figure}

The energy spectrum of the two-band model is given by $E_\pm({\bf k})=\pm\sqrt{|g(k_x,k_y)|^2+\hbar^2\tilde{v}^2k_z^2}$ (see figure~\ref{fig:Weylspectrum}).
It gives two linearly dispersing Weyl cones of opposite chiralities in the \BZ centered at $K^+_0=\Gamma=(0,0,0)$ and $K^-_0=M=(\pi,\pi,0)$. Near these points, the Hamiltonians are of the linear form $H(K_0^\pm+\delta{\bf k})=\hbar\delta{\bf k}^TV^\pm\vec\sigma+O(\delta k^2)$, where $\vec\sigma=(\sigma_x,\sigma_y,\sigma_z)$ are Pauli matrices acting on the $(\psi^\odot,\psi^\otimes)$ degrees of freedom. The velocity matrices are \begin{align}\hbar V^\pm=\left(\begin{array}{ccc}-t_1&\pm t_2&0\\-t_1&\mp t_2&0\\0&0&\hbar\tilde{v}\end{array}\right),\end{align} whose determinant's %$\det(\hbar V)=\pm2\hbar\tilde{v}t_1t_2$ 
sign decides the $\pm$ chirality of the Weyl fermion at $\Gamma$ and $M$, i.e.~the $\pm1$ Fermi surface Chern invariants~\cite{WanVishwanathSavrasovPRB11,Ashvin_Weyl_review,RMP}. %Expanding about the two Weyl points and ignoring higher order terms gives $H(X^{\pm} + \delta {\mathbf{k}})= -t_1 (\delta k_x + \delta k_y) \sigma_x  \mp t_2 (\delta k_x - \delta k_y) \sigma_y  + v \delta k_z \sigma_z$.
The \AFTR symmetries \eqref{AFTR} in the single-body picture are expressed under Fourier transformation as \begin{gather}\mathcal{T}_{11}\vec\psi_{\bf k}\mathcal{T}_{11}^{-1}=T_{11}({\bf k})\vec\psi_{-\bf k},\quad\mathcal{T}_{\bar{1}1}\vec\psi_{\bf k}\mathcal{T}_{\bar{1}1}^{-1}=T_{\bar{1}1}({\bf k})\vec\psi_{-\bf k},\nonumber\\T_{11}({\bf k})=\left(\begin{array}{ccc}0&-e^{i(k_x+k_y)}\\1&0\end{array}\right)\mathcal{K},\nonumber\\T_{\bar{1}1}({\bf k})=\left(\begin{array}{ccc}0&-e^{ik_y}\\e^{-ik_x}&0\end{array}\right)\mathcal{K},\label{AFTRk}\end{gather} where $\mathcal{K}$ is the complex conjugation operator. They satisfy the appropriate algebraic relations \eqref{AFTRalg} in momentum space \begin{gather}T_{11}(-{\bf k})T_{\bar{1}1}({\bf k})=T_{\bar{1}1}(-{\bf k})T_{11}({\bf k})=-e^{-ik_y}\nonumber\\T_{11}(-{\bf k})T_{\bar{1}1}({\bf k})^{-1}=T_{\bar{1}1}(-{\bf k})^{-1}T_{11}({\bf k})=e^{-ik_x}\end{gather} and the coupled wire model \eqref{BlochHam} is \AFTR symmetric \begin{align}T_{11}({\bf k})H({\bf k})&=H(-{\bf k})T_{11}({\bf k}),\nonumber\\T_{\bar{1}1}({\bf k})H({\bf k})&=H(-{\bf k})T_{\bar{1}1}({\bf k}).\label{WeylTBT11}\end{align} The Weyl points are at time reversal invariant momenta (\hypertarget{TRIM}{TRIM}) $K^\pm_0\equiv-K^\pm_0$ (modulo the reciprocal lattice $2\pi\mathbb{Z}^2$), and the \AFTR operators $T_{11}(K^\pm_0)=-i\sigma_y\mathcal{K}$ and $T_{\bar{1}1}(K^\pm_0)=\mp i\sigma_y\mathcal{K}$ square to minus one. Hence the Weyl points are not only protected by the non-vanishing Fermi surface Chern invariant but also the Kramers' theorem. In addition, the model is also $C_2$ symmetric \begin{align}C_2({\bf k})H({\bf k})=H(C_2{\bf k})C_2({\bf k})\label{WeylTBC2}\end{align} where the twofold symmetry \eqref{C2} is represented in the single-body picture by a diagonal matrix \begin{align}\mathcal{C}_2\vec\psi_{\bf k}\mathcal{C}_2^{-1}=C_2({\bf k})\vec\psi_{C_2\bf k},\quad C_2({\bf k})=\left(\begin{array}{*{20}c}i&0\\0&-ie^{-i(k_x+k_y)}\end{array}\right)\label{C2k}\end{align} (suppressing the screw phase $e^{-ik_za/2}$ in the continuum limit $a\to0$). It agrees with the fermion statistics \eqref{C2square} $C_2(-k_x,-k_y,k_z)C_2(k_x,k_y,k_z)=-1$, and the algebraic relations \eqref{C2Trelation} with the \AFTR operators \begin{align}C_2(-{\bf k})T_{11}({\bf k})&=-T_{11}(C_2{\bf k})^{-1}C_2({\bf k})\nonumber\\C_2(-{\bf k})T_{\bar{1}1}({\bf k})&=-T_{\bar{1}1}(C_2{\bf k})^{-1}C_2({\bf k})\end{align} for $C_2{\bf k}=(-k_x,-k_y,k_z)$.

\subsection{The anomalous Dirac semimetal}\label{sec:anomaly}
We notice that the coupled wire Dirac model \eqref{WeylTBHam} and its massless energy spectrum in figure~\ref{fig:Weylspectrum} are anomalous with respect to the \AFTR symmetries $\mathcal{T}_{11}$ and $\mathcal{T}_{\bar{1}1}$ as well as the $C_2$ symmetry if it is proper symmorphic and not a screw rotation. This means that it cannot be realized in a single-body three dimensional lattice system with the \AFTR or $C_2$ symmetries. In a sense, it is not surprising at all since the chiral Dirac strings that constitute \eqref{WeylTBHam} are themselves violating fermion doubling~\cite{Nielsen_Ninomiya_1981,NielsenNinomiyaPLB1981}. Here we further elaborate on the anomalous Dirac spectrum (figure~\ref{fig:Weylspectrum}) where the pair of Weyl points are separately located at two \TRIM $K^\pm_0$. We also comment on the non-trivial consequence of the anomaly and pave the path for later discussion on many-body interactions.

We begin with two 2D planes in momentum space parallel to $k_yk_z$ located at $k_x=\pm\pi/2$. They are represented by the two blue planes in figure~\ref{fig:Weylspectrum}. The \AFTR or $C_2$ symmetries require the Chern invariants \begin{align}\mathrm{Ch}_1=\frac{i}{2\pi}\int\mathrm{Tr}(P\partial_{k_y}P\partial_{k_z}P)dk_ydk_z\label{1stChern}\end{align} at $k_x=\pm\pi/2$ to be opposite, where $P({\bf k})=(\openone-H({\bf k})/|E({\bf k})|)/2$ is the projection operator onto the negative energy band. This is because the \AFTR symmetry is anti-unitary and preserves the orientation of the $k_yk_z$ plane, whereas $C_2$ is unitary but reverses the orientation of the $k_yk_z$ plane. (See appendix~\ref{sec:Chernapp} for a detailed proof.) On the other hand, the two Chern invariants along the two planes must differ by 1 because they sandwich a single Weyl point at $\Gamma$. This forces the Chern invariants to be a half integer $\mathrm{Ch}_1=\pm1/2$, which is anomalous.

While the $C_2$ anomaly can be resolved simply by doubling the unit cell and assuming it originates from a microscopic non-symmorphic screw axis, the \AFTR anomaly is stronger because the two antiferromagnetic combinations \eqref{AFTRalg} generate lattice translations and fix the unit cell size. There are three resolutions. \begin{enumerate}\item The \AFTR symmetries are broken by high energy degrees of freedom when $k_z$ is large. \item The spectrum in figure~\ref{fig:Weylspectrum} is the holographic 3D boundary spectrum of an \AFTR symmetric weak topological insulator in 4D. \item The spectrum is generated by strong many-body interaction non-holographically in 3D.\end{enumerate} Below we discuss the first two resolutions, and we leave the many-body interaction-enabled situation to section~\ref{sec:intenable}.

\subsubsection{Broken symmetries and coarse-graining}\label{sec:brokensymmetry}
The mapping between the original Dirac fermion model and the emergent Dirac fermion model from a coupled-wire construction can be qualitatively understood as a coarse-graining procedure. Here, the high-energy microscopic electronic degrees of freedom are integrated out. The procedure can be repeated indefinitely and resembles a real-space renormalization. For example, the gapless Dirac electronic structure of the coupled wire model can acquire a finite mass by symmetry-breaking dimerizations. These dimerizations can be arranged in a topological manner that spatially wind non-trivially around a collective vortex. These second-stage vortices can subsequently be assembled into an array similar to the previous construction except now with a longer lattice constant. The system again recovers a massless Dirac spectrum under inter-vortex electron tunneling in low-energy and long length scale. The mapping therefore establishes an equivalence between the continuous isotropic massless Dirac fermion and the semi-discrete anisotropic coupled Dirac wire model.

\begin{figure}[htbp]
\centering\includegraphics[width=0.45\textwidth]{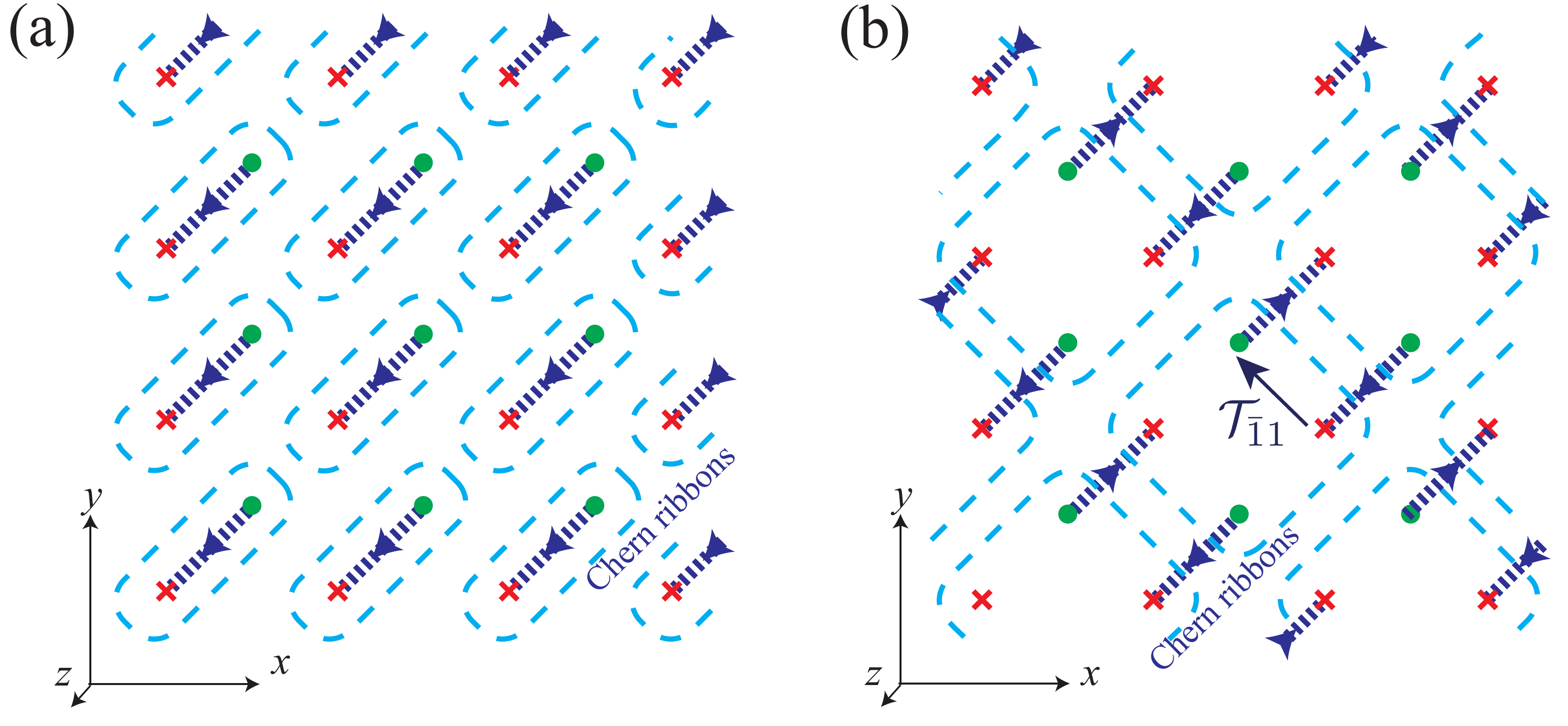}
\caption{Chiral Dirac channels ({\color{red}$\boldsymbol\times$} and {\color{green}$\bullet$}) realized on the edge of Chern insulating ribbons (dark blue directed lines) stacked along the $(\bar{1}10)$ normal direction.}\label{fig:Chernstack}
\end{figure}

\begin{figure}[htbp]
\centering\includegraphics[width=0.45\textwidth]{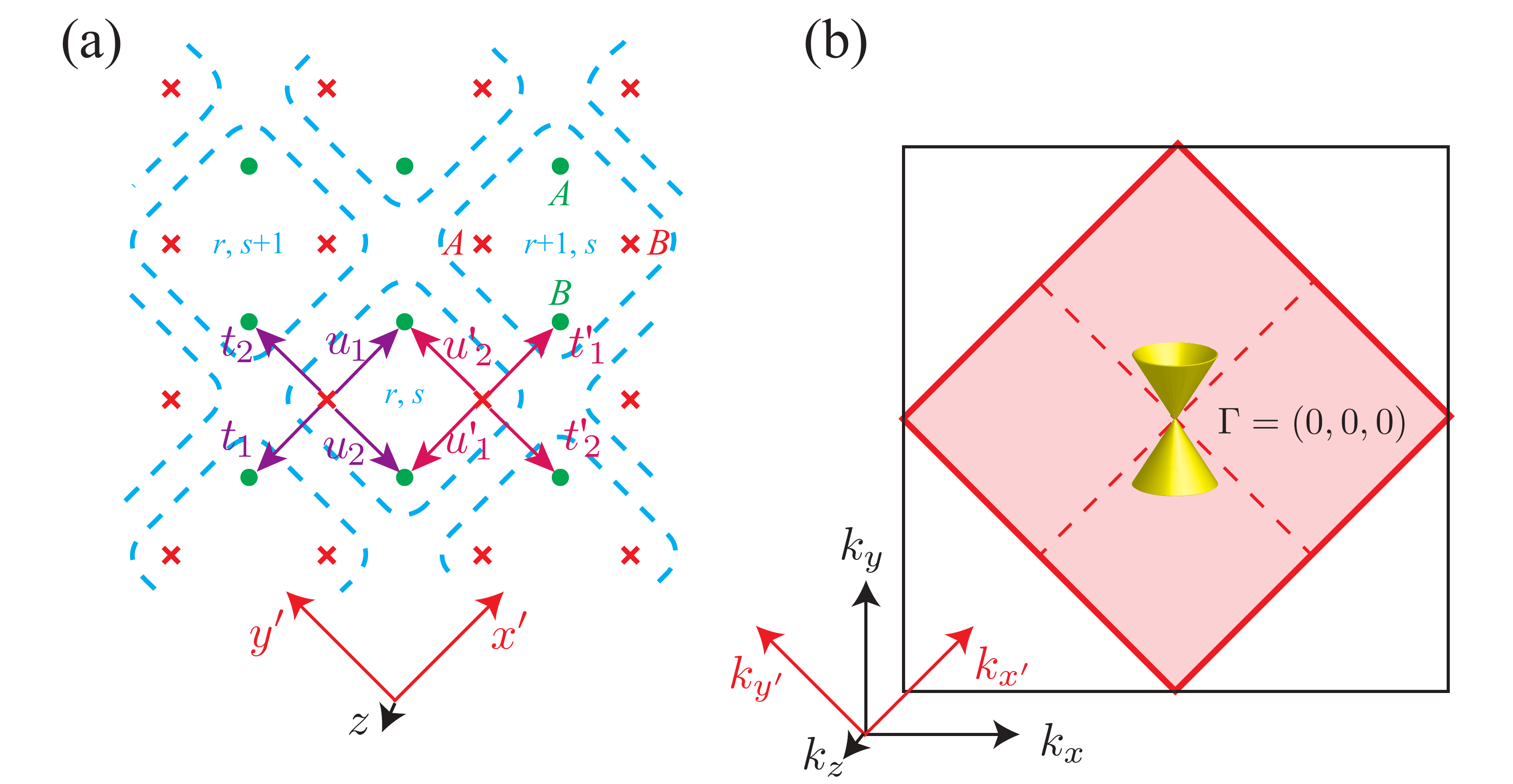}
\caption{(a) The massive AFTR and $C_2$ breaking coupled Dirac wire model. (b) The reduced Brillouin zone (BZ) after translation symmetry breaking where the two Weyl points collapse to a single Dirac point at $M$.}\label{fig:DiracTB}
\centering\includegraphics[width=0.48\textwidth]{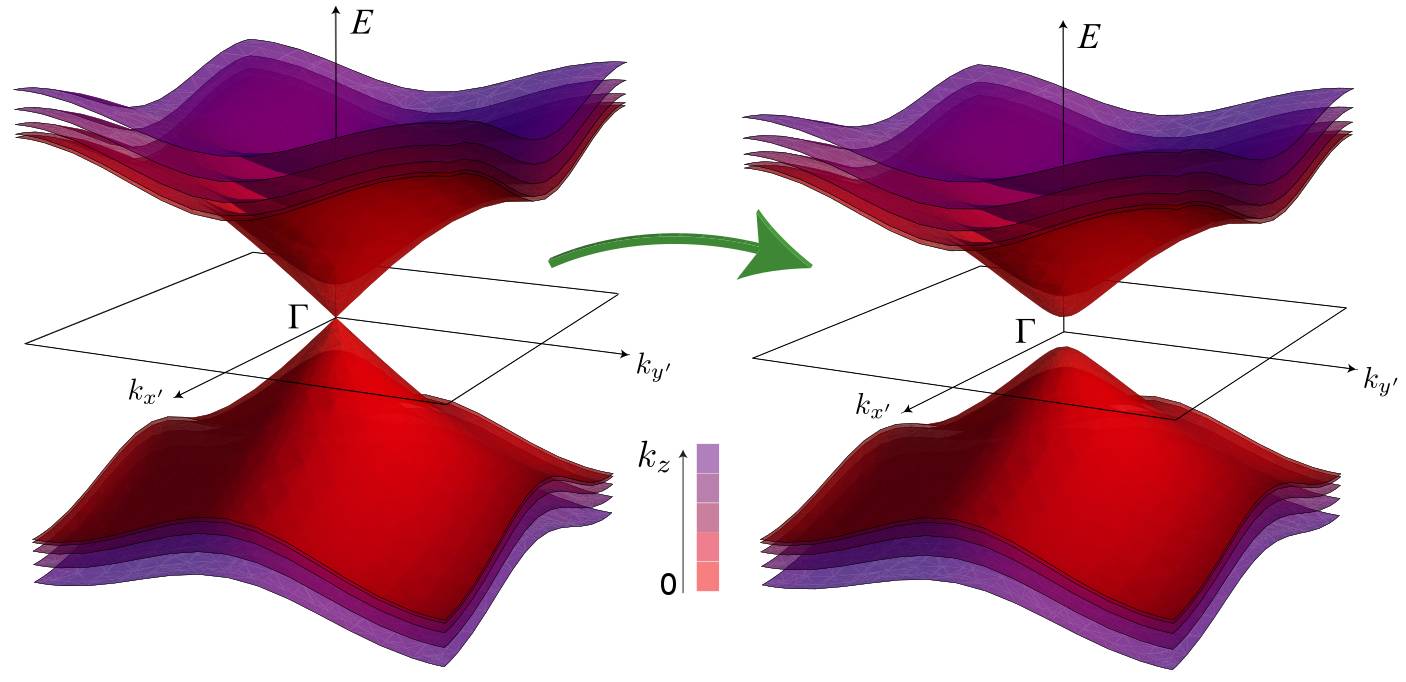}
\caption{Dirac mass gap $2|\Delta|$ introduced by AFTR and $C_2$ symmetry breaking dimerization $\Delta=\Delta_1+i\Delta_2$.}\label{fig:Diracmassjpg}
\end{figure}

In the present case when the chiral Dirac channels originate from vortex strings in an underlying microscopic Dirac insulator, the spatial modulation of mass parameters $m({\bf r})$ actually violate one of the \AFTR symmetries, $m({\bf r})^\ast\neq m({\bf r}+({\bf e}_x\pm{\bf e}_y)/2)$, where $\ast$ stands for complex conjugation. For instance, since all elliptic functions must contain at least two zeros and two poles in its periodic cell, the Jacobian elliptic mass function \eqref{Jacobielliptic} has longer periods than ${\bf e}_x$ and ${\bf e}_y$ in figure~\ref{fig:WeylTB}, and thus must break $\mathcal{T}_{11}$ or $\mathcal{T}_{\bar{1}1}$. The symmetry is broken only in the ultra-violet limit at large $k_z$ where the chiral Dirac line nodes meet the microscopic bulk band (see figure~\ref{fig:Diracstring}) at high energy $\sim|m({\bf r})|$. In fact, the above anomalous argument shows that {\em all} mass parameter configurations that produce the 3D vortex lattice array (figure~\ref{fig:vortexlattice}) must either (a) break both the \AFTR symmetries $\mathcal{T}_{11}$ and $\mathcal{T}_{\bar{1}1}$, or (b) preserve one but violate translation so that the unit cell is enlarged and the two Weyl points collapse onto each other in momentum space. (See figure~\ref{fig:Chernstack} and \ref{fig:DiracTB}.)

For instance, the microscopic system can be connected to a stack of Chern insulating ribbons (or lowest Landau levels) with alternating chiralities shown in figure~\ref{fig:Chernstack}. Instead of being supported by vortices of Dirac mass, the chiral Dirac wires are now realized as edge modes of Chern insulating strips. Each 2D ribbon (represented by thick dashed dark blue lines) is elongated in the out-of-paper $z$-direction but is finite along the $(110)$ direction and holds counter-propagating boundary chiral Dirac channels. The dark blue arrows represent the orientations of the Chern ribbons that accommodate the boundary Dirac channels with the appropriate propagating directions. Here the Chern ribbon pattern in figure~\ref{fig:Chernstack}(a) breaks both \AFTR axes. The pattern in figure~\ref{fig:Chernstack}(b) preserves $\mathcal{T}_{\bar{1}1}$. However, translation symmetry is also broken and the coupled Dirac wire model now has an enlarged unit cell (light blue dashed boxes) that consists of two pairs of counter-propagating chiral Dirac channels. All Chern ribbon patterns must break the $C_2$ symmetry about a Dirac wire because each wire is connected to one and only one Chern ribbon in a particular direction.

Now we go back to the vortex lattice generated by the Jacobian elliptic Dirac mass function $m({\bf r})$ in \eqref{Jacobielliptic} and consider its symmetries. For this purpose, we use the symmetry properties of the (rescaled) Jacobian elliptic function~\cite{ReinhardtWalker10} \begin{align}&\mathrm{sd}(x+iy)=-\mathrm{sd}(x+1+iy)=-\mathrm{sd}(x+iy+i)\nonumber\\&\mathrm{sd}\left(x+iy+\frac{1+i}{2}\right)=-i\frac{C}{\mathrm{sd}(x+iy)}\label{sdprop}\\&\mathrm{sd}(-x-iy)=-\mathrm{sd}(x+iy)\nonumber\end{align} where $C$ is some unimportant real constant that depends on the modulus of $\mathrm{sd}$ and will never appear in the mass function $m({\bf r})=m_0\mathrm{sd}(x+iy)/|\mathrm{sd}(x+iy)|$. We see from the minus sign in the first equation that the Jacobian elliptic function, and consequently the mass function, have primitive periods ${\bf e}_x\pm{\bf e}_y$ and therefore have a unit cell of size 2 (see figure~\ref{fig:DiracTB}(a)). Choosing $m_0=|m_0|e^{i\pi/4}$, we see from the second equation that $\mathcal{T}_{11}$ (or $\mathcal{T}_{\bar{1}1}$) is preserved (resp.~broken) \begin{align}m\left({\bf r}+\frac{{\bf e}_x\pm{\bf e}_y}{2}\right)=\pm m({\bf r})^\ast,\label{massT11}\end{align} and thus the parent Dirac Hamiltonian \eqref{DiracHam} is $\mathcal{T}_{11}$-symmetric \begin{align}\hat{T}H_{\mathrm{Dirac}}\left(-{\bf k},{\bf r}+\frac{{\bf e}_x+{\bf e}_y}{2}\right)\hat{T}^{-1}=H_{\mathrm{Dirac}}({\bf k},{\bf r}),\end{align} for $\hat{T}=is_y\mathcal{K}$. Lastly, the third property of \eqref{sdprop} entails the mass function $m({\bf r})=-m(C_2{\bf r})$ is odd under $C_2$, and consequently the parent Dirac Hamiltonian is (screw) rotation symmetric \begin{align}\hat{C}_2H_{\mathrm{Dirac}}(C_2{\bf k},C_2{\bf r})\hat{C}_2^{-1}=H_{\mathrm{Dirac}}({\bf k},{\bf r}),\label{massC2}\end{align} where $\hat{C}_2=is_z\mu_z$ (or microscopically $e^{-ik_za/2}is_z\mu_z$) anticommuting with the mass terms $m_1\mu_x+m_2\mu_y$ in $H_{\mathrm{Dirac}}$ (see \eqref{DiracHam}), and $C_2{\bf k}=(-k_x,-k_y,k_z)$, $C_2{\bf r}=(-x,-y,z)$.

Remembering that the coupled wire model \eqref{WeylTBHam} (figure~\ref{fig:WeylTB}) descended from a vortex lattice of the microscopic parent Dirac Hamiltonian \eqref{DiracHam}, the Dirac mass $m({\bf r})$ actually allows the model to carry fewer symmetries than the low-energy effective Hamiltonian \eqref{WeylTBHam} suggests. Now that the translation symmetry is lowered, the \BZ is reduced (see figure~\ref{fig:DiracTB}(b)) so that the two Weyl points now coincide at the origin $\Gamma$. This recovers an unanomalous Dirac semimetallic model \eqref{DiracHam0} around $(k_{x'},k_{y'})=(0,0)$. The fourfold degenerate Dirac point is protected and pinned at $\Gamma$ due to the remaining \AFTR symmetry $\mathcal{T}_{11}$ -- which takes the role of a spinful time reversal ($\hat{T}^2=-1$) in the continuum limit -- and the $C_2$ (screw) symmetry about the $z$-axis. However, if any of these symmetries is further broken, the fourfold degeneracy of the Dirac point is not protected (c.f.~the original continuum Dirac model \eqref{DiracHam}). Figure~\ref{fig:DiracTB}(a) shows a dimerized coupled Dirac wire model that introduces a finite mass for the Dirac fermion. We label the Dirac fermion operators as $\psi_{r,s}^{\mu,\sigma}$, for $\sigma=\odot,\otimes$ the chirality, $\mu=A,B$ the new sublattice label, and $(r,s)$ label the coordinates of the unit cell according to the $45^\circ$-rotated $x',y'$-axes. \begin{align}\mathcal{H}'=&\sum_{r,s}\sum_{\mu=A,B}\hbar\tilde{v}\left({\psi_{r,s}^{\mu,\odot}}^\dagger k_z\psi_{r,s}^{\mu,\odot}-{\psi_{r,s}^{\mu,\otimes}}^\dagger k_z\psi_{r,s}^{\mu,\otimes}\right)\nonumber\\&+iu_1{\psi_{r,s}^{A,\odot}}^\dagger\psi_{r,s}^{A,\otimes}-iu'_1{\psi_{r,s}^{B,\odot}}^\dagger\psi_{r,s}^{B,\otimes}+h.c.\nonumber\\&-u_2{\psi_{r,s}^{B,\odot}}^\dagger\psi_{r,s}^{A,\otimes}+u'_2{\psi_{r,s}^{A,\odot}}^\dagger\psi_{r,s}^{B,\otimes}+h.c.\label{DiracTBHam}\\&-it_1{\psi_{r-1,s}^{A,\odot}}^\dagger\psi_{r,s}^{A,\otimes}+it'_1{\psi_{r+1,s}^{B,\odot}}^\dagger\psi_{r,s}^{B,\otimes}+h.c.\nonumber\\&+t_2{\psi_{r,s+1}^{B,\odot}}^\dagger\psi_{r,s}^{A,\otimes}-t'_2{\psi_{r,s-1}^{A,\odot}}^\dagger\psi_{r,s}^{B,\otimes}+h.c.\nonumber\end{align} For instance, the model is identical to the \AFTR and $C_2$ symmetric one in \eqref{WeylTBHam} when $t_j=t'_j=u_j=u'_j$ for $j=1,2$. However, when the symmetries are broken, these hopping parameters do not have to agree.

The Bloch band Hamiltonian after Fourier transformation is \begin{gather}H({\bf k})=\left(\begin{array}{*{20}c}\hbar\tilde{v}k_z\openone&h(k_{x'},k_{y'})\\h(k_{x'},k_{y'})^\dagger&-\hbar\tilde{v}k_z\openone\end{array}\right),\label{DiracBloch}\\h(k_{x'},k_{y'})=\left(\begin{array}{*{20}c}iu_1-it_1e^{-ik_{x'}}&u'_2-t'_2e^{-ik_{y'}}\\-u_2+t_2e^{ik_{y'}}&-iu'_1+it'_1e^{ik_{x'}}\end{array}\right)\nonumber\end{gather} where the $2\times2$ identity matrix $\openone$ and $h(k_{x'},k_{y'})$ acts on the sublattice $\mu=A,B$ degrees of freedom, and $-\pi\leq k_{x'},k_{y'}\leq\pi$ are the rotated momenta. We perturb about the Dirac fixed point by introducing the dimerizations $\Delta_j$ \begin{align}t_j=t'_j=u_j-\Delta_j=u'_j-\Delta_j\end{align} for $j=1,2$. About the $\Gamma=(0,0,0)$ point, \begin{align}H(\Gamma+\delta{\bf k})=&\hbar\tilde{v}\delta k_z\sigma_z-t_1\delta k_{x'}\sigma_x-t_2\delta k_{y'}\sigma_y\mu_x\nonumber\\&-\Delta_1\sigma_y\mu_z+\Delta_2\sigma_y\mu_y+O(\delta k^2).\label{DiracHamwire}\end{align} See figure~\ref{fig:Diracmassjpg} for its massive spectrum.

\begin{figure}[htbp]
\centering\includegraphics[width=0.4\textwidth]{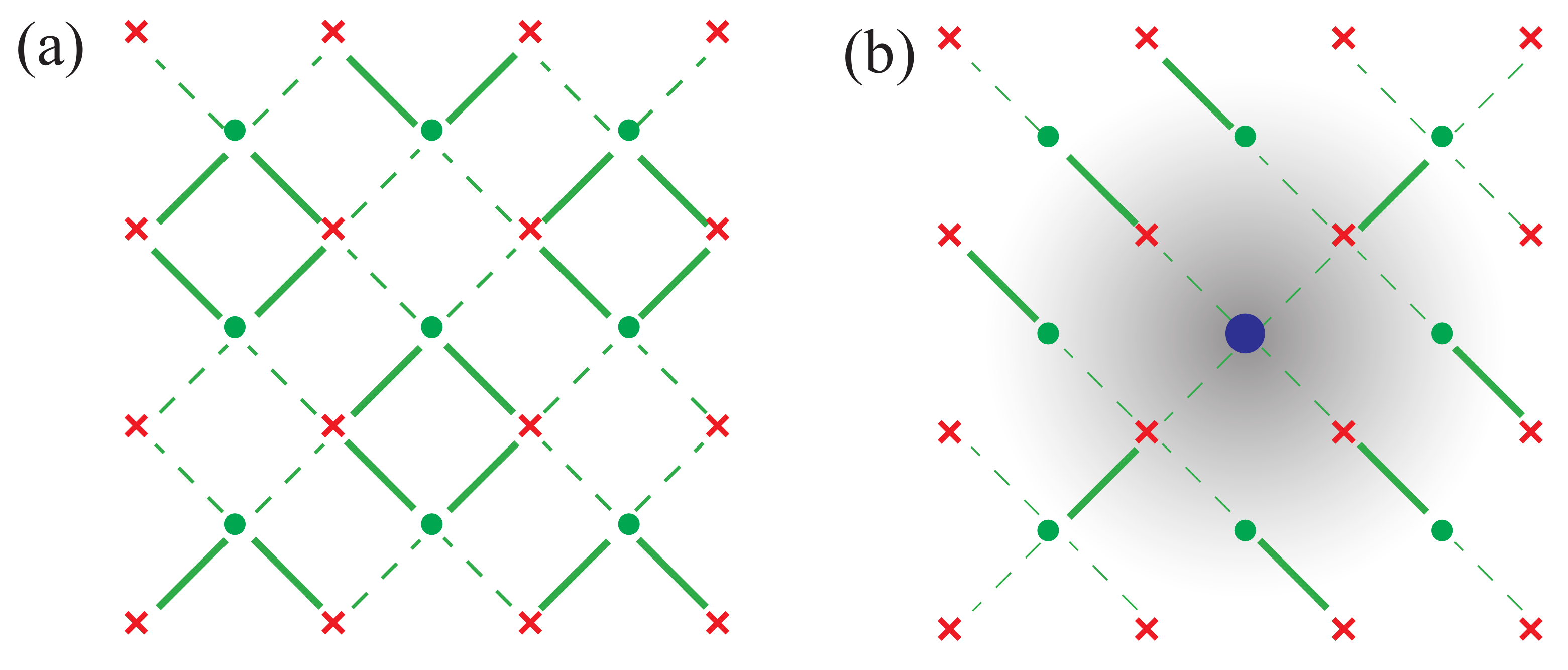}
\caption{(a) Dimerized model of a massive Dirac fermion. (b) Vortex of dimerizations $\Delta=\Delta_1+i\Delta_2$ that leaves behind a massless localized chiral Dirac channel (blue dot).}\label{fig:dimerization}
\end{figure}

Here the \AFTR symmetry $\mathcal{T}_{11}$ and the twofold rotation $\mathcal{C}_2$ are represented in the single-body picture by \begin{align}T_{11}({\bf k})&=\left(\begin{smallmatrix}0&0&-e^{ik_x}&0\\0&0&0&-1\\1&0&0&0\\0&e^{ik_x}&0&0\end{smallmatrix}\right)\mathcal{K},\nonumber\\C_2({\bf k})&=\left(\begin{smallmatrix}i&0&0&0\\0&ie^{-i(k_x+k_y)}&0&0\\0&0&-ie^{-ik_x}&0\\0&0&0&-ie^{-ik_y}\end{smallmatrix}\right)\end{align} (again suppressing the $C_2$ screw phase $e^{-ik_za/2}$ in the continuum limit $a\to0$). In the small $k_x,k_y$-limit, $T_{11}(0)=-i\sigma_y\mathcal{K}$ and $C_2(0)=i\sigma_z$. It is straightforward to check that the dimerization $\Delta_2$ preserves $\mathcal{T}_{11}$ while both $\Delta_1,\Delta_2$ breaks $C_2$.

Since the coupled wire model \eqref{DiracHamwire} and the parent continuum Dirac model \eqref{DiracHam} have the same matrix and symmetry structure, we can apply the same construction we discussed before to the new coarse-grained model \eqref{DiracHamwire}. For instance, the non-competing dimerizations $\Delta({\bf r})=\Delta_1({\bf r})+i\Delta_2({\bf r})$ can spatially modulate and form vortices in a longer length scale. Figure~\ref{fig:dimerization}(b) shows a dimerization pattern that corresponds to a single vortex in $\Delta$. The solid (dashed) lines represent strong (resp.~weak) backscattering amplitudes. In the fully dimerized limit where the dashed bonds vanish, all Dirac channels are gapped except the one at the center (showed as a blue dot). In the weakly dimerized case, there is a collective chiral Dirac channel whose wave function is a superposition of the original channels and is exponentially localized at the $\Delta$-vortex core, but now with a length scale longer than that of the original $m$-vortex lattice. These collective chiral Dirac $\Delta$-vortices can themselves form a coupled array, like \eqref{WeylTBHam}, and give a Dirac semimetal of even longer length scale. The single-body coupled vortex construction is therefore a coarse-graining procedure that recovers equivalent emergent symmetries at each step. \begin{align}\begin{diagram}\mbox{Dirac semimetal}&\pile{\rTo^{\mbox{\small mass vortices}}\\\lTo_{\mbox{\small coupled wire model}}}&\mbox{chiral Dirac strings}\end{diagram}\end{align}

\subsubsection{Holographic projection from 4D}\label{sec:holproj4D}
The coupled wire model \eqref{WeylTBHam} with two \AFTR axes can be supported by a weak topological insulator (\hypertarget{WTI}{WTI}) in four dimensions. Instead of realizing the chiral Dirac channels using mass vortices of a 3D Dirac semimetal, they can be generated as edge modes along the boundaries of 2D Chern insulators (or lowest Landau levels). The 4D \WTI is constructed by stacking layers of Chern insulators parallel to the $zw$-plane along the $x$ and $y$ directions. The Chern layers $L_{\bf r}$, labeled by the checkerboard lattice vector ${\bf r}=r_x{\bf e}_x+r_y{\bf e}_y$ on the $xy$-plane, have alternating orientations so that $\mathrm{Ch}[L_{\bf r}]=1$ if $r_x,r_y$ are integers and $\mathrm{Ch}[L_{\bf r}]=-1$ if $r_x,r_y$ are half-integers.  The model therefore carries both \AFTR symmetries $\mathcal{T}_{11}$ and $\mathcal{T}_{\bar{1}1}$ as well as the $C_2$ rotation about $zw$, and when cleaved along a 3D hyper-surface normal to $w$, it generates the array of alternating chiral Dirac channels in figure~\ref{fig:WeylTB}.

The 4D \WTI model can also be regarded as a stack of 3D antiferromagnetic topological insulators (\hypertarget{AFTI}{AFTI})~\cite{MongEssinMoore10}. Restricting to the 3D hyperplane normal to $-{\bf e}_x+{\bf e}_y$, this model consists of alternating Chern insulating layers parallel to the $wz$-plane stacked along the ${\bf e}_x+{\bf e}_y$ direction. This 3D model describes an \AFTI with a non-trivial $\mathbb{Z}_2$ index. For instance along the boundary surfaces normal to $w$ or $z$ that preserve the antiferromagnetic symmetry $\mathcal{T}_{11}$, the model leaves behind a 2D array of alternating chiral Dirac wires. The uniform nearest wire backscattering term $t_1$ (see \eqref{WeylTBHam}) introduces a linear dispersion along the $11$-direction and gives rise to a single massless surface Dirac cone spectrum at a \TRIM on the boundary of the surface \BZ where $\mathcal{T}_{11}^2=-1$. The 4D \WTI model is identical to stacking these 3D \AFTI along the $\bar{1}1$-off-diagonal direction $-{\bf e}_x+{\bf e}_y$. A more detailed discussion on coupled wire constructions of a 4D strong and weak topological insulator can also be found in Ref.~\onlinecite{ParkTeoGilbertappearsoon}.

%A single wire with a chiral Dirac mode running through it is not allowed. Here we discuss what are the possible sources of chiral Dirac modes in our model. There are two possible sources, this three-dimensional array of chiral wires is the hypersurface of a four-dimensional Topological Insulator. The other possibility is that there are strips of Pfaffians, and each chiral Dirac mode (central charge c=1, conductance $\sigma$ = 1) is a combination of two adjacent edge modes of the Pfaffian strip (central charge c=1/2, conductance $\sigma$ = 1/2) as shown in fig.x. \textcolor{red}{insert a figure later}. Both of these cases preserve both of the TR operators $\mathcal{T}_{1 \bar{1}}$ and $\mathcal{T}_{1 {1}}$.

\subsection{Surface Fermi arcs}\label{sec:fermiarc1}

\subsubsection{AFTR breaking surfaces}\label{sec:fermiarcAFTRbreaking}

\begin{figure}[htbp]
\centering\includegraphics[width=0.4\textwidth]{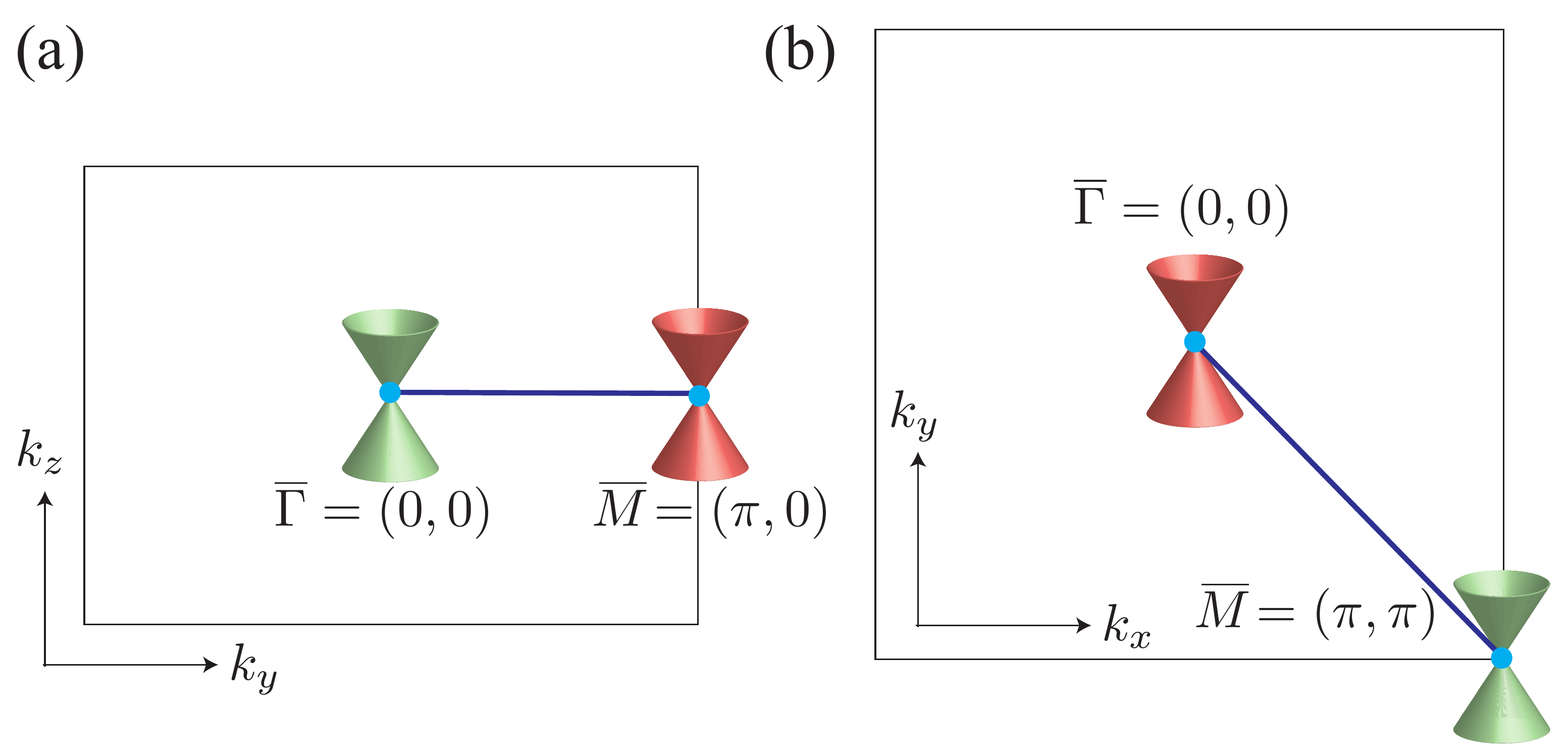}
\caption{Fermi arcs (blue lines) joining projected Weyl points on the surface Brillouin zones along (a) the $(100)$ surface and (b) the $(001)$ surface.}\label{fig:fermiarc1}
\end{figure}
We discuss the surface states of the coupled Dirac wire model \eqref{WeylTBHam}. Similar to the boundary surface of a translation symmetry protected Dirac semimetal (or more commonly called a Weyl semimetal), there are Fermi arcs connecting the surface-projected Weyl points~\cite{WanVishwanathSavrasovPRB11,Ashvin_Weyl_review,RMP}. First we consider the $(100)$ surface normal to $x$-axis (see figure~\ref{fig:WeylTB}). We assume the boundary cuts between unit cells and set the Fermi energy at $\varepsilon_f=0$. At $k_z=0$ and given a fixed $k_y\in(-\pi,\pi)$, the tight-binding model \eqref{BlochHam} is equivalent to the Su-Schriffer-Heeger model~\cite{SSH} or a 1D class AIII topological insulator~\cite{SchnyderRyuFurusakiLudwig08,Kitaevtable08} along the $x$-direction protected by the chiral symmetry $\sigma_zH(k_x)=-H(k_x)\sigma_z$. It is characterized by the winding number \begin{align}w(k_y)&=\frac{i}{2\pi}\int_{-\pi}^\pi \frac{1}{g(k_x,k_y)}\frac{\partial g(k_x,k_y)}{\partial k_x}dk_x\\&=\left(1+\mathrm{sgn}(k_y t_1/t_2)\right)/2.\nonumber\end{align} When $t_1,t_2$ have the same (or opposite) sign, the quasi-1D model is topological along the positive (resp.~negative) $k_y$-axis and thus carries a boundary zero mode. This corresponds to the Fermi line joining the two surface projected Weyl points at $\overline{\Gamma}$ and $\overline{M}$ (see figure~\ref{fig:fermiarc1}(a)). As the zero modes have a fixed chirality according to $\sigma_z$, they propagate uni-directionally with the dispersion $E(k_z)=\hbar\tilde{v}k_z\sigma_z$. The cleaving surface breaks \AFTR and $C_2$ symmetries, and so does the Fermi arc in figure~\ref{fig:fermiarc1}(a). For instance, any one of the \AFTR symmetries maps the boundary surface to an inequivalent one that cuts through unit cells instead of between them. As a result, the Fermi arc will connect the Weyl points along the opposite side of the $k_y$-axis for this surface. 

The $(010)$ surface Fermi arc structure is qualitatively equivalent to that of the $(100)$ surface. The $(110)$ and $(1\bar{1}0)$ surfaces that cleave along the diagonal and off-diagonal axes (see figure~\ref{fig:WeylTB}) respectively preserve the \AFTR symmetries $\mathcal{T}_{11}$ and $\mathcal{T}_{\bar{1}1}$. There are no protected surface Fermi arcs because the two bulk Weyl points project onto the same point on the surface Brillouin zone. Lastly, we consider the $(001)$ surface normal to the $z$-axis, which is the direction of the chiral Dirac strings that constitute the coupled wire model. A chiral Dirac channel cannot terminate on the boundary surface. In a single-body theory, it must bend and connect with an adjacent counter-propagating one. Although the $(001)$ plane is closed under the $C_2$ as well as both the \AFTR symmetries, the surface bending of Dirac channels must violate at least one of them. Here we consider the simplest case where the counter-propagating pair of Dirac channels within a unit cell re-connects on the boundary surface. This boundary is equivalent to a domain wall interface separating the Dirac semimetal \eqref{WeylTBHam} from an insulator where Dirac channels backscatters to their counter-propagating partner within the same unit cell. 

The domain wall Hamiltonian takes the form of a differential operator
\begin{align}\hat{\mathcal{H}}=&\sum_{m,j}-i\hbar\tilde{v}\left({\psi_{m,j}^\odot}^\dagger\partial_z\psi_{m,j}^\odot-{\psi_{m,j}^\otimes}^\dagger\partial_z\psi_{m,j}^\otimes\right)\label{WeylTBHamwall}\\&+it_1\left({\psi_{m,j}^\odot}^\dagger\psi_{m,j}^\otimes+\theta(z){\psi_{m-1,j-1}^\odot}^\dagger\psi_{m,j}^\otimes\right)+h.c.\nonumber\\&+t_2\theta(z)\left({\psi_{m-1,j}^\odot}^\dagger\psi_{m,j}^\otimes+{\psi_{m,j-1}^\odot}^\dagger\psi_{m,j}^\otimes\right)+h.c.\nonumber\end{align} by replacing $k_z\leftrightarrow-i\partial_z$ in \eqref{WeylTBHam}. Here $\theta(z)$ can be the unit step function or any function that asymptotically approaches 1 for $z\to\infty$ or 0 for $z\to-\infty$. The model therefore describes the Dirac semimetal \eqref{WeylTBHam} for positive $z$, and an insulator for negative $z$ where Dirac channels are pair annihilated within a unit-cell by $t_1$. After a Fourier transformation, the Bloch Hamiltonian $\hat{H}(k_x,k_y)$ is identical to \eqref{BlochHam} by replacing $k_z\leftrightarrow-i\partial_z$ and $g(k_x,k_y,z)=it_1(1+\theta(z)e^{-i(k_y+k_x)})+t_2\theta(z)(e^{-ik_x}+e^{-ik_y})$. Given any fixed $k_x,k_y$, the differential operator $\hat{H}(k_x,k_y)$ is identical to the Jackiw-Rebbi model~\cite{JackiwRebbi76}. Deep in the insulator, $g(k_x,k_y,z\to-\infty)=it_1$. There is an interface zero mode at the surface domain wall if $g$ changes sign, i.e.~if $g(k_x,k_y,z\to\infty)=|g|e^{i\varphi}$ has argument $\varphi=-\mathrm{sign}(t_1)\pi/2$. When $\varepsilon_f=0$, the zero modes trace out a Fermi arc that connects the two surface projected Weyl points (see figure~\ref{fig:fermiarc1}(b)).

\begin{figure}[htbp]\centering\includegraphics[width=0.4\textwidth]{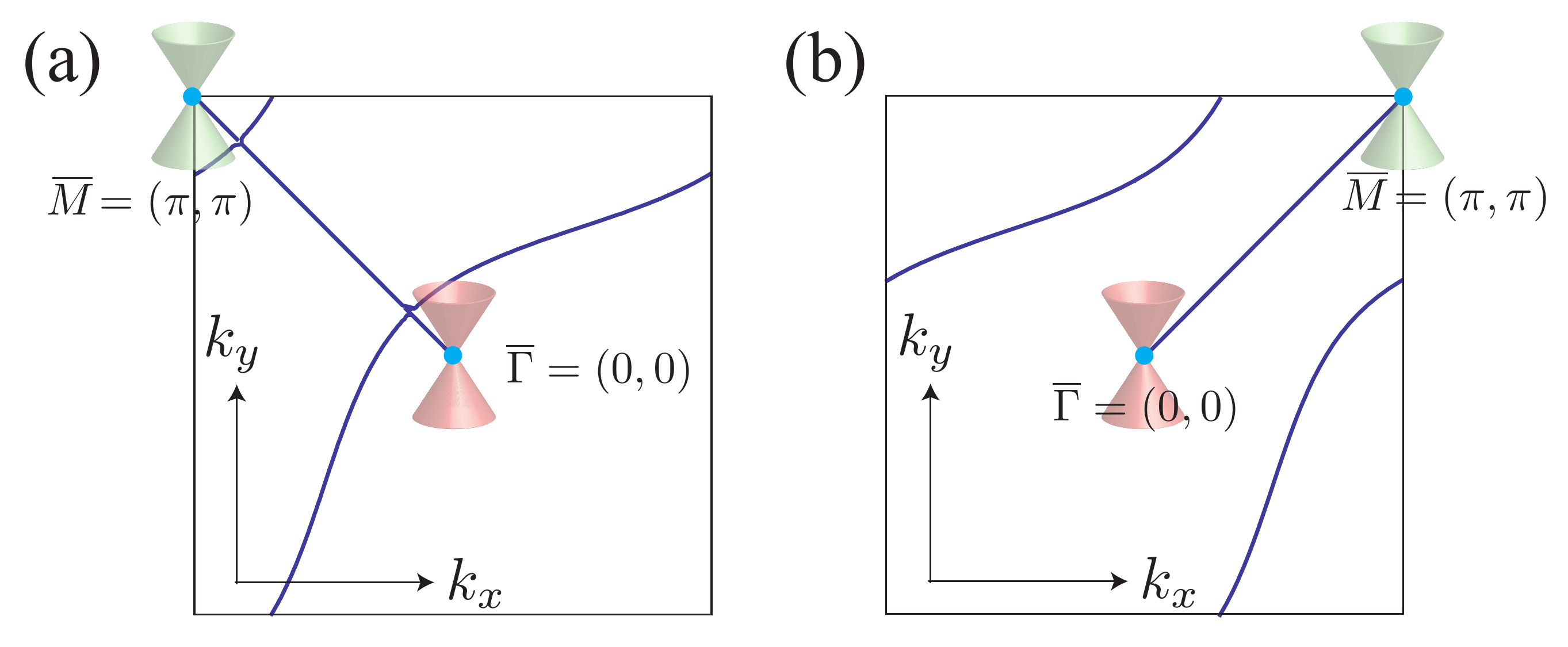}\caption{Fermi arcs (blue lines) on the $(001)$ surface with alternative boundary conditions (a) $g(k_x,k_y)=-it_1$ and (b) $g(k_x,k_y)=-t_2e^{-ik_y}$ in the insulating domain, for $t_2/t_1=2$.}\label{fig:fermiarc2}\end{figure}

We notice that in the insulating phase (or on the boundary surface), Dirac wires can be backscattered with a different phase and dimerized out of the unit cell. These different boundary conditions correspond to distinct surface Fermi arc patterns. Figure~\ref{fig:fermiarc2} shows two alternatives. (a) shows the zero energy arcs when intra-cell backscattering reverses sign $t_1\to-t_1$ in the insulating domain. (b) shows a case when the dimerization is taken along the off-diagonal axis. These inequivalent boundary conditions differ by some three dimensional integer quantum Hall states, which correspond to additional chiral Fermi arcs that wrap non-trivial cycles around the 2D toric surface Brillouin zone.

\subsubsection{AFTR preserving surfaces}\label{sec:fermiarcAFTRpreserving}

We also notice that the Fermi arc structures in figures~\ref{fig:fermiarc1}(b) and \ref{fig:fermiarc2} are allowed because both the \AFTR symmetries $\mathcal{T}_{11}$, $\mathcal{T}_{\bar{1}1}$ and the $C_2$ symmetry are broken by the insulating domain. Any dimerization that preserves only one of $\mathcal{T}_{11}$ and $\mathcal{T}_{\bar{1}1}$ necessarily breaks translation symmetry, and corresponds to an enlarged unit cell and a reduced Brillouin zone (c.f.~figure~\ref{fig:Chernstack} and \ref{fig:DiracTB}). As a result, the two Weyl points would now collapse onto the same $\overline{\Gamma}$ point. Any momentum plane that contains the $k_z$-direction and avoids the $\Gamma$ point must have trivial Chern invariant, because it could always be deformed (while containing the $k_z$-direction and avoiding the $\Gamma$ point) to the reduced Brillouin zone boundary, where its Chern invariant would be killed by the \AFTR symmetry. %There would therefore be no protected surface Fermi arcs.

However, the trivial bulk Chern invariant does not imply the absence of surface state. This can be understood by looking at the surface boundary in real space. Here, we assume the Dirac strings that constitute the coupled wire model \eqref{WeylTBHam} are supported by vortices of an underlying Dirac mass (see figure~\ref{fig:vortexlattice} and eq.\eqref{DiracHam}). The semimetallic coupled wire model terminates along the $xy$-plane against vacuum, which is modeled by the Dirac insulator $H_{\mathrm{vacuum}}=\hbar v{\bf k}\cdot\vec{s}\mu_z+m_0\mu_x$, say with $m_0>0$. Recall from \eqref{massT11} that the Dirac mass vortex configuration \eqref{Jacobielliptic} is \AFTR symmetric along the $\mathcal{T}_{11}$-directions. The Dirac insulating vacuum is symmetric under local \TR as well as continuous translation. It however breaks the screw rotation symmetry $\hat{C}_2=is_z\mu_z$, but we here only focus on the \AFTR symmetry.

\begin{figure}[htbp]
\centering\includegraphics[width=0.3\textwidth]{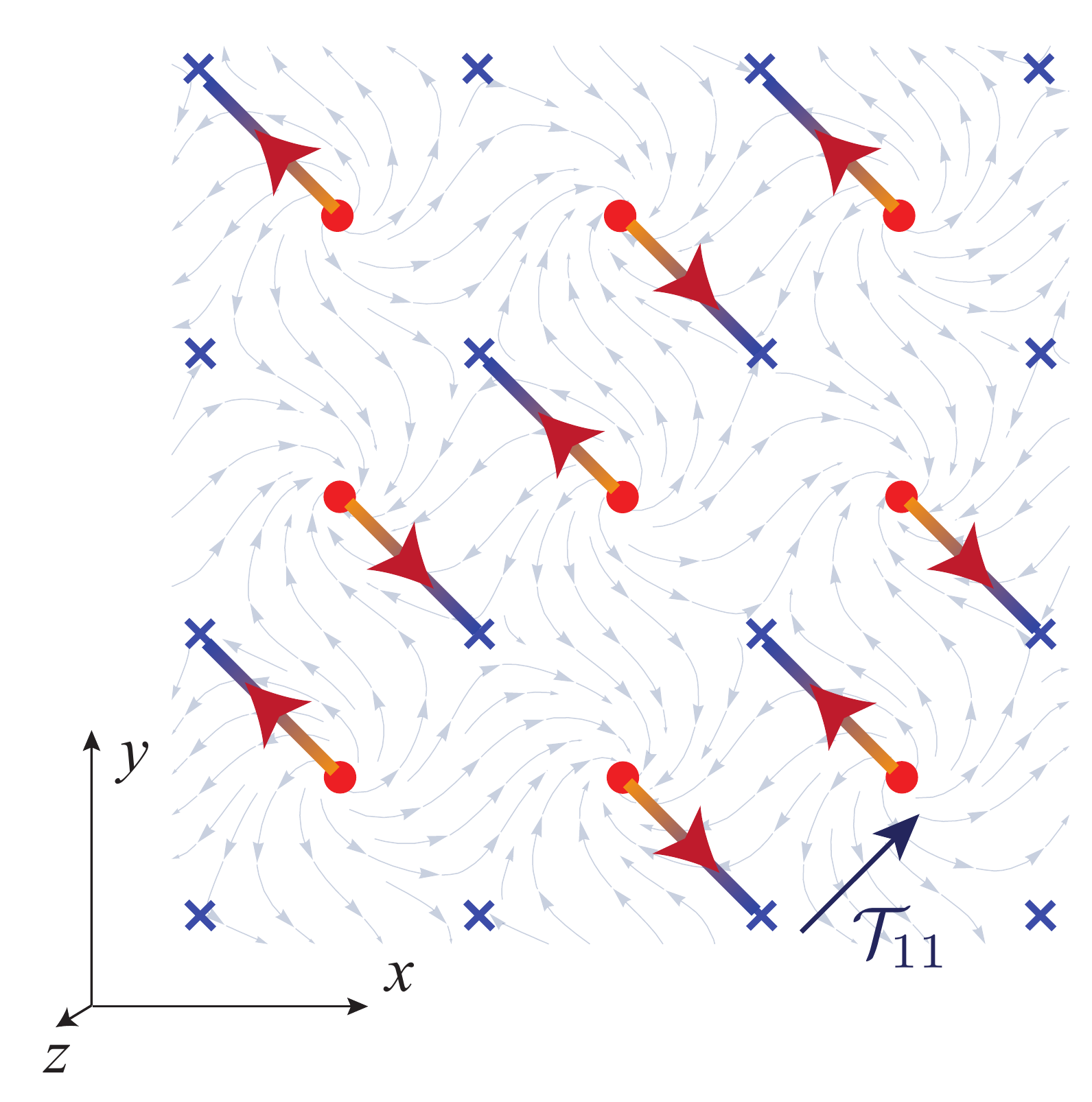}
\caption{Surface chiral Dirac channels of the coupled wire model \eqref{WeylTBHam} terminated along the $xy$ plane.}\label{fig:SurfaceStates1bdy}
\end{figure}

The surface boundary supports chiral Dirac channels that connect the chiral Dirac strings in the semimetallic bulk that are normal to the surface. The surface channels are shown in figure~\ref{fig:SurfaceStates1bdy}. The {\color{blue}$\times$} ({\color{red}$\bullet$}) represent chiral vortices in the bulk that direct electrons away from (resp.~onto) the surface. The vector field represents the Dirac mass $m({\bf r})=m_x({\bf r})+im_y({\bf r})$ modulation in the semimetallic bulk near the surface. The surface Dirac line channels~\cite{TeoKane} -- shown by directed lines connecting the bulk Dirac strings {\color{blue}$\times$}, {\color{red}$\bullet$} -- are located where the \TR symmetric Dirac mass $m_x$ changes sign across the surface boundary and the \TR breaking Dirac mass $m_y$ flips sign across the line channels along the surface. In other words, they are traced out of points on the surface where $m_x<0$ and $m_y=0$. Each of these surface channels carries a chiral Dirac electronic mode that connects the bulk chiral Dirac vortices. They can couple through inter-channel electron tunneling, but the collective gapless surface state cannot be removed from low-energy by dimerization without breaking the \AFTR symmetry $\mathcal{T}_{11}$. %The surface state is, in a sense, half of that of a weak topological insulator (or 3D stack of quantum spin Hall layers). 

{}

\section{Many-body interacting variations}\label{sec:interaction}

We discuss the effect of strong many-body interactions in a Dirac semimetal in three dimensions. Before we do so, it is worth stepping back and reviewing the two dimensional case in order to illustrate the issue and idea that will be considered and generalized in three dimensions. The massless Dirac fermion with $H=\hbar v(k_xs_y-k_ys_x)$ that appears on the surface of a topological insulator~\cite{HasanKane10,QiZhangreview11,HasanMoore11,RMP} is protected by time reversal (TR) and charge $U(1)$ symmetries and is anomalous. This means that there is no single-body energy gap opening mass term that preserves the symmetries, and there is no single-body fermionic lattice model in two dimensions that supports a massless Dirac fermion without breaking the symmetries. Neither of these statements hold true in the many-body setting. The surface Dirac fermion can acquire a \TR and charge $U(1)$ preserving many-body interacting mass.~\cite{WangPotterSenthilgapTI13,ChenFidkowskiVishwanath14,MetlitskiKaneFisher13b,BondersonNayakQi13} Consequently, this also enables a massless symmetry preserving Dirac fermion in a pure 2D system without holographically relying on a semi-infinite 3D topological bulk. For instance, one can take a quasi-2D topological insulator slab with finite thickness and remove the Dirac fermion on one of the two surfaces by introducing an interacting mass gap. This leaves a single massless Dirac fermion on the opposite surface without breaking symmetries.

A massless Dirac fermion in three dimensional semimetallic materials can be protected in the single-body picture by screw rotation, time reversal and charge $U(1)$ symmetries (see reviews Ref.~\onlinecite{Ashvin_Weyl_review,RMP,ArmitageMeleVishwanath16} and section~\ref{sec:DiracSemimetal}). From a theory point of view, it can be supported on the 3D boundary of a 4D weak topological insulator, where the two Weyl fermions are located at distinct time reversal invariant momenta (recall figure~\ref{fig:Weylspectrum} and section~\ref{sec:holproj4D} for the antiferromagnetic case). In this case, the massless fermions are protected by translation, time reversal and charge $U(1)$ symmetries. In this section, we address the following issues. (1) We show by explicitly constructing an exactly solvable coupled wire model that the 3D Dirac fermion can acquire a many-body interacting mass while preserving all symmetries. (2) We show in principle that an antiferromagnetic time reversal (AFTR) symmetric massless 3D Dirac system with two Weyl fermions separated in momentum space can be enabled by many-body interactions without holographically relying on a higher dimensional topological bulk.

\begin{figure}[htbp]
\centering\includegraphics[width=0.3\textwidth]{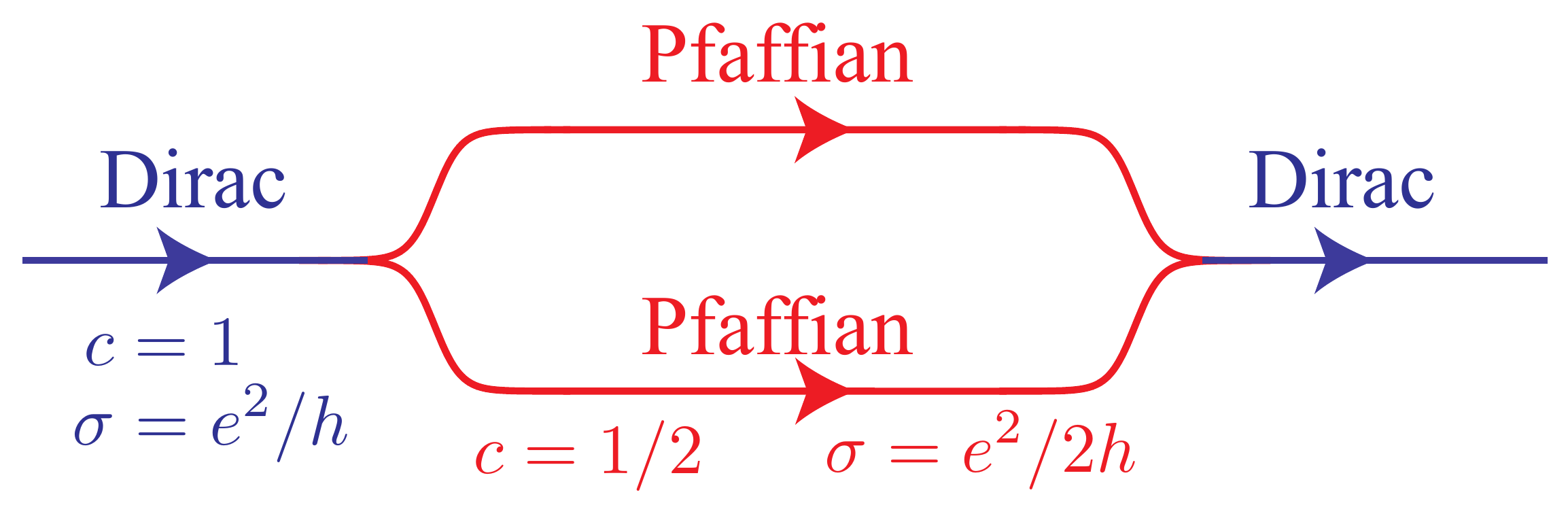}
\caption{Gluing and splitting a pair of chiral Pfaffian 1D channels into and from a chiral Dirac channel.}\label{fig:glueingsplitting}
\end{figure}

We begin with the Dirac semimetallic coupled wire model in figure~\ref{fig:Weylspectrum} and \eqref{WeylTBHam}. In particular, we focus on many-body interactions that facilitate the fractionalization of a $(1+1)$D chiral Dirac channel \begin{align}\mathrm{Dirac}=\mathrm{Pfaffian}\otimes\mathrm{Pfaffian}\label{fractionalization}\end{align} (see also figure~\ref{fig:glueingsplitting}). In a sense, each chiral Pfaffian channel carries half of the degrees of freedom of the Dirac. For instance, it has half the electric and thermal conductances, which are characterized by the filling fraction $\nu=1/2$ and the chiral central charge $c=1/2$ in \eqref{conductance}. Throughout this paper, we refer to the low-energy effective \CFT -- that consists of an electrically charged $U(1)_4$ bosonic component, say moving in the $R$ direction, and a neutral Majorana fermion component moving in the opposite $L$ direction -- simply as a Pfaffian \CFT \begin{align}\mathrm{Pfaffian}=U(1)_4\otimes\overline{\mathrm{Ising}}.\label{PfaffianCFT}\end{align} (In this article, we follow the level convention for $U(1)$ in the \CFT community~\cite{bigyellowbook}. The same theory may be more commonly referred to as $U(1)_8$ in the fractional quantum Hall community. For clarification, see Lagrangian \eqref{Pfaffian} and \eqref{LFQHCS}.)

While this is not the focus of this article, here we clarify and disambiguate the three ``Pfaffian" fractional quantum Hall (\hypertarget{FQH}{FQH}) states that commonly appear in the literature. All these $(2+1)$D states are theorized at filling fraction $\nu=1/2$, although being applied to $\nu=5/2$ in materials, and have identical electric transport properties. However, they have distinct thermal Hall transport behaviors. They all have very similar anyonic quasiparticle (\hypertarget{QP}{QP}) structures. For instance, they all have four Abelian and two non-Abelian \QP (up to the electron). On the other hand, the charge $e/4$ non-Abelian Ising anyons of the three states have different spin-exchange statistics. First, the gapless boundary of the Moore-Read Pfaffian \FQH state~\cite{MooreRead,ReadMoore,GreiterWenWilczekPRL91,GreiterWenWilczek91} can be described by the $(1+1)$D chiral \CFT $U(1)_4\otimes\mathrm{Ising}$ where the charged boson and neutral fermion sectors are co-propagating. It therefore carries the chiral central charge $c=1+1/2=3/2$, which dictates the thermal Hall response \eqref{conductance}. Second, the ``anti-Pfaffian" \FQH state~\cite{LevinHalperinRosenow07,LeeRyuNayakFisher07} is the particle-hole conjugate of the Moore-Read Pfaffian state. Instead of half-filling the lowest Landau level by electrons, one can begin with the completely filled lowest Landau level, and half-fill it with holes. In a sense the anti-Pfaffian state is obtained by subtracting the completely filled lowest Landau level by a Moore-Read Pfaffian state. Along the boundary, the $(1+1)$D \CFT $U(1)_{1/2}\otimes\overline{U(1)_4\otimes\mathrm{Ising}}$ consists of the forward propagating chiral Dirac $U(1)_{1/2}$ sector that corresponds to the lowest Landau level, and the backward propagating Moore-Read Pfaffian $\overline{U(1)_4\otimes\mathrm{Ising}}$. Here $\overline{\mathcal{C}}$ can be interpreted as the time-reversal conjugate of the chiral \CFT $\mathcal{C}$. The thermal transport is governed by the edge chiral central charge $c=1-3/2=-1/2$, which has an opposite sign from the filling fraction. Thus, unlike the Moore-Read Pfaffian state, the net electric and thermal currents now travel in opposite directions along the edge. Lastly, the recently proposed particle-hole symmetric (\hypertarget{PHS}{PHS}) Pfaffian state~\cite{Son15,BarkeshliMulliganFisher15,WangSenthil16}, which is going to be the {\em only} Pfaffian \FQH state considered in this article (see Ref.~\onlinecite{KaneSternHalperin17} for a coupled wire construction), has the chiral edge \CFT \eqref{PfaffianCFT}. As the electrically charged boson and neutral fermion sectors are counter-propagating, the net thermal edge transport is governed by the chiral central charge $c=1-1/2=1/2$. The chiral $(1+1)$D \PHS Pfaffian \CFT \eqref{PfaffianCFT} is also present along the line interface separating a \TR symmetric $\mathcal{T}$-Pfaffian~\cite{ChenFidkowskiVishwanath14} domain and a \TR breaking magnetic domain on the surface of a 3D topological insulator. (Similar constructions can be applied to alternative \TR symmetric topological insulator surface states~\cite{WangPotterSenthilgapTI13,MetlitskiKaneFisher13b,BondersonNayakQi13}, but they will not be considered in this article.) Other than their thermal transport properties, the three Pfaffian \FQH state can also be distinguished by the charge $e/4$ Ising anyon, which has spin $h=1/8$, $-1/8$ or $0$ for the Moore-Read Pfaffian, anti-Pfaffian or \PHS Pfaffian states respectively. 

Since we will not be considering the Moore-Read Pfaffian or its particle-hole conjugate anti-Pfaffian state, we will simply refer to the \PHS Pfaffian state as the Pfaffian state. The low-energy effective chiral $(1+1)$D \CFT takes the decoupled form between the boson and fermion \begin{align}\mathcal{L}_{\mathrm{Pfaffian}}&=\mathcal{L}_{\mathrm{charged}}+\mathcal{L}_{\mathrm{neutral}}\label{Pfaffian}\\&=\frac{8}{2\pi}\partial_t\phi_R\partial_x\phi_R+v(\partial_x\phi_R)^2\nonumber\\&\;\;\;+i\gamma_L(\partial_t-\tilde{v}\partial_x)\gamma_L\nonumber\end{align} where we have set $\hbar=1$. Here $\phi_R$ is the free chiral $U(1)_4$ boson. It generates the $(1+1)$D theory $\mathcal{L}_{\mathrm{charged}}$, which is identical to the boundary edge theory of the $(2+1)$D bosonic Laughlin $\nu=1/8$ fractional quantum Hall state described by the topological Chern-Simon theory~\cite{WenZee92,Wenedgereview} \begin{align}\mathcal{L}_{2+1}=\frac{K}{4\pi}\alpha\wedge d\alpha+et\alpha\wedge dA\label{LFQHCS}\end{align} with $K=8$ and $t=2$. The $U(1)_4$ \CFT carries the electric conductance $\sigma=tK^{-1}t=1/2$ in units of $2\pi e^2=e^2/h$ and a thermal conductance characterized by the chiral central charge $c=c_R=1$. Primary fields are of the form of (normal ordered) chiral vertex operators $:e^{im\phi_R}:$, for $m$ an integer, and carries charge $q=m/4$ in units of $e$ and conformal scaling dimension (i.e.~conformal spin) $h=h_R=m^2/16$. We summarize and abbreviate the operator product expansion \begin{align}e^{im_1\phi_R(z)}e^{im_2\phi_R(w)}=e^{i(m_1+m_2)\phi_R(w)}(z-w)^{m_1m_2/8}+\ldots\end{align} by the Abelian fusion rule \begin{align}e^{im_1\phi_R}\times e^{im_2\phi_R}=e^{i(m_1+m_2)\phi_R},\end{align} where $z\sim\tau+ix$ is the complex space-time parameter and $\tau=i\pi vt/2$ is the Euclidean time.

$\gamma_L^\dagger=\gamma_L$ is the free Majorana fermion. It generates the $(1+1)$D theory $\mathcal{L}_{\mathrm{neutral}}$, which is equivalent to a chiral component of the critical Ising \CFT or the boundary edge theory of the $(2+1)$D Kitaev honeycomb model~\cite{Kitaev06} in its B-phase with \TR breaking (i.e.~a chiral $p_x+ip_y$ superconductor coupled with a $\mathbb{Z}_2$ gauge theory). It carries trivial electric conductance but contributes to a finite thermal conductance characterized by the chiral central charge $c=-c_L=-1/2$. The Ising \CFT has primary fields $1$, $\gamma_L$ and $\sigma_L$, where the twist field (or Ising anyon) $\sigma_L$ carries the conformal spin $h=-h_L=-1/16$. Again, we abbreviate the operator product expansions \begin{gather}\gamma_L(\bar{z})\gamma_L(\bar{w})=\frac{1}{\bar{z}-\bar{w}}+\ldots\nonumber\\\sigma_L(\bar{z})\gamma_L(\bar{w})=\frac{\sigma_L(\bar{w})}{(\bar{z}-\bar{w})^{1/2}}+\ldots\nonumber\\\sigma_L(\bar{z})\sigma_L(\bar{w})=\frac{1}{(\bar{z}-\bar{w})^{1/8}}+(\bar{z}-\bar{w})^{3/8}\gamma_L(\bar{w})\nonumber\end{gather} by the fusion rule \begin{gather}\gamma_L\times\gamma_L=1,\quad\sigma_L\times\gamma_L=\sigma_L\nonumber\\\sigma_L\times\sigma_L=1+\gamma_L,\end{gather} where $\bar{z}\sim\tau-ix$ is the complex space-time parameter and $\tau=i\tilde{v}t$ is the Euclidean time. 

General primary fields of the Pfaffian \CFT decompose into the $U(1)_4$ part and the Ising part. They take the form \begin{align}1_m=e^{im\phi_R},\quad\psi_m=e^{im\phi_R}\gamma_L,\quad\sigma_m=e^{im\phi_R}\sigma_L.\label{Pfaffianfields}\end{align} The conformal spins and fusion rules also decompose so that \begin{align}h_{1_m}=\frac{m^2}{16},\quad h_{\psi_m}=\frac{m^2}{16}+\frac{1}{2},\quad h_{\sigma_m}=\frac{m^2-1}{16}\end{align} modulo 1, $q_m=m/4$ in units of $e$, and \begin{gather}1_{m_1}\times1_{m_2}=\psi_{m_1}\times\psi_{m_2}=1_{m_1+m_2}\nonumber\\1_{m_1}\times\psi_{m_2}=\psi_{m_1+m_2}\nonumber\\1_{m_1}\times\sigma_{m_2}=\psi_{m_1}\times\sigma_{m_2}=\sigma_{m_1+m_2}\nonumber\\\sigma_{m_1}\times\sigma_{m_2}=1_{m_1+m_2}+\psi_{m_1+m_2}.\end{gather} The $2\pi$ monodromy phase $\mathcal{M}^{XY}_Z=R^{XY}_ZR^{YX}_Z$ between primary fields $X$ and $Y$ with a fixed overall fusion channel $Z$ can be deduced by the {\em ribbon identity}~\cite{Kitaev06} \begin{align}e^{2\pi ih_Z}=\vcenter{\hbox{\includegraphics[width=0.5in]{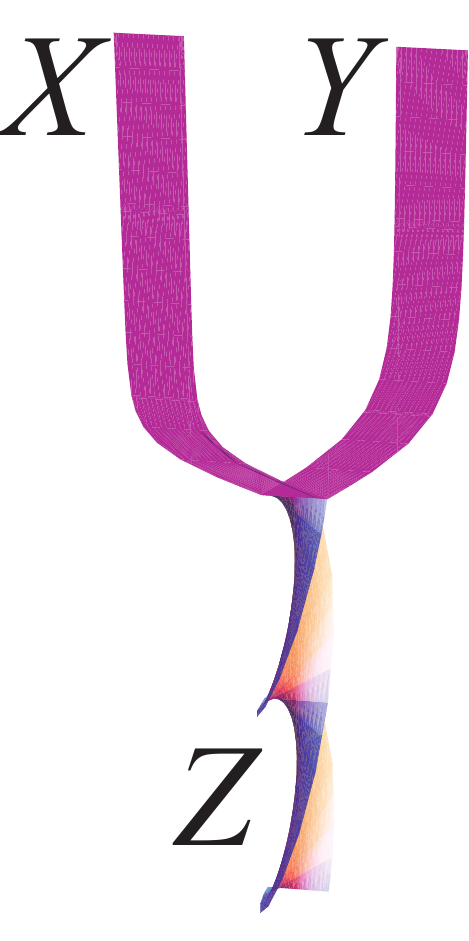}}}=\vcenter{\hbox{\includegraphics[width=0.5in]{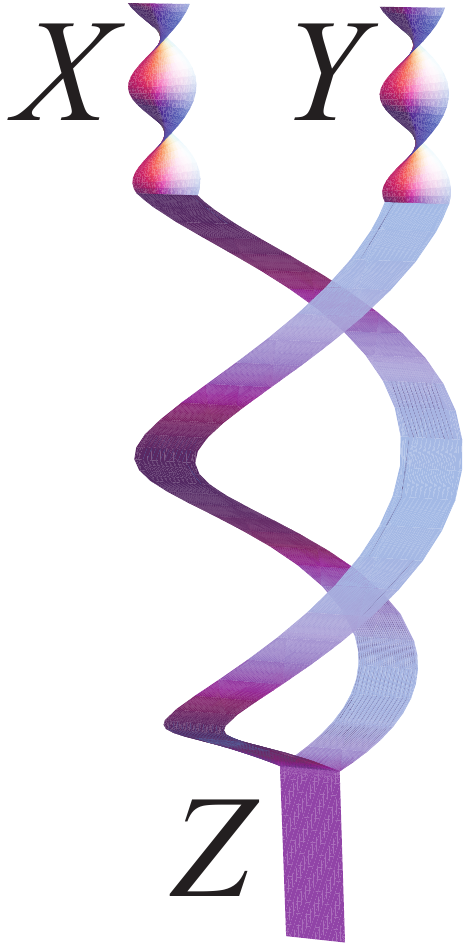}}}=\mathcal{M}^{XY}_Ze^{2\pi i(h_X+h_Y)}\label{ribbonapp}\end{align} for $h_{X,Y,Z}$ the conformal spins for primary fields $X,Y,Z$. Unlike the gauge dependent $\pi$-exchange phase $R^{XY}_Z$, the $2\pi$-monodromy phase $\mathcal{M}^{XY}_Z=e^{2\pi i(h_Z-h_X-h_Y)}$ is gauge independent and physical.

The electronic quasiparticle is the composition $\psi_{\mathrm{el}}=e^{-i4\phi_R}\gamma_L$ so that it is fermionic and has electric charge $-1$ in units of $e$. Since electron is the fundamental building block of the system, locality of $\psi_{\mathrm{el}}$ only allows primary fields $X$ that have trivial monodromy $\mathcal{M}^{X,\psi_{\mathrm{el}}}=1$ with the electron. As a result, this restricts $1_m,\psi_m$ to even $m$ and $\sigma_m$ to odd $m$. Lastly, the coupled wire models constructed later will involve the Pfaffian channels that propagate in both forward and backward directions. We will denote the backward case by $\overline{\mathrm{Pfaffian}}$, whose Lagrangian density is the time reversal of \eqref{Pfaffian}, i.e.~replacing $R\leftrightarrow L$, $i\leftrightarrow-i$ and $\partial_t\leftrightarrow-\partial_t$. 

\subsection{Gluing and splitting}\label{sec:gluing}

A pair of co-propagating Pfaffian \CFT can be ``glued" together into a single chiral Dirac electronic channel. We first consider the decoupled pair $\mathcal{L}_0=\mathcal{L}_{\mathrm{Pfaffian}}^A+\mathcal{L}_{\mathrm{Pfaffian}}^B$, where $\mathcal{L}_{\mathrm{Pfaffian}}^{A/B}$ is the Lagrangian density of one of the two Pfaffian \CFT labeled by $A,B$. The pair of Majorana fermions can compose an electrically neutral Dirac fermion $d_L=(\gamma^A_L+i\gamma^B_L)/\sqrt{2}$, which can then be bosonized $d_L\sim e^{i\phi^\sigma_L}$, for $\phi^\sigma_L$ the chiral $\overline{U(1)_{1/2}}$ boson. The bare Lagrangian now becomes the multi-component $U(1)_4^A\otimes U(1)_4^B\otimes\overline{U(1)_{1/2}}$ boson \CFT \begin{align}\mathcal{L}_0=\frac{1}{2\pi}\partial_t\boldsymbol{\phi}^TK\partial_x\boldsymbol{\phi}+\partial_x\boldsymbol{\phi}^TV\partial_x\boldsymbol{\phi},\label{881}\end{align} where $\boldsymbol{\phi}=(\phi_R^A,\phi_R^B,\phi^\sigma_L)$, $K$ is the $3\times3$ diagonal matrix $K=\mathrm{diag}(8,8,-1)$, and $V$ is some non-universal velocity matrix. A primary field is a vertex operator $e^{i{\bf m}\cdot\boldsymbol{\phi}}$ labeled by an integral vector ${\bf m}=(m^A,m^B,\tilde{m})$. It carries conformal spin $h_{\bf m}={\bf m}^TK^{-1}{\bf m}/2$ and electric charge $q_{\bf m}={\bf t}^TK^{-1}{\bf m}$ in units of $e$, where ${\bf t}=(2,2,0)$ is the charge vector. As ${\bf n}=(1,-1,4)$ is an electrically neutral null vector (i.e.~${\bf n}^TK{\bf n}=0$ and ${\bf t}\cdot{\bf n}=0$), it corresponds to the charge $U(1)$ preserving backscattering coupling \begin{align}\delta\mathcal{H}=-u\cos\left({\bf n}^TK\boldsymbol{\phi}\right)=-u\cos\left(8\phi^A_R-8\phi^B_R-4\phi^\sigma_L\right)\label{glueingH}\end{align} that gaps~\cite{Haldane95} and annihilates a pair of counter-propagating boson modes. The interacting Hamiltonian can also be expressed in terms of many-body backscattering of the Pfaffians' primary fields \begin{align}\delta\mathcal{H}=-u:\left(d_L^\dagger d_R\right)^4:+h.c.\end{align} where $d_R=1_2^A1_{-2}^B$ is the electrically neutral Dirac fermion composed of the pair of oppositely charged semions in the two Pfaffian sectors.

In strong coupling, the gapping Hamiltonian introduces an interacting mass and the ground state expectation value $\langle\Phi\rangle=n\pi/2$, for $n$ an integer and $\Phi=2\phi^A_R-2\phi^B_R-\phi^\sigma_L$. In low energy, it leaves behind the chiral boson combination $\tilde\phi_R=2\phi_R^A+2\phi_R^B$, which has trivial operator product (i.e.~commutes at equal time) with the order parameter $\Phi$. The low-energy theory after projecting out the gapped sectors becomes \begin{align}\mathcal{L}_0-\delta\mathcal{H}\longrightarrow\mathcal{L}_{\mathrm{Dirac}}=\frac{1}{2\pi}\partial_t\tilde\phi_R\partial_x\tilde\phi_R+v(\partial_x\tilde\phi_R)^2\end{align} which is identical to the bosonized Lagrangian density of a chiral Dirac fermion. For instance, the vertex operator $\psi_R^{\mathrm{el}}\sim e^{i\tilde\phi_R}\sim 1_2^A1_2^B$ has the appropriate spin and electric charge of an electronic Dirac fermion operator ($h=1/2$ and $q=1$ in units of $e$). Notice that the vertex operator $e^{i\tilde\phi_R/2}$ has $-1$ monodromy with the local electronic $\psi_R^{\mathrm{el}}$ and therefore is not an allowed excitation in the fermionic theory.

We notice in passing that the gluing potential \eqref{glueingH} facilitates an anyon condensation process~\cite{BaisSlingerlandCondensation}, where the maximal set of mutually local neutral bosonic anyon pairs \begin{align}\begin{array}{*{20}c}1_{4m}^A1_{-4m}^B,\psi_{4m}^A\psi_{-4m}^B,\\\psi_{4m+2}^A1_{-4m-2}^B,1_{4m+2}^A\psi_{-4m-2}^B,\sigma_{4m+1}^A\sigma_{-4m-1}^B\end{array}\label{condensebosons}\end{align} is condensed, where $m$ is an arbitrary integer. All primary fields that are non-local (i.e.~with non-trivial monodromy) with any of the condensed bosons in \eqref{condensebosons} are confined. Any two primary fields that differ from each other by a condensed boson in \eqref{condensebosons} are now equivalent. The condensation therefore leaves behind the electronic Dirac fermion \begin{align}\psi^{\mathrm{el}}_R=\psi^A_4\equiv\psi^B_4\equiv1_2^A1_2^B\end{align} and its combinations. 

\begin{figure}[htbp]
\centering\includegraphics[width=0.5\textwidth]{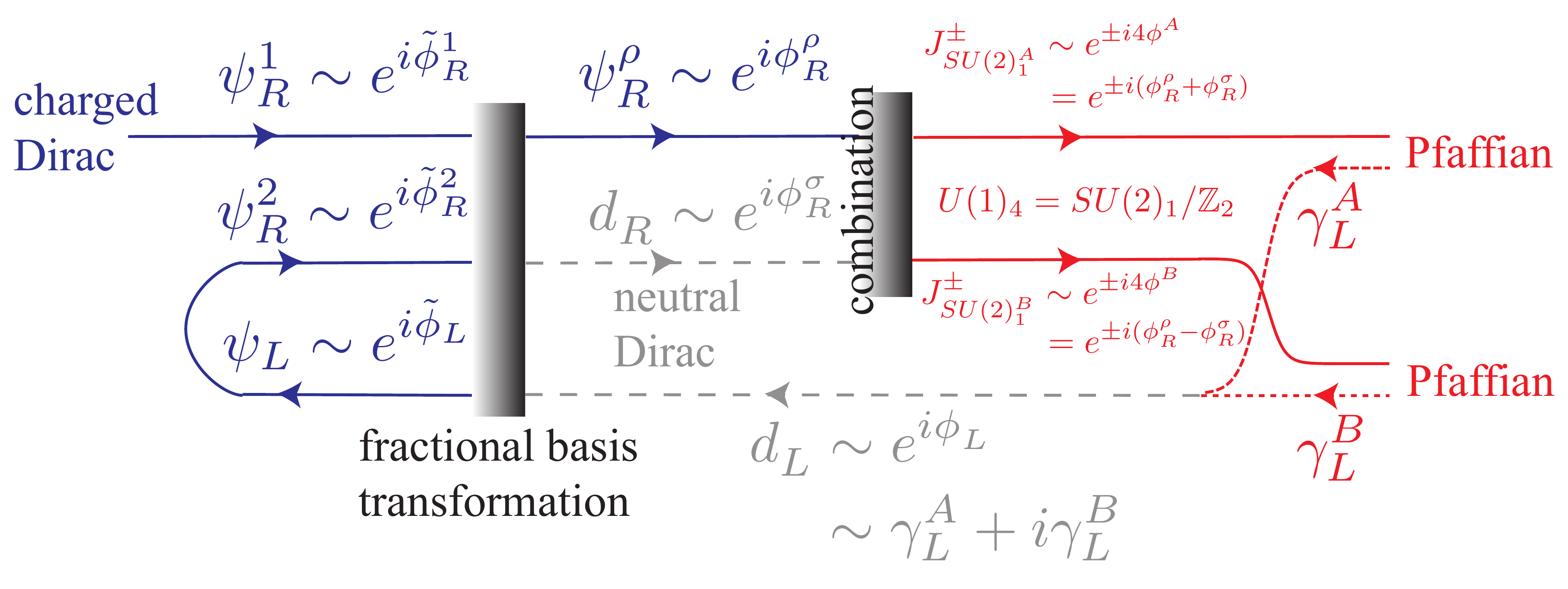}
\caption{Schematics of splitting a chiral Dirac channel into a pair of Pfaffian channels.}\label{fig:fractionalization}
\end{figure}

On the other hand, a chiral Dirac channel can be decomposed into a pair of chiral Pfaffian channels (see figure~\ref{fig:fractionalization} for a summary). First, perhaps from some channel re-construction, we append to the chiral Dirac channel an additional pair of counter-propagating Dirac modes. This can be realized by pulling a parabolic electronic/hole band from the conduction/valence band to the Fermi level, or introducing non-linear dispersion to the original chiral channel. In low-energy, the three Dirac fermion modes can be bosonized $\psi^{1,2}_R\sim e^{i\tilde\phi_R^{1,2}}$, $\psi_L\sim e^{-i\tilde\phi_L}$ and they are described by the multicomponent boson Lagrangian \begin{align}\widetilde{\mathcal{L}}_{\mathrm{Dirac}}=\frac{1}{2\pi}\partial_t\widetilde{\boldsymbol\phi}^T\tilde{K}\partial_x\widetilde{\boldsymbol\phi}+\partial_x\widetilde{\boldsymbol\phi}^T\tilde{V}\partial_x\widetilde{\boldsymbol\phi}\label{3Dirac}\end{align} for $\widetilde{\boldsymbol\phi}=(\tilde\phi_R^1,\tilde\phi_R^2,\tilde\phi_L)$, $\tilde{K}$ is the diagonal matrix $\tilde{K}=\mathrm{diag}(1,1,-1)$, and $\tilde{V}$ is some non-universal velocity matrix. A general composite excitation can be expressed by a vertex operator $e^{i{\bf m}\cdot\widetilde{\boldsymbol\phi}}$, for ${\bf m}$ an integral 3-vector, with spin $h_{\bf m}=|{\bf m}|^2/2$ and electric charge $q_{\bf m}={\bf m}^T\tilde{K}\tilde{\bf t}$ in units of $e$, where $\tilde{\bf t}=(1,1,1)$ is the charge vector.

Next we perform a {\em fractional} basis transformation \begin{align}\begin{array}{*{20}l}\phi^\rho_R=\tilde\phi^1_R+\tilde\phi^2_R+\tilde\phi_L\\\phi^\sigma_R=\tilde\phi^1_R-\frac{1}{2}\tilde\phi^2_R+\frac{1}{2}\tilde\phi_L\\\phi^\sigma_L=\tilde\phi^1_R+\frac{1}{2}\tilde\phi^2_R+\frac{3}{2}\tilde\phi_L\end{array}.\label{fracbasistrans0}\end{align} While the $\tilde{K}$ matrix is invariant under the transformation, the charge vector changes to $\tilde{\bf t}\to(1,0,0)$. $\psi^\rho_R\sim e^{i\phi^\rho_R}$ is the local electronic Dirac fermion that carries spin $1/2$ and electric charge $e$, and $d_{R/L}\sim e^{i\phi^\sigma_{R/L}}$ are counter-propagating electrically neutral Dirac fermions. As the $\tilde{K}$ matrix is still diagonal, these fermions have trivial mutual $2\pi$-monodromy and are local with respect to each other. However, it is important to notice that the neutral Dirac fermions $d_{R/L}$ actually consist of fractional electronic components.

Now we focus on the two $R$-moving Dirac channels. By pairing the Dirac fermions, they form two independent $SU(2)_1$ Kac-Moody current operators~\cite{bigyellowbook} \begin{align}J_3^{A/B}(z)&=i2\sqrt{2}\partial_z\phi^{A/B}_R(z)\label{SU2current}\\J_\pm^{A/B}(z)&=\frac{J_1^{A/B}(z)\pm iJ_2^{A/B}(z)}{\sqrt{2}}=e^{\pm i4\phi^{A/B}_R(z)}\nonumber\end{align} where $4\phi^A_R=\phi^\rho_R+\phi^\sigma_R$ and $4\phi^B_R=\phi^\rho_R-\phi^\sigma_R$. Both $SU(2)_1$ sectors are electrically charged so that the bosonic vertex operators $J_\pm^{A/B}$ carries charge $\pm e$. They obey the $SU(2)$ current algebra at level 1 \begin{align}J^\lambda_{\mathsf{i}}(z)J^{\lambda'}_{\mathsf{j}}(w)=\frac{\delta^{\lambda\lambda'}\delta_{\mathsf{ij}}}{(z-w)^2}+\sum_{\mathsf{k}=1}^3\frac{i\sqrt{2}\delta^{\lambda\lambda'}\varepsilon_{\mathsf{ijk}}}{z-w}J^\lambda_{\mathsf{k}}(w)+\ldots\label{SU2algebra}\end{align} for $\lambda,\lambda'=A,B$. It is crucial to remember that $J_\pm^A\sim\psi^\rho_Rd_R$ and $J_\pm^B\sim\psi^\rho_Rd_R^\dagger$ contains the fractional Dirac components $d_R$. Thus, the primitive local bosons are actually pairs of the current operators, i.e.~$e^{i8\phi^{A/B}_R}$. Equivalently, this renormalizes the compactification radius of the boson $4\phi^{A/B}_R$ so that in a closed periodic space-time geometry, we only require electronic Cooper pair combinations such as the charge $2e$ local operators \begin{gather}e^{i8\phi^A_R}=e^{i(4\tilde\phi^1_R+\tilde\phi^2_R+3\tilde\phi_L)}\sim(\psi^1_R)^4\psi^2_R(\psi_L^\dagger)^3\nonumber\\e^{i8\phi^B_R}=e^{i(3\tilde\phi^2_R+\tilde\phi_L)}\sim(\psi^2_R)^3\psi_L^\dagger\end{gather} to be periodic. The incorporation of anti-periodic boundary condition for $J_\pm^{A/B}=e^{\pm i4\phi^{A/B}_R}$ results in the $\mathbb{Z}_2$-orbifold theory~\cite{Ginsparg88,DijkgraafVafaVerlindeVerlinde99} $U(1)_4=SU(2)_1/\mathbb{Z}_2$ for both $A$ and $B$ sectors. For instance, the primitive twist fields are given by $e^{\pm i\phi^{A/B}_R}$, which have $-1$ monodromy phase with $J_\pm^{A/B}$. 

At this point, including the $L$-moving neutral Dirac sector, we have recovered the muticomponent boson $\boldsymbol\phi=(\phi^A_R,\phi^B_R,\phi^\sigma_L)$ described by the Lagrangian \eqref{881}. Lastly, we simply have to decompose the remaining neutral Dirac into Majorana components, $d_L=(\gamma^A_L+i\gamma^B_L)/\sqrt{2}$. The $A$ and $B$ Pfaffian sectors can then be independently generated by the charged $U(1)_4$ boson $\phi^{A/B}_R$ and the neutral Majorana fermion $\gamma^{A/B}_L$. As a consistency check, the charge $e$ fermionic (normal ordered) combinations defined in \eqref{Pfaffianfields} \begin{align}\psi_4^A&\sim e^{i4\phi^A_R}\gamma_L^A\sim e^{i\tilde\phi^1_R}+e^{i(3\tilde\phi^1_R+\tilde\phi^2_R+3\tilde\phi_L)}\label{psi4def1}\\\psi_4^B&\sim e^{i4\phi^B_R}\gamma_L^B\sim e^{i(-\tilde\phi^1_R+\tilde\phi^2_R-\tilde\phi_L)}-e^{i(\tilde\phi^1_R+2\tilde\phi^2_R+2\tilde\phi_L)}\nonumber\end{align} are in fact local quasi-electronic. %(The minus sign in the bosonized expression for $\psi_4^A$ comes from the Klein factors defined later in \eqref{ETcomm0} and \eqref{ETcomm1}).

Unlike in the gluing case where there is a gapping Hamiltonian \eqref{glueingH} that pastes a pair of Pfaffians into a Dirac, here in the splitting case we have simply performed some kind of fractional basis transformation that allows us to express Dirac as a pair of Pfaffians. In fact, one can check that the energy-momentum tensor of the Dirac theory \eqref{3Dirac} is identical to that of a pair of Pfaffians \eqref{Pfaffian}. However, this does not mean the Pfaffian primary fields are natural stable excitations. In fact, as long as there is a pair of co-propagating Pfaffian channels, all primary fields except the non-fractionalized electronic ones are unstable against the gluing Hamiltonian $\delta\mathcal{H}$ in \eqref{glueingH} and are generically gapped. In order for the Pfaffian \CFT to be stabilized, one has to suppress $\delta\mathcal{H}$. A possible way is to somehow spatially separate the pair. This issue is addressed in the subsection below using many-body interaction in the coupled wire model of a Dirac semimetal (or the \PHS Pfaffian \FQH state in Ref.~\onlinecite{KaneSternHalperin17}).

%\subsubsection{\texorpdfstring{$AB$}{AB} symmetry}
%The bi-partition $\mathrm{Dirac}=\mathrm{Pfaffian}^A\otimes\mathrm{Pfaffian}^B$ carries a flip symmetry that exchanges the two Pfaffian sectors. When the two Pfaffian channels are spatially separated (see figure~\ref{fig:glueingsplitting}), the flip symmetry is simply the twofold rotation that exchanges the two parallel channels along their center-of-mass axis. We will use this flip operation to generate the $C_2$ symmetry in the interacting coupled wire model in the next subsection. Focusing on a single wire, the twofold symmetry flips \begin{align}C_2\phi^{A/B}_RC_2^{-1}=\phi^{B/A}_R+\frac{\pi}{8},\quad C_2\phi^\sigma_LC_2^{-1}=\phi^\sigma_L+\frac{\pi}{2}.\end{align} The constants phases ensures the $2\pi$ rotation $C_2^2$ associates the appropriate $-1$ twist phase 
%$e^{2\pi ih_X}$ determined by the spin $h_X$ of the primary field $X$. 

\subsection{Symmetry preserving massive interacting model}\label{sec:interactionmodels}

We begin with the 3D array of chiral Dirac strings in figure~\ref{fig:vortexlattice}. In section~\ref{sec:DiracSemimetal}, we showed that the single-body coupled wire model \eqref{WeylTBHam} described a Dirac semimetal with two Weyl fermions (see figure~\ref{fig:Weylspectrum}). The system had emergent antiferromagnetic time reversal (AFTR) symmetries $\mathcal{T}_{11}$ and $\mathcal{T}_{\bar{1}1}$ along the diagonal and off-diagonal axes (see \eqref{WeylTBT11}). Together they generate an emergent lattice translation symmetry with a 2-wire unit cell, and separate the two Weyl points in the Brillouin zone. The symmetries are lowered beyond the effective model when the microscopic high-energy degrees of freedom are included. For example, the mass function \eqref{Jacobielliptic} that supports the Dirac vortex string lattice has a 4-wire periodic unit cell and only preserves one of the \AFTR symmetries $\mathcal{T}_{11}$ (see \eqref{massT11}). With the lowered translation symmetry, the two Weyl points now coincide at the same momentum. Inter-species (or inter-valley) mixing is forbidden by the remaining \AFTR symmetry and a (screw) twofold rotation symmetry $C_2$ about $z$ (see \eqref{WeylTBC2} and \eqref{massC2}). Previously in section~\ref{sec:brokensymmetry}, we introduced symmetry breaking wire dimerizations in \eqref{DiracTBHam} that led to a massive Dirac insulator. In this section, we construct many-body gapping interactions that preserves the two \AFTR symmetries $\mathcal{T}_{11}$ and $\mathcal{T}_{\bar{1}1}$, the $C_2$ symmetry, as well as charge $U(1)$ conservation. 

\begin{figure}[htbp]
\centering
(a)\includegraphics[width=0.25\textwidth]{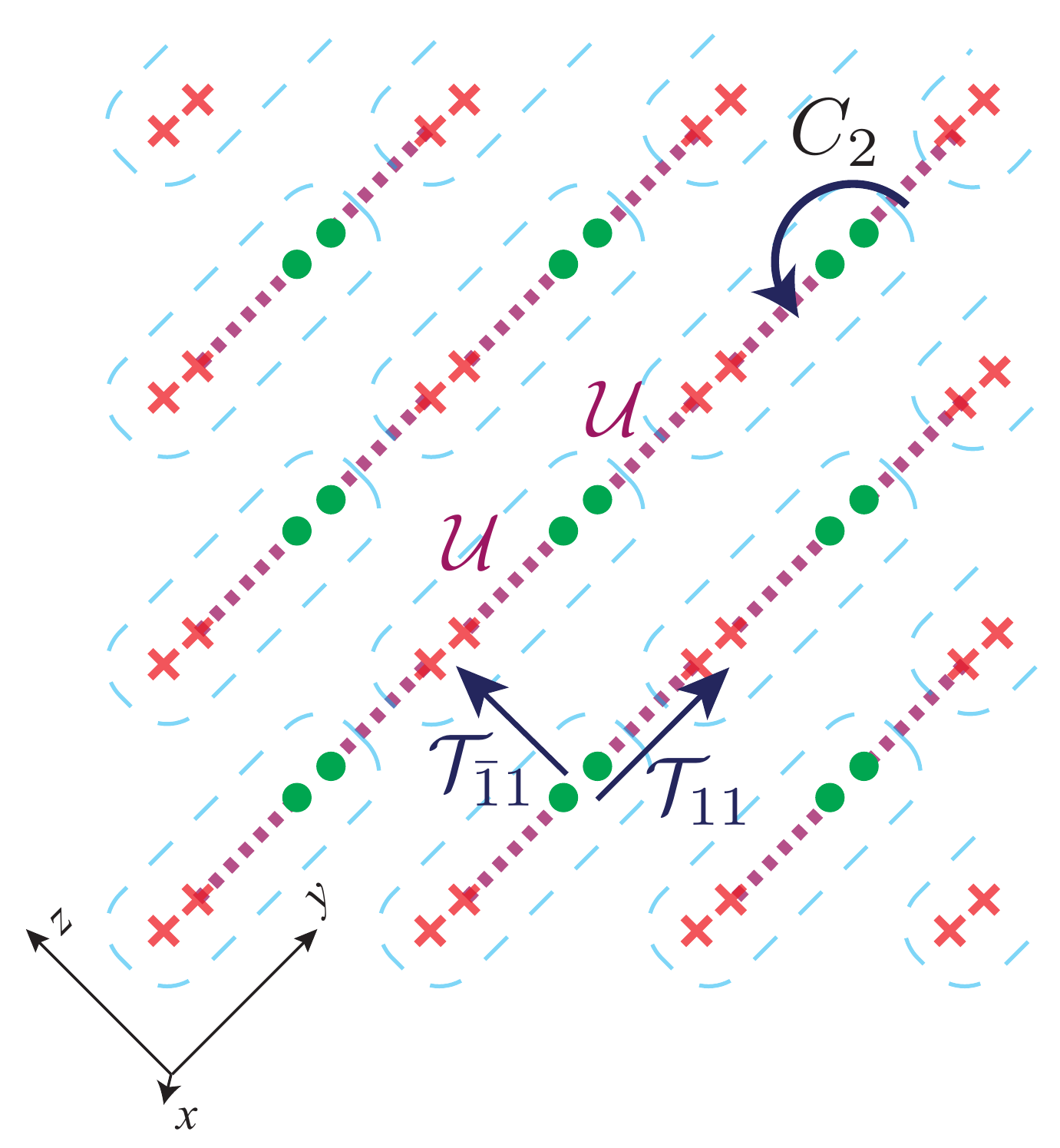}
(b)\includegraphics[width=0.45\textwidth]{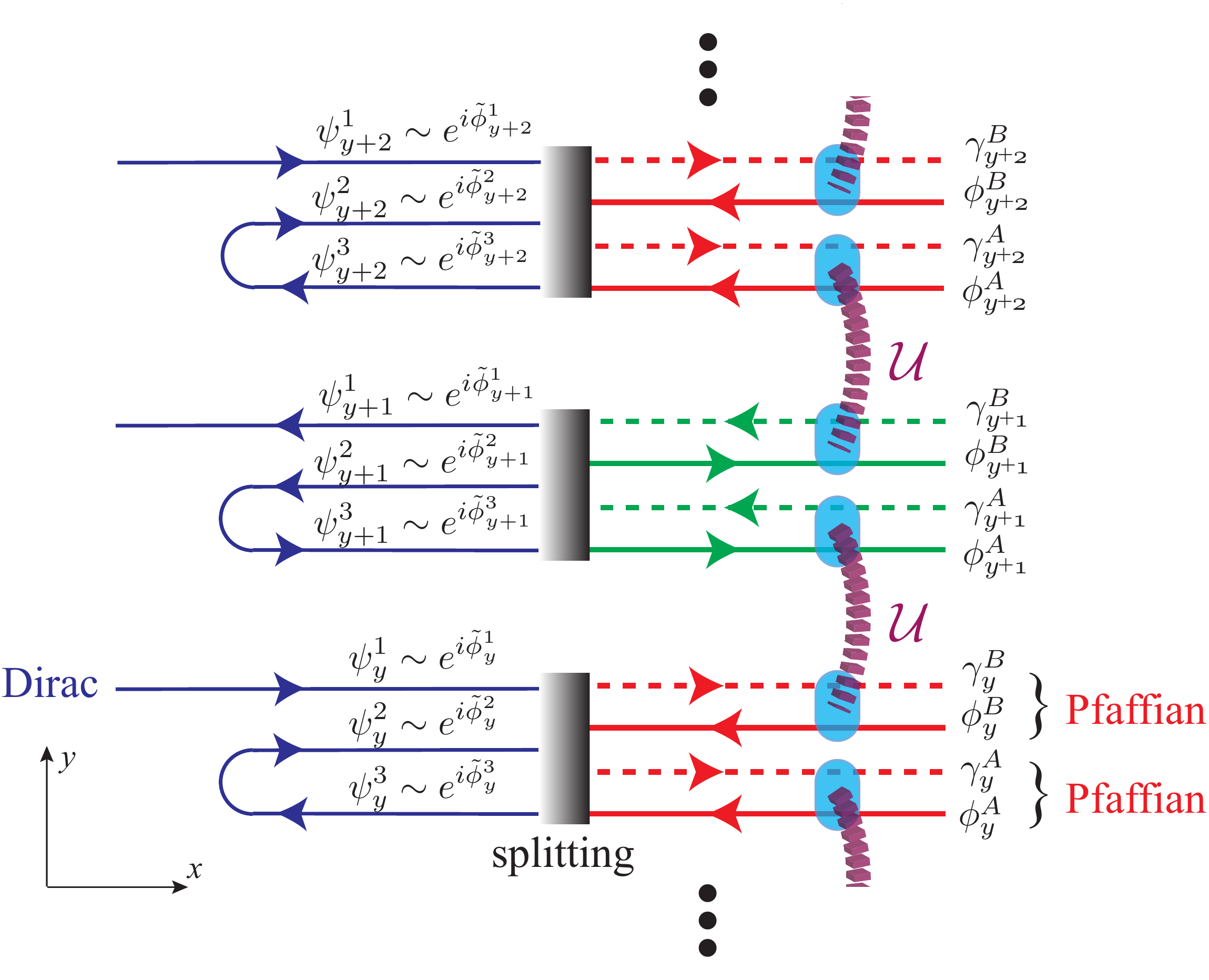}
\caption{Symmetry preserving many-body gapping interaction. (a) Each {\color{red}$\boldsymbol\times$}/{\color{green}$\bullet$} represents a chiral Pfaffian channel into/out-of paper. Purple dashed line represents many-body gapping interaction $\mathcal{U}$ in \eqref{mbdint}. (b) Coupled wire model on a single layer along the diagonal axis.}\label{fig:gappinginteraction}
\end{figure}

The many-body gapping scheme is summarized in figure~\ref{fig:gappinginteraction}. From the previous subsection, we saw that each chiral Dirac channel can be decomposed into a pair of independent Pfaffian channels. They can then be backscattered in opposite directions to neighboring wires. Figure~\ref{fig:gappinginteraction}(a) shows a particular dimerization pattern of the Pfaffian channels that preserves the symmetries. In this case, the many-body backscattering interaction $\mathcal{U}$ is directed along the diagonal axis. In the limit when $\mathcal{U}$ is much stronger than the single-body electron tunneling in the previous semimetallic model \eqref{WeylTBHam}, the system decomposes into decoupled diagonal layers and it suffices to consider the interaction on a single layer. For convenience, we here change our spatial coordinates so that the diagonal axis is now labeled by $y$ and the wires now propagate along $x$.

Focusing on a single diagonal layer, the system in the non-interacting limit first consists of a 2D array of chiral Dirac strings with alternating propagating directions (see the left side of figure~\ref{fig:gappinginteraction}(b)). We notice that this is identical to the starting point of the coupled wire construction of the topological insulator Dirac surface state considered by Mross, Essin and Alicea in Ref.~\onlinecite{MrossEssinAlicea15}. For instance, the alternating Dirac channels there were supported between magnetic strips with alternating orientations on the topological insulator surface, and an uniform nearest-channel electron tunneling recovered the massless 2D Dirac spectrum protected by the \AFTR symmetry. They then proceeded to propose symmetry preserving many-body gapping interactions facilitated by adding 2D \FQH strips between the channels. While this reconstruction trick can be applied on the 2D surface of a topological insulator, it is not feasible in our 3D situation and would require drastic modification of the bulk semimetal. Instead, here we propose an alternative gapping scheme that does not involve additional topological phases. In other words, we are going to construct a 3D gapped and layered topological phase solely from interacting electronic Dirac wires.

First, in order to implement the splitting described in the previous subsection, we assume each Dirac string consists of two Dirac channels going in one direction and a third Dirac channel going the opposite direction (see the left side of figure~\ref{fig:gappinginteraction}(b)). We denote the electronic Dirac fermions on the $y^{\mathrm{th}}$ wire by $\boldsymbol\psi_y=(\psi_y^1,\psi_y^2,\psi_y^3)$ and bosonize \begin{align}\psi_y^{1,2}(x)\sim e^{i\tilde\phi^{1,2}_y(x)},\quad\psi_y^3(x)\sim e^{-i\tilde\phi^3_y(x)}.\label{bosondef}\end{align} The sliding Luttinger liquid\cite{OHernLubenskyToner99,EmeryFradkinKivelsonLubensky00,VishwanathCarpentier01,SondhiYang01,MukhopadhyayKaneLubensky01} Lagrangian density is \begin{align}\mathcal{L}_{\mathrm{layer}}=\sum_{y=-\infty}^\infty\frac{(-1)^y\tilde{K}_{jk}}{2\pi}\partial_t\tilde\phi_y^j\partial_x\tilde\phi_y^k+\tilde{V}_{jk}\partial_x\tilde\phi_y^j\partial_x\tilde\phi_y^k\label{Llayer}\end{align} where $\tilde{K}=(\tilde{K}_{jk})_{3\times3}=\mathrm{diag}(1,1,-1)$, $\tilde{V}$ is some non-universal velocity matrix, and repeating species indices $j,k$ are summed over. The boson operators obey the equal-time commutation relation (\hypertarget{ETCR}{ETCR}) \begin{align}&\left[\tilde\phi_y^j(x),\tilde\phi_{y'}^{j'}(x')\right]=c^{jj'}_{yy'}(x-x')\nonumber\\=&i\pi(-1)^y\delta_{yy'}\tilde{K}^{jj'}\mathrm{sgn}(x'-x)\label{ETcomm0}\\&+i\pi(-1)^y\delta_{yy'}S^{jj'}\nonumber\\&+i\pi(-1)^{\mathrm{max}\{y,y'\}}\mathrm{sgn}(y-y')\Sigma^{jj'}\sigma_z^{y-y'+1}\nonumber\end{align}
%\begin{align}\left[\tilde\phi_y^j(x),\tilde\phi_{y'}^{j'}(x')\right]=&i\pi(-1)^{\mathrm{max}\{y,y'\}}\Big[\delta_{yy'}\tilde{K}^{jj'}\mathrm{sgn}(x'-x)\nonumber\\&+\delta_{yy'}\mathrm{sgn}(j-j')+\mathrm{sgn}(y-y')\Big]\label{ETcomm0}\end{align} 
where $\mathrm{sgn}(s)=s/|s|=\pm1$ for $s\neq0$ and $\mathrm{sgn}(0)=0$, \begin{align}S=\left(\begin{smallmatrix}0&1&-1\\-1&0&1\\1&-1&0\end{smallmatrix}\right),\quad\Sigma=\left(\begin{smallmatrix}1&1&-1\\1&1&-1\\-1&-1&1\end{smallmatrix}\right),\label{Kleinfactors}\end{align} and $\sigma_z=\pm1$. The introduction of the specific Klein factors $S^{jj'}$, $\Sigma^{jj'}$ and the undetermined sign $\sigma_z$ are necessary for the correct representations of the $\mathcal{T}_{11}$ and $\mathcal{C}_2$ symmetries in the bosonization setting, and these choices will be justified below. The first line of \eqref{ETcomm0} is equivalent to the commutation relation between conjugate fields \begin{align}\left[\tilde\phi_y^j(x),\partial_{x'}\tilde\phi_{y'}^{j'}(x')\right]=2\pi i(-1)^y\delta_{yy'}\tilde{K}^{jj'}\delta(x-x')\label{ETcomm00}\end{align} which is set by the ``$p\dot{q}$" term in $\mathcal{L}_{\mathrm{layer}}$. The alternating signs $(-1)^y$ in \eqref{ETcomm00} and \eqref{Llayer} changes the propagating directions from wire to wire. The second and third line of \eqref{ETcomm0} guarantee the correct anticommutation relations $\{e^{\pm i\tilde\phi^j_y},e^{\pm i\tilde\phi^{j'}_{y'}}\}=0$ between Dirac fermions along distinct channels $j\neq j'$ or distinct wires $y\neq y'$. The reason the $\tilde{C}_2$ matrix is defined in this form will become clear in the fractional basis discussed later in \eqref{bosonC2Pf}.

The anti-unitary \AFTR symmetry along the diagonal $\mathcal{T}_{11}$ direction transforms the bosons according to \begin{align}\mathcal{T}_{11}\tilde\phi^j_y\mathcal{T}_{11}^{-1}=-\tilde\phi^j_{y+1}+\frac{1+(-1)^y}{2}\tilde{K}^{jj}\pi.\label{bosonTR11}\end{align} The unitary $\mathcal{C}_2$ rotation takes \begin{gather}\mathcal{C}_2\tilde\phi^j_y\mathcal{C}_2^{-1}=\left(\tilde{C}_2\right)^j_{j'}\tilde\phi^{j'}_{-y}+(-1)^yv^j\frac{\pi}{2},\label{bosonC2}\\\tilde{C}_2=\left(\begin{smallmatrix}1&2&2\\2&1&2\\-2&-2&-3\end{smallmatrix}\right),\quad{\bf v}=\left(\begin{smallmatrix}v^1\\v^2\\v^3\end{smallmatrix}\right)=\left(\begin{smallmatrix}3\\-3\\1\end{smallmatrix}\right).\nonumber\end{gather} Moreover, we choose the representation so that the sign $\sigma_z$ in the \ETCR \eqref{ETcomm0} is preserved by the \AFTR operator but is flipped by the $C_2$ symmetry, \begin{align}\mathcal{T}_{11}\sigma_z\mathcal{T}_{11}^{-1}=\sigma_z,\quad\mathcal{C}_2\sigma_z\mathcal{C}_2^{-1}=-\sigma_z.\end{align} 

The \ETCR \eqref{ETcomm0} is consistent with the \AFTR symmetry. This means that evaluating $\mathcal{T}_{11}\left[\tilde\phi_y^j(x),\tilde\phi_{y'}^{j'}(x')\right]\mathcal{T}_{11}^{-1}$ by taking the \AFTR operator inside the commutator \begin{align}&\left[\mathcal{T}_{11}\tilde\phi_y^j(x)\mathcal{T}_{11}^{-1},\mathcal{T}_{11}\tilde\phi_{y'}^{j'}(x')\mathcal{T}_{11}^{-1}\right]\nonumber\\&=\left[\tilde\phi_{y+1}^j(x),\tilde\phi_{y'+1}^{j'}(x')\right]=c^{jj'}_{y+1,y'+1}(x-x')\end{align} yields the same outcome as taking the \TR of the purely imaginary scalar \begin{align}\mathcal{T}_{11}c^{jj'}_{yy'}(x-x')\mathcal{T}_{11}^{-1}=-c^{jj'}_{yy'}(x-x').\end{align} The \ETCR \eqref{ETcomm0} is also consistent with the $\mathcal{C}_2$ symmetry \begin{align}(\tilde{C}_2)^{j_1}_{j'_1}c^{j'_1j'_2}_{-y_1,-y_2}(x_1-x_2)(\tilde{C}_2)^{j_2}_{j'_2}=\mathcal{C}_2c^{j_1j_2}_{y_1y_2}(x_1-x_2)\mathcal{C}_2^{-1}.\label{ETcommC2consistent}\end{align} This is because the Klein factors \eqref{Kleinfactors} are $C_2$ symmetric \begin{align}\tilde{C}_2S\tilde{C}_2^T=S,\quad\tilde{C}_2\Sigma\tilde{C}_2^T=\Sigma.\end{align} Notice that the undetermined sign $\sigma_z$, which is odd under $\mathcal{C}_2$, in \eqref{ETcomm0} is essential for the \ETCR to be consistent with $C_2$.

The last term in the \AFTR operation \eqref{bosonTR11} makes sure \begin{align}\mathcal{T}_{11}^2\tilde\phi_y^j(x)\mathcal{T}_{11}^{-2}=\tilde\phi_{y+2}^j+(-1)^y\tilde{K}^{jj}\pi,\end{align} which is necessary for $\mathcal{T}^2_{11}=(-1)^F\mathrm{translation}(2{\bf e}_y)$. Here the fermion parity operator is $(-1)^F=e^{i\pi\sum_{yj}N_y^j}$, where \begin{align}N_y^j=\int\frac{dx}{2\pi}\partial_x\tilde\phi_y^j(x)\label{numop}\end{align} is the number operator. The vector ${\bf v}$ in the $\mathcal{C}_2$ operation \eqref{bosonC2} satisfies $(\delta^j_{j'}+(\tilde{C}_2)^j_{j'})v^{j'}/2=\tilde{K}^{jj}$, and consequently \begin{align}\mathcal{C}_{2}^2\tilde\phi_y^j(x)\mathcal{C}_{2}^{-2}=\tilde\phi_y^j+(-1)^y\tilde{K}^{jj}\pi,\end{align} which is consistent with $\mathcal{C}_2^2=(-1)^F$. Lastly, it is straightforward to check that the symmetry representations \eqref{bosonTR11} and \eqref{bosonC2} are compatible with the algebraic relation \eqref{C2Trelation}, i.e. \begin{align}&\mathcal{C}_2\mathcal{T}_{11}\tilde\phi^j_y\mathcal{T}_{11}^{-1}\mathcal{C}_2^{-1}\\&=(-1)^F\mathcal{T}_{11}^{-1}\mathcal{C}_2\tilde\phi^j_y\mathcal{C}_2^{-1}\mathcal{T}_{11}(-1)^{-F}.\nonumber\end{align}

%The, at first sight, obscured expression of the $3\times3$ rotation matrix $C_2$ in \eqref{bosonC2} actually takes a much simply form under the fractional basis transformation considered in the previous subsection~\ref{sec:gluing}. We defer this simplification a bit later, but at this point, we notice that the Klein fators $S$ and $\Sigma$ in the \ETCR \eqref{ETcomm0} are chosen to be consistent with the $C_2$ symmetry, \begin{align}C_2SC_2^T=S,\quad C_2\Sigma C_2^T=\Sigma.\end{align} 

Following the splitting scheme summarized in figure~\ref{fig:fractionalization}, we again define a fractional basis transformation (c.f.~\eqref{fracbasistrans0}) \begin{align}\begin{pmatrix}\phi^\rho_y\\\phi^{\sigma1}_y\\\phi^{\sigma2}_y\end{pmatrix}=\left(\begin{array}{*{20}c}1&1&1\\1&-1/2&1/2\\1&1/2&3/2\end{array}\right)\left(\begin{array}{*{20}c}\tilde\phi^1_y\\\tilde\phi^2_y\\\tilde\phi^3_y\end{array}\right)\label{fracbasistrans}\end{align} for each wire, so that $\psi^\rho_y\sim e^{i\phi^\rho_y}$ is a Dirac fermion carrying electric charge $e$, $d^{\sigma1}_y\sim e^{i\phi^{\sigma1}_y}$ ($d^{\sigma2}_y\sim e^{i\phi^{\sigma2}_y}$) is an electrically neutral Dirac fermion propagating in the same (resp.~opposite) direction as $\psi^\rho_y$.

For convenience, sometimes we combine the transformed bosonized variables into $\boldsymbol\phi_y=(\phi^1_y,\phi^2_y,\phi^3_y)=(\phi^A_y,\phi^B_y,\phi^{\sigma2}_y)$, which is related to the original local ones in \eqref{Llayer} by $\phi^J_y=G^J_j\tilde\phi^j_y$ where \begin{align}G=\begin{pmatrix}1/2&1/8&3/8\\0&3/8&1/8\\1&1/2&3/2\end{pmatrix}.\end{align} The \AFTR symmetry operation \eqref{bosonTR11} becomes \begin{align}\mathcal{T}_{11}\phi^I_y\mathcal{T}_{11}^{-1}=-\phi^I_{y+1}+\frac{1+(-1)^y}{2}\pi\kappa^I\label{bosonTR11Pf}
%\mathcal{T}_{11}\phi^{A/B}_y\mathcal{T}_{11}^{-1}&=-\phi^{A/B}_{y+1}+\frac{1+(-1)^y}{2}\pi,\nonumber\\\mathcal{T}_{11}\phi^{\sigma 2}_y\mathcal{T}_{11}^{-1}&=-\phi^{\sigma 2}_{y+1}.
\end{align} where $\kappa^I=G^I_j\tilde{K}^{jj}$ which is $1/4$ for $I=1,2$ and $0$ for $I=3$. The $C_2$ transformation \eqref{bosonC2} becomes \begin{gather}\mathcal{C}_2\phi^I_y\mathcal{C}_2^{-1}=\left(C_2\right)^I_J\phi^J_{-y}+(-1)^yG^I_jv^j\frac{\pi}{2},\label{bosonC2Pf}\\C_2=G\tilde{C}_2G^{-1}=\left(\begin{smallmatrix}0&1&0\\1&0&0\\0&0&-1\end{smallmatrix}\right),\quad G{\bf v}=\left(\begin{smallmatrix}3/2\\-1\\3\end{smallmatrix}\right).\nonumber\end{gather} The $3\times3$ $C_2$ matrix takes a much simpler form here using the fractional basis than in \eqref{bosonC2}. In fact, the original $\tilde{C}_2$ matrix in the local basis in \eqref{bosonC2} was defined so that $C_2=G\tilde{C}_2G^{-1}$ would act according to \eqref{bosonC2Pf}. Roughly speaking, ignoring the constant phases $G{\bf v}$, the $C_2$ symmetry switches $\phi^A_y\leftrightarrow\phi^B_{-y}$ and sends $\phi^{\sigma2}_y\to-\phi^{\sigma2}_{-y}$.

Next, we combine these co-propagating pair of fermions to form two $SU(2)_1$ current algebras (c.f.~\eqref{SU2current} and \eqref{SU2algebra}) \begin{align}&J_3^{A/B}(y,w)=i2\sqrt{2}\partial_w\phi^{A/B}_y(w)\nonumber\\&J_\pm^{A/B}(y,w)=e^{\pm i4\phi^{A/B}_y(w)}%\\&4\phi^A_y(w)=\phi^\rho_y(w)+\phi^{\sigma1}_y(w),\quad 4\phi^B_y(w)=\phi^\rho_y(w)-\phi^{\sigma1}_y(w)\nonumber
\end{align} where $w\sim\tau+(-1)^yx$ is the complex spacetime parameter. As a reminder, the charge $\pm e$ bosons $J^{A/B}_\pm$ are non-electronic fractional operators, although they carry non-fractional statistics.

The remaining counter-propagating neutral Dirac fermion can be decomposed into real and imaginary components \begin{align}d_y^{\sigma}(w)\sim\cos\phi^{\sigma2}_y(w)+i\sin\phi^{\sigma2}_y(w).\end{align} Majorana fermions can be constructed by multiplying these components with ``Jordan-Wigner" string \begin{align}\gamma^A_y&\sim\cos\phi^{\sigma2}_y\prod_{y'>y}(-1)^{N_{y'}^2+N_{y'}^3},\nonumber\\\gamma^B_y&\sim\sin\phi^{\sigma2}_y\prod_{y'>y}(-1)^{N_{y'}^2+N_{y'}^3},\label{MFdef}\end{align} where $N^j_y$ are the number operators defined in \eqref{numop}, so that they obey mutual fermionic statistics $\{\gamma^\lambda_y(x),\gamma^{\lambda'}_{y'}(x')\}=\delta^{\lambda\lambda'}\delta_{yy'}\delta(x-x')$, for $\lambda,\lambda'=A,B$. Similar to the charge  $\pm e$ bosons $J^{A/B}_\pm$, the electrically neutral Dirac fermion $d_y^\sigma$ and consequently the Majorana fermions $\gamma^{A/B}_y$ are also non-electronic fractional operators. This $AB$-decomposition splits each Dirac wire into a pair of decoupled Pfaffian sectors (see figure~\ref{fig:gappinginteraction}(b)).

Before we move on to the symmetric interaction, some further elaborations are needed for the number operators $N_y^j$ and their corresponding fermion parity operators $e^{i\pi N_y^j}$. In our construction, the counter-propagating pair of channels with $j=2,3$ are appended to the original one with $j=1$ to make the Pfaffian fractionalization feasible. We choose the Hilbert space so that the two additional fermion parity operators agree, $e^{i\pi N_y^2}=e^{i\pi N_y^3}$. However, we allow fluctuations to the combined parity $e^{i\pi(N_y^2+N_y^3)}$ and only require it squares to the identity, $e^{2\pi i(N_y^2+N_y^3)}=1$. In other words, $e^{i\pi(N_y^2+N_y^3)}=e^{-i\pi(N_y^2+N_y^3)}$ and it does not matter which one we take as $(-1)^{N_y^2+N_y^3}$ in the ``Jordan-Wigner" string in \eqref{MFdef}. This convention will also be useful later in seeing that the many-body interaction is exactly solvable and symmetry preserving. Extra care is sometimes required. For example, unlike the original Dirac channel where the parity is simply $(-1)^{N_y^1}=e^{\pm i\pi N_y^1}$ because $e^{2\pi iN_y^1}=1$, the individual parity operators $(-1)^{N_y^{2,3}}$ of these additional channels are not well-defined because $e^{2\pi iN_y^{2,3}}\neq1$, i.e.~$e^{i\pi N_y^{2,3}}\neq e^{-i\pi N_y^{2,3}}$. Also, although $e^{2\pi i(N_y^2+N_y^3)}=1$, one cannot in general modify a boson angle parameter simply by $\Theta\to\Theta+2\pi i(N_y^2+N_y^3)$ because $\Theta$ and the number operators may not commute. For instance, using the Baker-Campbell-Hausdorff formula and the \ETCR \eqref{ETcomm0}, $e^{i4\phi^{A/B}}$ and $e^{i4\phi^{A/B}+2\pi i(N_y^2+N_y^3)}$ are off by a minus sign.

The Pfaffian fractionalization is stabilized by the inter-wire many-body backscattering interaction (see figure~\ref{fig:gappinginteraction}(b)) \begin{align}\mathcal{U}&=-u\sum_{y=-\infty}^\infty\cos\phi^{\sigma2}_{y+1}\sin\phi^{\sigma2}_y\cos\left(4\phi^A_{y+1}-4\phi^B_y\right)\nonumber\\
&=-u\sum_{y=-\infty}^\infty(-1)^yi\gamma^A_{y+1}\gamma^B_y\cos\left(\Theta_{y+1/2}\right),\label{mbdint}\end{align} for $\Theta_{y+1/2}(x)=4\phi^A_{y+1}(x)-4\phi^B_y(x)+\pi(N_{y+1}^2+N_{y+1}^3)$. Previously in \eqref{psi4def1}, we saw that the combinations $\psi_4^A\sim e^{i4\phi^A}\gamma^A$ and $\psi_4^B\sim e^{i4\phi^B}\gamma^B$ can be decomposed into products of electron operators. Similarly, each interaction in the first line of \eqref{mbdint} can be decomposed into products in the form of $e^{\pm i(\phi^{\sigma2}_{y+1}\pm4\phi^A_{y+1})}e^{\pm i(\phi^{\sigma2}_y\pm4\phi^B_y)}$ (with some scalar $U(1)$ coefficient), where the exponents $\phi^{\sigma2}\pm4\phi^{A/B}$ are linear integral combinations of $\tilde\phi^j$. Thus, the interaction can be re-written in terms of backscatterings of local electronic operators. However, we will omit the electronic expression as \eqref{mbdint} is more useful in discussing ground state and symmetries. 

$\mathcal{U}$ describes a symmetry-preserving exactly solvable model. Using the \ETCR \eqref{ETcomm0} it is straightforward to check that the (normal ordered) order parameters \begin{align}\mathcal{O}_{y+1/2}^F(x)=i\gamma^A_{y+1}(x)\gamma^B_y(x),\quad\mathcal{O}_{y+1/2}^\Theta(x)=e^{i\Theta_{y+1/2}(x)}\label{orderparameters}\end{align} mutually commute, i.e. $\left[\mathcal{O}_{y+1/2}^{F/\Theta}(x),\mathcal{O}_{y'+1/2}^{F/\Theta}(x')\right]=0$. Therefore, the model is exactly solvable, and its ground states are characterized by the ground state expectation values (\hypertarget{GEV}{GEV}) of the order parameters \begin{align}l_0\langle\mathcal{O}_{y+1/2}^F\rangle=(-1)^y\langle\mathcal{O}_{y+1/2}^\Theta\rangle=\pm1\end{align} so that the interacting energy $\langle\mathcal{U}\rangle$ is minimized, where $l_0$ is some non-universal microscopic length scale. Pinning the \GEV $\langle\Theta_{y+1/2}\rangle=n_{y+1/2}\pi$, for $n_{y+1/2}\in\mathbb{Z}$, gaps all degrees of freedom in the charged $U(1)_4^{A/B}=SU(2)_1^{A/B}$ sector. The remaining neutral fermions are gapped by the decoupled Majorana backscattering \begin{align}\delta\mathcal{H}_{\mathrm{Majorana}}=u\sum_{y=-\infty}^\infty(-1)^yi\langle\mathcal{O}^\Theta_{y+1/2}\rangle\gamma^A_{y+1}\gamma^B_y.\label{MajHam}\end{align} It is worth noting that a $\pi$-kink excitation of $\langle\Theta_{y+1/2}\rangle$ flips the Majorana mass in \eqref{MajHam} and therefore bounds a zero energy Majorana bound state~\cite{Kitaevchain}. A $\pi$-kink at $x_0$ can be created by the vertex operators $e^{\pm i\phi^A_{y+1}(x_0)}$ or $e^{\pm i\phi^B_y(x_0)}$ which carry $\pm1/4$ of an electric charge. (Recall the bosonic vertices $e^{i4\phi^{A/B}_y}$ carry charge $e$.) This $e/4$ excitation therefore corresponds to the Ising anyon in the Pfaffian \FQH state.

From the \AFTR symmetry action \eqref{bosonTR11Pf}, one can show that the Majorana fermions \eqref{MFdef} transform according to \begin{align}\mathcal{T}_{11}\gamma^A_y\mathcal{T}_{11}^{-1}=\gamma^A_{y+1},\quad\mathcal{T}_{11}\gamma^B_y\mathcal{T}_{11}^{-1}=-\gamma^B_{y+1}.\end{align} Therefore the fermion order parameter $\mathcal{O}^F_{y+1/2}=i\gamma^A_{y+1}\gamma^B_y$ \eqref{orderparameters} is translated under the antiunitary symmetry \begin{align}\mathcal{T}_{11}\mathcal{O}^F_{y+1/2}\mathcal{T}_{11}^{-1}=\mathcal{O}^F_{y+3/2}.\label{T11OF}\end{align} The boson angle parameter $\Theta_{y+1/2}$ defined below \eqref{mbdint} changes to $-\Theta_{y+3/2}-(-1)^y\pi$ under \AFTR, and therefore the boson order parameter $\mathcal{O}^\Theta_{y+1/2}=e^{i\Theta_{y+1/2}}$ is flipped and translated \begin{align}\mathcal{T}_{11}\mathcal{O}^\Theta_{y+1/2}\mathcal{T}_{11}^{-1}=-\mathcal{O}^\Theta_{y+3/2}.\label{T11OT}\end{align} Together, \eqref{T11OF} and \eqref{T11OT} show that the many-body interaction $\mathcal{U}$ in \eqref{mbdint} is \AFTR symmetric.

The $C_2$ action \eqref{bosonC2} flips the number operator $\mathcal{C}_2(N_y^2+N_y^3)\mathcal{C}_2^{-1}=-N_{-y}^2-N_{-y}^3$, and therefore the parity operators appear in the ``Jordan-Wigner" string \eqref{MFdef} are $C_2$ symmetric, $\mathcal{C}_2(-1)^{N_y^2+N_y^3}\mathcal{C}_2^{-1}=(-1)^{N_{-y}^2+N_{-y}^3}$. With the help of the $C_2$ action \eqref{bosonC2Pf} in the fractional basis, one sees that $\mathcal{C}_2\cos\phi^\sigma_y\mathcal{C}_2^{-1}=(-1)^{y+1}\sin\phi^\sigma_{-y}$ and $\mathcal{C}_2\sin\phi^\sigma_y\mathcal{C}_2^{-1}=(-1)^{y+1}\cos\phi^\sigma_{-y}$ and thus the Majorana fermions \eqref{MFdef} transform according to \begin{align}\mathcal{C}_2\gamma^A_y\mathcal{C}_2^{-1}=(-1)^{y+1}\gamma^B_{-y}(-1)^{F_{2+3}},\\\mathcal{C}_2\gamma^B_y\mathcal{C}_2^{-1}=(-1)^{y+1}\gamma^A_{-y}(-1)^{F_{2+3}},\nonumber\end{align} where $(-1)^{F_{2+3}}=\prod_{y=-\infty}^\infty(-1)^{N_y^2+N_y^3}$ is the total fermion parity of channel 2 and 3. This shows the fermion order parameter is odd under $C_2$ \begin{align}&\mathcal{C}_2\mathcal{O}^F_{y+1/2}\mathcal{C}_2^{-1}\nonumber\\&=i(-1)^{y+2}\gamma^B_{-y-1}(-1)^{F_{2+3}}(-1)^{y+1}\gamma^A_{-y}(-1)^{F_{2+3}}\nonumber\\&=-i\gamma^A_{-y}\gamma^B_{-y-1}=-\mathcal{O}^F_{-y-1/2}.\label{C2OF}\end{align} On the other hand, one can also show from the $C_2$ action \eqref{bosonC2Pf} that the boson angle parameter changes as $\mathcal{C}_2\Theta_{y+1/2}\mathcal{C}_2^{-1}=-\Theta_{-y-1/2}-(-1)^y\pi$ and therefore the boson order parameter $\mathcal{O}^\Theta_{y+1/2}=e^{i\Theta_{y+1/2}}$ is conjugated and flipped under $C_2$ \begin{align}\mathcal{C}_2\mathcal{O}^\Theta_{y+1/2}\mathcal{C}_2^{-1}=-{\mathcal{O}^\Theta_{-y-1/2}}^\dagger.\label{C2OT}\end{align} When combined together, the minus signs in \eqref{C2OF} and \eqref{C2OT} cancel and they show that the many-body interaction $\mathcal{U}$ in \eqref{mbdint} preserves $C_2$.

Now that we have introduced symmetry preserving gapping interactions on a single diagonal layer, we can extend it to the entire 3D structure by transferring \eqref{mbdint} to all layers using the off-diagonal \AFTR operator $\mathcal{T}_{\bar{1}1}$ (see figure~\ref{fig:gappinginteraction}(a)). The resulting state belongs to a topological phase in three dimensions with an excitation energy gap. It preserves both \AFTR symmetries $\mathcal{T}_{11}$ and $\mathcal{T}_{\bar{1}1}$ as well as the (screw) $C_2$ symmetry. The choice of writing this paper with a specific model with these symmetries was intentional, we wanted to work out the simplest example with specific symmetries explicitly for illustrative reasons instead of doing a more general classification type argument. We leave the SPT-SET correspondences for general symmetries as an open question, but we expect that the methods presented in this work can be useful in exploring them.

\subsection{Antiferromagnetic stabilization}\label{sec:AFMstabilization}
The exactly-solvable many-body interacting model \eqref{mbdint} (see also figure~\ref{fig:gappinginteraction}) shows that the Dirac semimetal \eqref{WeylTBHam} can acquire a many-body mass gap without breaking symmetries. However, it is not clear how dominant or stable the topological phase described by \eqref{mbdint} is. There are alternative interactions that lead to other metallic or insulating phases that preserve or break symmetries. The scaling dimensions and the relevance of the interaction terms~\cite{Fradkinbook,Tsvelikbook} can be tuned by the velocity matrix $V_{jk}$ in \eqref{Llayer} that is affected by forward scattering interactions among co-propagating channels. Instead of considering energetics, we focus on a topological deliberation -- inspired by the coupled wire construction of quantum Hall states~\cite{KaneMukhopadhyayLubensky02,TeoKaneCouplewires} -- that can drastically reduce the number of possible interactions and may stabilize the desired interactions when applied to materials.

The coupled wire model considered so far assumes all electronic Dirac modes at the Fermi level have zero momentum $k_x=0$. This is convenient for the purpose of constructing an exactly solvable model because momentum is automatically conserved by the backscattering interactions. However, this also allows a huge collection of competing interactions. We propose the application of a commensurate modulation of magnetic field to restrict interactions that conserve momentum. There are multiple variations to the application, which depend on the details of the Dirac material and the Dirac vortices. To illustrate the idea, we present one possible simple scenario.

\begin{figure}[htbp]
\centering\includegraphics[width=0.45\textwidth]{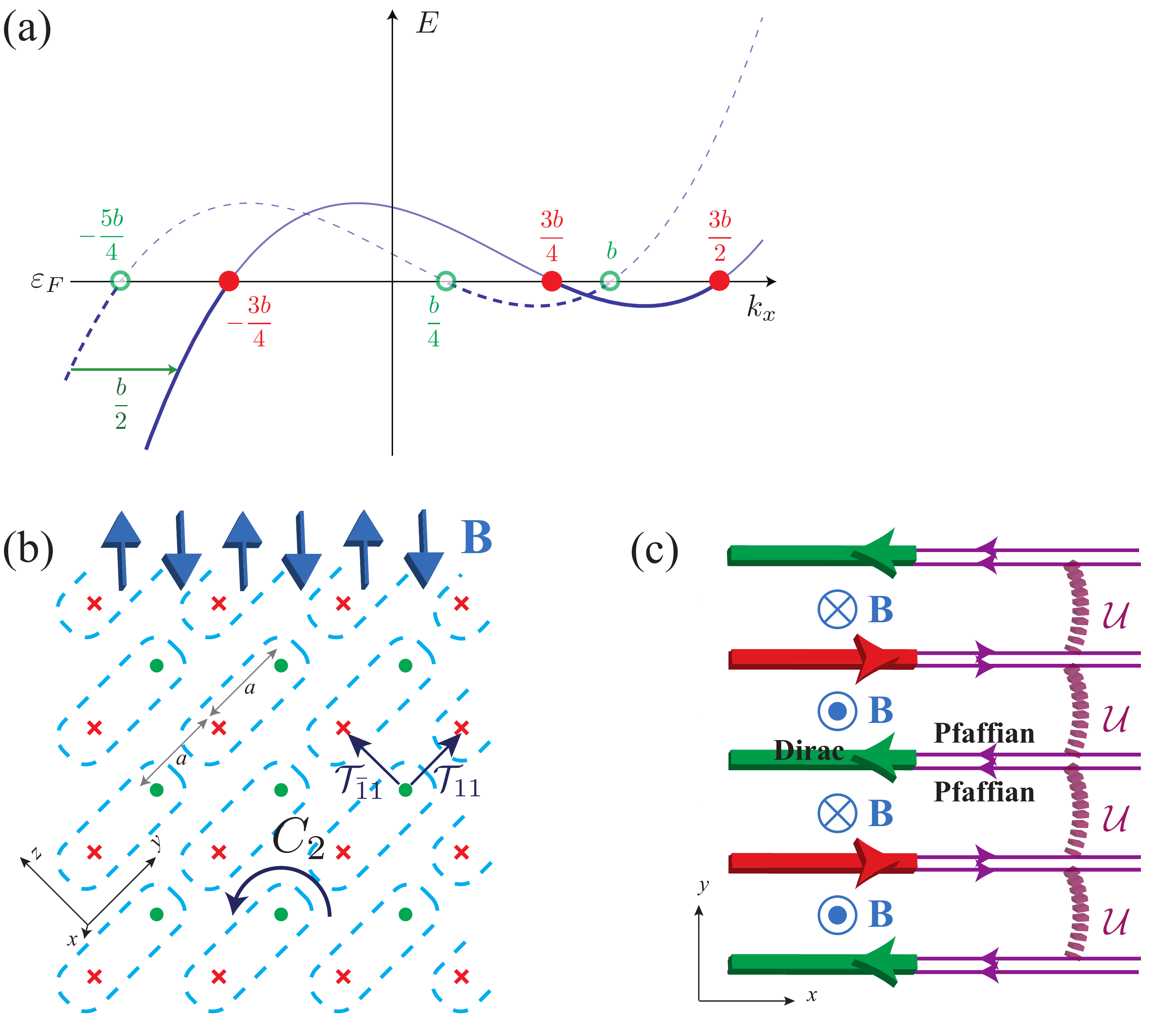}
\caption{(a) The energy dispersion $E_{y=2l}(k_x)$ with (solid curve) or without (dashed curve) the alternating magnetic field. (b) The alternating magnetic field configuration that preserves the AFTR and $C_2$ symmetries. (c) The alternating magnetic field across a single layer along the $xy$ plane.}\label{fig:afm}
\end{figure}

First we go back to a single Dirac wire and consider a non-linear dispersion \begin{align}E^0_{y=2l}(k_x)&=\frac{\hbar v}{b^2}(k_x-k_F^1)(k_x-k_F^2)(k_x-k_F^3),\nonumber\\E^0_{y=2l+1}(k_x)&=-\frac{\hbar v}{b^2}(k_x+k_F^1)(k_x+k_F^2)(k_x+k_F^3),\end{align} where $v$ and $b$ are some non-universal velocity and wave number parameters. We assume $k_F^2<k_F^3<k_F^1$ so that when the Fermi energy is at $\varepsilon_F=0$, there are two right (left) moving modes at $k_x=k_F^1,k_F^2$ and one left (resp.~right) moving one at $k_X=k_F^3$ along an even (resp.~odd) wire. This matches the three-channel Dirac wire \eqref{3Dirac} used in the splitting scheme in section~\ref{sec:gluing}. We assume the three Fermi wave numbers satisfy a commensurate condition \begin{align}2k_F^1+k_F^2-3k_F^3=0,\label{kcomm}\end{align} and we set \begin{align}b=2(k_F^3-k_F^1-k_F^2).\label{bcomm1}\end{align} The dashed band in figure~\ref{fig:afm}(a) shows one commensurate energy dispersion along an even wire.

Next, we consider a spatially modulating magnetic field ${\bf B}({\bf r})=B({\bf r}){\bf e}_{11}$, where \begin{align}B({\bf r})=\sum_{m=-\infty}^\infty B_m\sin\left[\pi\frac{\sqrt{2}(2m+1)}{a}{\bf e}_{\bar{1}1}\cdot{\bf r}\right],\end{align} ${\bf e}_{11}=({\bf e}_y+{\bf e}_z)/\sqrt{2}$ and ${\bf e}_{\bar{1}1}=(-{\bf e}_y+{\bf e}_z)/\sqrt{2}$, that preserves both the \AFTR and $C_2$ symmetries, \begin{align}B({\bf r}+a{\bf e}_y)=B({\bf r}+a{\bf e}_z)=B(C_2{\bf r})=-B({\bf r})\end{align} (see figure~\ref{fig:afm}(b) for the 3D field configuration). Moreover, we assume the field is commensurate with the Fermi wave numbers so that the magnetic flux per unit length across the $xy$ layer between adjacent wires (see figure~\ref{fig:afm}(c)) is \begin{align}\frac{\Phi_B}{L}=\frac{\phi_0}{2\pi}b\label{bcomm2}\end{align} where $L$ is the wire length, $\phi_0=hc/e$ is the magnetic flux quantum. Equivalently, the average magnetic field strength in the normal $z$-direction between adjacent wires is $|\overline{B_z}|=|\overline{B}|/\sqrt{2}=(\hbar c/ea)b$, where $a$ is the displacement between adjacent counter-propagating wires. We choose the vector potential $A_x(y,z)=[(-1)^y+(-1)^z-1]|\overline{B_z}|a/2$ and $A_y=A_z=0$ along the $(y,z)^{\mathrm{th}}$ wire. 

Along a wire on the $xy$ plane where $z=0$, the three electronic Dirac channels are now bosonized by \begin{align}\psi_y^{1,2}(x)&\sim e^{i[(-1)^y(k_F^{1,2}x+bx/2)+\tilde\phi_y^{1,2}(x)]},\\\psi_y^3(x)&\sim e^{i[(-1)^y(k_F^3x+bx/2)-\tilde\phi_y^3(x)]},\nonumber\end{align} where the momenta are shifted by $k_F^j\to k_F^j+(e/\hbar c)A_x$. The phase oscillation $e^{ikx}$ is canceled in an interaction term only when momentum is conserved, or otherwise the interaction would drop out after the integration over $x$. It is straightforward to check that the Majorana fermions \eqref{MFdef}, which contain the operators $e^{\pm i\phi^\sigma}$, have zero momentum because of the Fermi wave number commensurate condition \eqref{kcomm}. In addition, the boson backscattering $\cos(4\phi^A_{y+1}-4\phi^B_y)$ in \eqref{mbdint} preserves momentum because the magnetic field is also commensurate (see \eqref{bcomm1} and \eqref{bcomm2}).

\section{Interaction-enabled Dirac/Weyl semimetal}\label{sec:intenable}
%Recall interaction enable topological phases
So far in this section, we have been discussing the gapping of the Dirac semimetal while preserving the \AFTR and $C_2$ symmetries. In this subsection, we focus on an opposite aspect of the symmetric many-body interaction -- the enabling of a (semi)metallic phase that is otherwise forbidden by symmetries in the single-body setting. We noticed in subsection~\ref{sec:anomaly} that the pair of momentum-separated Weyl points in figure~\ref{fig:Weylspectrum} is anomalous. In fact, it is well-known already that Weyl nodes~\cite{Murakami2007,WanVishwanathSavrasovPRB11,YangLuRan11,burkovBalenstPRL11,Ashvin_Weyl_review}, if separated in momentum space, must come in multiples of four in a lattice translation and time reversal symmetric three dimensional non-interacting system. 

This no go theorem can be rephrased into a feature. \begin{enumerate}\item If the low energy excitations of a \TR symmetric lattice (semi)metal in three dimensions consists of one pair of momentum-separated Weyl nodes, then the system must involve many-body interaction.\end{enumerate} We refer to this \TR and lattice translation symmetric strongly-correlated system as an interaction-enabled topological Dirac (semi)metal. We assume the Weyl nodes are fixed at two \TR invariant momenta, and therefore they are stable against symmetry-preserving deformations. Otherwise, if the Weyl nodes are not located at high symmetry points, they can be moved and pair annihilated. Also, as explained in the beginning of section~\ref{sec:DiracSemimetal} and contrary to the more common contemporary terminology, we prefer to call the (semi)metal ``Dirac" rather than ``Weyl" because of the doubling. Perhaps more importantly, we propose the following conjecture. \begin{enumerate}\addtocounter{enumi}{1}\item Beginning with the interaction-enabled Dirac (semi)metal, {any} single-body symmetry-breaking mass must lead to a 3D gapped topological phase that cannot be adiabatically connected to a band insulator.\end{enumerate} We suspect this statement can be proven by a filling argument similar to that of Hasting-Oshikawa-Lieb-Schultz-Mattis~\cite{LiebSchultzMattis61,Oshikawa00,Hastings04}, and may already be available in Ref.~\onlinecite{WatanabePoVishwanath17} by Watanabe, Po and Vishwanath. This conjecture applies to the coupled wire situation where the gapped phase is long-range entangled and supports fractional excitations. Its topological order is out of the scope of this article, but will be presented in a future work~\cite{SirotaRazaTeoappearsoon}. In a broader perspective, this type of statements may provide connections between strongly-interacting and non-interacting phases and help understanding quantum phase transitions of long-range entangled 3D phases from that of single-body band insulating ones.

Before discussing the three dimensional case, we make the connection to a few known interaction-enabled topological phases with or without an energy gap in low dimensions. First, zero energy Majorana fermions $\gamma_j=\gamma_j^\dagger$ in a true zero dimensional non-interacting (spinless) \TR symmetric system must bipartite into an equal number of positive chiral ones $\mathcal{T}\gamma_j\mathcal{T}^{-1}=+\gamma_j$ and negative chiral ones $\mathcal{T}\gamma_l\mathcal{T}^{-1}=-\gamma_l$. Fidkowski and Kitaev showed in Ref.~\onlinecite{FidkowskiKitaev10} that under a combination of two-body interactions, eight Majoranas with the same chirality can acquire a \TR preserving mass and be removed from low energy. This leaves behind a collection of zero energy Majoranas that have a non-trivial net chirality of eight. Second, all $(1+1)$D \TR symmetric topological BDI superconductors~\cite{SchnyderRyuFurusakiLudwig08,Kitaevtable08,QiHughesRaghuZhang09,HasanKane10,QiZhangreview11,RMP} must break inversion because the zero energy Majorana boundary modes must have opposite chiralities at opposite ends. The Fidkowski-Kitaev interaction however allows one to construct a non-trivial $(1+1)$D topological model that preserves both \TR and inversion but at the same time supports four protected Majorana zero modes at each end~\cite{LapaTeoHughes14}. Third, a single massless Dirac fermion in $(2+1)$D is anomalous in a (spinful) \TR and charge $U(1)$ preserving non-interacting lattice system. On the other hand, it can be enabled by many-body interactions. For instance, when one of the two opposing surfaces of a topological insulator slab is gapped by symmetry-preserving interactions~\cite{WangPotterSenthilgapTI13,MetlitskiKaneFisher13b,ChenFidkowskiVishwanath14,BondersonNayakQi13}, a single massless Dirac fermion is left behind on the opposite surface as the only gapless low energy degrees of freedom of the quasi-$(2+1)$D system. Similar slab construction can be applied to the superconducting case, and interactions can allow any copies of massless Majorana fermions to manifest in $(2+1)$D with the presence of (spinful) \TR symmetry.

On the contrary, there are anomalous gapless fermionic states that cannot be enabled even by strong interactions. Chiral fermions that only propagate in a single direction cannot be realized in a true $(1+1)$D lattice system. They can only be supported as edge modes of $(2+1)$D topological phases such as quantum Hall states~\cite{Wenedgereview} or chiral $p_x+ip_y$ superconductors~\cite{Volovik99,ReadGreen}. Otherwise, they would allow heat transfer~\cite{KaneFisher97,Cappelli01,Kitaev06} from a low temperature reservoir to a high temperature one, thereby violating the second law of thermodynamics. Similarly, a single massless Weyl fermion can only be present as the $(3+1)$D boundary state of a $(4+1)$D topological bulk~\cite{ZhangHu01,BernevigChernHuToumbasZhang02,QiHughesZhang08,SchnyderRyuFurusakiLudwig08,Kitaevtable08}. It cannot exist in a true $(3+1)$D lattice system~\cite{Nielsen_Ninomiya_1981,NielsenNinomiyaPLB1981}, or otherwise under a magnetic field there would be unbalanced chiral fermions propagating along the field direction that constitute the ABJ-anomaly~\cite{Adler69,BellJackiw69,NielsenNinomiya83}.

\begin{figure}[htbp]
\centering\includegraphics[width=0.47\textwidth]{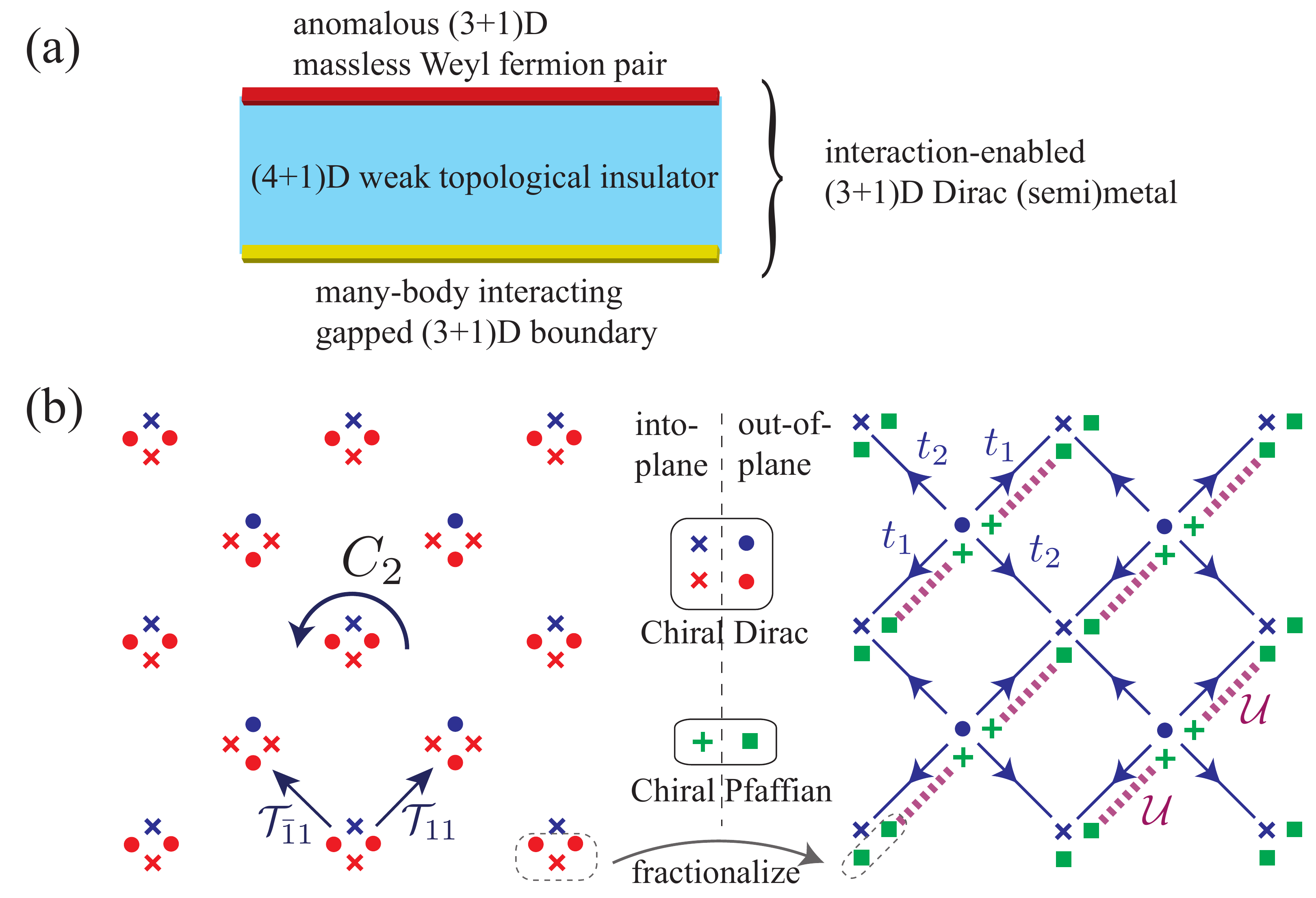}
\caption{(a) A quasi-$(3+1)$D interaction-enabled Dirac (semi)metal constructed by a 4D slab of WTI. (b) Coupled wire model of an anomalous Dirac (semi)metal enabled by interaction with $C_2$ rotation and both AFTR $\mathcal{T}_{11},\mathcal{T}_{\bar{1}1}$ symmetries.}\label{fig:intenable}
\end{figure}

In this section, we focus on the simplest anomalous gapless fermionic states in $(3+1)$D that can be enabled by interactions. As eluded in section~\ref{sec:holproj4D}, a weak topological insulator in $(4+1)$D can support the anomalous energy spectrum in figure~\ref{fig:Weylspectrum} on its boundary so that a pair of opposite Weyl points sit at two distinct \TRIM on the boundary Brillouin zone. A 4D \WTI slab, where the fourth spatial dimension is open and the other three are periodic, has two $(3+1)$D boundaries and each carries a pair of Weyl fermions. The coupling between the two pairs of Weyl fermions are suppressed by the system thickness and bulk energy gap. By introducing symmetry-preserving gapping interactions on the bottom surface, the anomalous gapless fermionic state is left behind on the top surface and is enabled in this quasi-$(3+1)$D setting (see figure~\ref{fig:intenable}(a)).

Inspired by this construction, we propose a true $(3+1)$D coupled wire model which has the anomalous energy spectrum in figure~\ref{fig:Weylspectrum} and preserves the \AFTR symmetries in both $\mathcal{T}_{11}$ and $\mathcal{T}_{\bar{1}1}$ directions as well as the $C_2$ (screw) rotation symmetry. The model is summarized in figure~\ref{fig:intenable}(b). It consists of a checkerboard array of electronic wires, where each wire has two chiral Dirac channels propagating into-paper and another two propagating out-of-paper. Contrary to the model considered in section~\ref{sec:DiracSemimetal}, here the net chirality on each wire cancels and therefore the wires are true $(1+1)$D systems without being supported by a higher dimensional bulk. Using the splitting scheme described in section~\ref{sec:gluing}, along each wire, one can fractionalize a group of three Dirac channels {\color{red}$\bullet\bullet\times$} ({\color{red}$\times\times\bullet$}) into a pair of co-propagating chiral Pfaffian channels {\color{green}$\blacksquare\blacksquare$} (resp.~{\color{green}$++$}). The two Pfaffian channels then can be backscattered in opposite directions using the many-body interaction $\mathcal{U}$ (dashed purple lines) described in section~\ref{sec:interactionmodels}. This introduces an excitation energy gap that removes three Dirac channels per wire from low energy. Lastly, single-body backscatterings $t_1,t_2$ (solid directed blue lines) among the remaining Dirac channels {\color{blue}$\bullet\times$} described in \eqref{WeylTBHam} and figure~\ref{fig:WeylTB} give rise to the low-energy Weyl spectrum in figure~\ref{fig:Weylspectrum}. Since the many-body interaction $\mathcal{U}$ and the single-body backscatterings $t_1,t_2$ preserve the $C_2$ rotation and both \AFTR symmetries $\mathcal{T}_{11}$ and $\mathcal{T}_{\bar{1}1}$, the model describes an interaction-enabled anomalous (semi)metal that is otherwise forbidden in a non-interacting non-holographic setting. 

The non-local anti-ferromagnetic nature of the time reversal symmetry is built-in in the present coupled wire model. We speculate in passing that a local conventional \TR symmetric Dirac (semi)metallic phase consisting of a single pair of momentum-space-separated Weyl nodes may also be enabled by interaction. On one hand, the \AFTR symmetry could be restored to a local \TR symmetry by ``melting" the checkerboard wire array. On the other hand, there could also be an alternative wire configuration that facilitates a coupled wire model with a local conventional \TR symmetry.

Lastly, we gap the interaction-enabled Dirac semimetallic model (figure~\ref{fig:intenable}) by a symmetry-breaking single-body mass. This can be achieved by introducing electronic backscattering terms that dimerize the remaining Dirac channels {\color{blue}$\bullet\times$}, and were described by \eqref{DiracTBHam} in section~\ref{sec:brokensymmetry}. The resulting state is an insulating $(3+1)$D topological phase with long-range entanglement. For instance, each diagonal layer gapped by the many-body interaction $\mathcal{U}$ has the identical topological order of the $\mathcal{T}$-Pfaffian surface state of a topological insulator. 

\section{Fractional surface states}\label{sec:fracsurface}

In section~\ref{sec:fermiarc1}, we discussed the surface states of the single-body coupled Dirac wire model \eqref{WeylTBHam} (see also figure~\ref{fig:WeylTB}). In particular, we showed in figure~\ref{fig:SurfaceStates1bdy} that an \AFTR symmetry preserving surface hosts open chiral Dirac channels, which connect and leak into the 3D (semi)metallic bulk. Earlier in this section, we discussed the effects of many-body interaction, which leads to two possible phases: (a) a gapped topological phase (see section~\ref{sec:interactionmodels}) that preserves one of the two \AFTR symmetries, say $\mathcal{T}_{11}$, and (b) a gapless interaction-enabled Dirac semimetal (see section~\ref{sec:intenable}) that preserves the $C_2$ rotation and both \AFTR symmetries $\mathcal{T}_{11}$ and $\mathcal{T}_{\bar{1}1}$. Here, we describe the boundary states of the two interacting phases on a surface closed under the symmetries.

\begin{figure}[htbp]
\centering\includegraphics[width=0.48\textwidth]{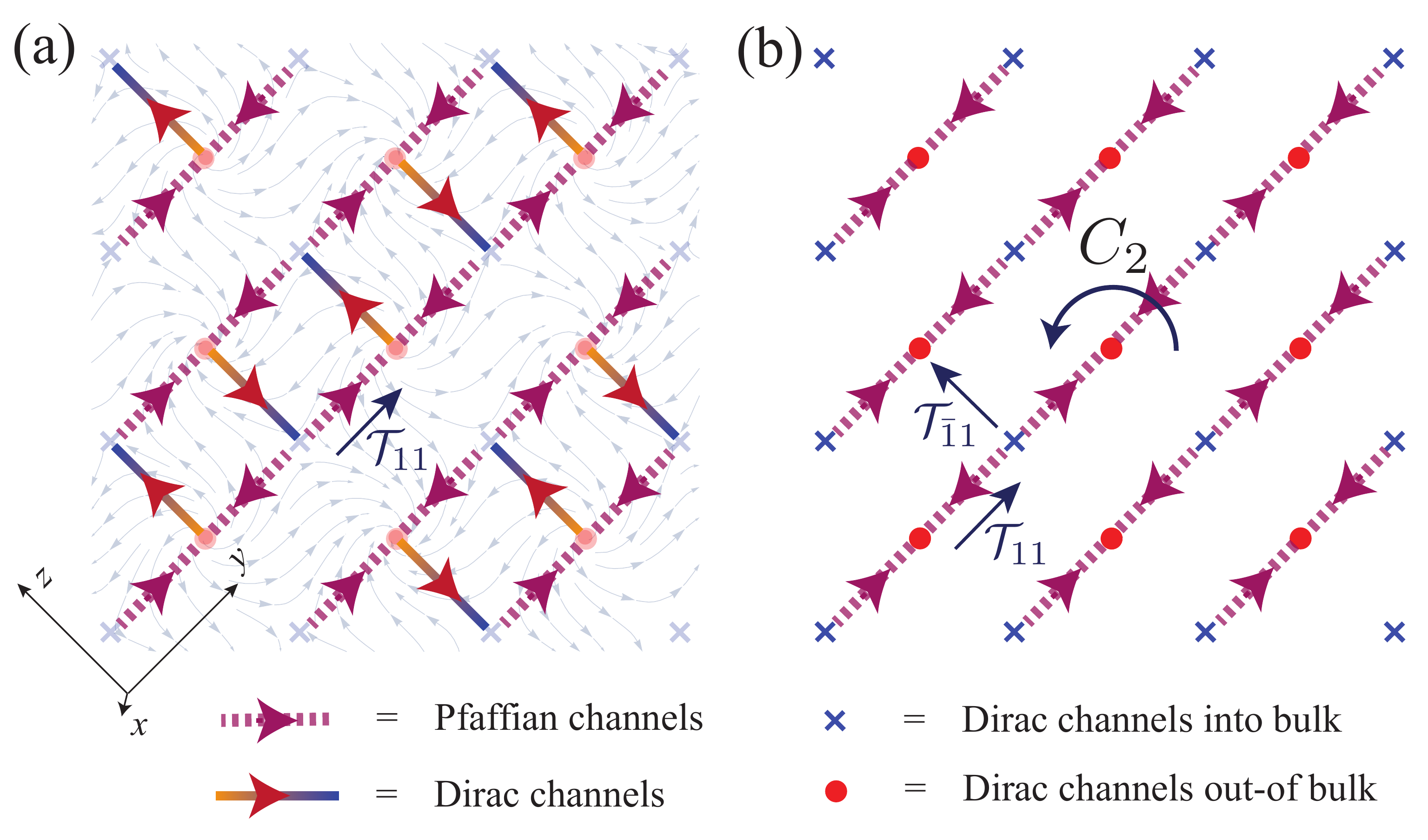}
\caption{Fractional surface states of (a) a 3D Dirac insulator gapped by many-body interaction that preserves $\mathcal{T}_{11}$, and (b) a 3D gapless interaction-enabled Dirac semimetal that preserves $\mathcal{T}_{11}$, $\mathcal{T}_{\bar{1}1}$ and $C_2$.}\label{fig:SurfaceStates}
\end{figure}

First, we consider the coupled wire model with the many-body interaction \eqref{mbdint} (see also figure~\ref{fig:gappinginteraction}) and a boundary surface along the $yz$-plane perpendicular to the wires. The surface network of fractional channels is shown in figure~\ref{fig:SurfaceStates}(a). We assume the bulk chiral Dirac wires ({\color{blue}$\times$}{\color{red}$\bullet$}) are supported as vortices of Dirac mass in the bulk (recall \eqref{DiracHam}), where the texture of the mass parameters is represented by the underlying vector field. The model is juxtaposed along the $yz$- boundary plane against the trivial Dirac insulating state $H_{\mathrm{vacuum}}=\hbar v{\bf k}\cdot\vec{s}\mu_z+m_0\mu_x$, which models the vacuum. The line segments on the surface plane where the Dirac mass $m_0\mu_x$ changes sign host chiral Dirac channels (c.f.~subsection~\ref{sec:fermiarcAFTRpreserving}).

Unlike the single-body (semi)metallic case in figure~\ref{fig:SurfaceStates1bdy} where the surface Dirac channels connects with the bulk ones, now the many-body interacting bulk is insulating and does not carry low-energy gapless excitations. Thus, the surface Dirac channels here cannot leak into the bulk and must dissipate to other low-energy degrees of freedom on the surface. The many-body interwire backscattering interaction in \eqref{mbdint} (and figure~\ref{fig:gappinginteraction}) leaves behind chiral Pfaffian channels on the surface. These fractional channels connect back to the surface Dirac channels in pairs. The surface network of chiral channels preserves the \AFTR $\mathcal{T}_{11}$ symmetry. However, the low-energy surface state is not protected. Electronic states can be localized by dimerizing the Pfaffian channels in the $z$ (or $\bar{1}1$) direction.

Second, we consider the interaction-enabled Dirac semimetallic model summarized in figure~\ref{fig:intenable}(b) in section~\ref{sec:intenable} and again let it terminate along the symmetry preserving $yz$-plane perpendicular to the wires. The surface gapless channels are shown in figure~\ref{fig:SurfaceStates}(b). Here, the semimetallic bulk preserves $C_2$ as well as the two \AFTR symmetries $\mathcal{T}_{11}$ and $\mathcal{T}_{\bar{1}1}$. The bulk array of wires are true $(1+1)$D systems and are not supported as edge modes or vortices of a higher dimensional bulk. The pair of into-paper Dirac modes are bent into the pair of out-of-paper ones along each wire at the terminal. Similar to the previous case, the many-body bulk interwire backscattering interaction leaves behind surface chiral Pfaffian channels. Through the mode bending at the wire terminal, these Pfaffian channels join in pairs and connect to the chiral Dirac channels in the bulk that constitute the Dirac semimetal. In this case, the surface state is protected by $C_2$, $\mathcal{T}_{11}$ and $\mathcal{T}_{\bar{1}1}$, and is forced to carry fractional gapless excitations as a consequence and signature of the anomalous symmetries. For instance, the charge $e/4$ Ising-like quasiparticle and the charge $e/2$ semion can in principle be detected by shot noise tunneling experiments. These gapless fractional excitations however are localized on the surface because the Dirac (semi)metallic bulk only supports gapless electronic quasiparticles.

\section{Conclusion and Discussion}\label{sec:conclusion}
Dirac and Weyl (semi)metals have generated immense theoretical and experimental interest. On the experimental front, this is fueled by an abundant variety of material classes and their detectable \ARPES and transport signatures. On the theoretical front, Dirac/Weyl (semi)metal is the parent state that, under appropriate perturbations, can give birth to a wide range of topological phases, such as topological (crystalline) insulators and superconductors. In this work, we explored the consequences of a specific type of strong many-body interaction based on a coupled-wire description. In particular, we showed that (i) a 3D Dirac fermion can acquire a finite excitation energy gap in the many-body setting while preserving the symmetries that forbid a single-body Dirac mass, and (ii) interaction can enable an anomalous antiferromagnetic time-reversal symmetric topological (semi)metal whose low-energy gapless degrees of freedom are entirely described by a pair of non-interacting electronic Weyl nodes separated in momentum space. A brief conceptual summary was presented in section~\ref{sec:introsummary} and will not be repeated here. Instead, we include a short discussion on what are the broad implications of this work and then discuss possible future directions:

\textit{Theoretical impact:} We believe this work is a first step towards a duality between quantum critical transitions of short-range entangled symmetry-protected topological phases and long-range entangled symmetry-enriched topological phases. The 3D topological order in these phases is completely different from 2D topological order as it can have both point and line-like excitations and a much richer structure ~\cite{SirotaRazaTeoappearsoon}. There have been field-theoretical descriptions, along the lines of BF and Chern-Simons theories, of 3D topological phases that support these richer structures, such as loop braiding. However, there have been only very few exact solvable examples and none of them patch the field-theoretical descriptions and microscopic electronic systems. The construction presented in this work opens a new direction towards making such a connection. The models are exactly solvable, and they originate from a microscopic Dirac electronic system with local 2-body interactions. They also have potential impact on numerical modelling. For example, the interacting coupled wire model can be approached by a lattice electronic model, which forgoes exact solvability but potentially leads to new critical transitions between topological phases in 3D.

\textit{High-energy impact:} For a single pair of Weyl nodes with opposite chirality, time-reversal symmetry (TRS) must be broken. Hence, for time-reversal symmetric systems, at least 4 Weyl nodes are required. In this paper, we have shown that, as enabled by many-body interactions, an electronic system can support a single pair of Weyl nodes in low-energy without violating TRS (c.f.~subsection~\ref{sec:intenable}). Such a material, if it exists, can be verified experimentally by \ARPES, and as a non-trivial consequence, our results assert that such a material must encode long-range entanglement. The existence of a single pair of massless Weyl fermions without TRS breaking in 3+1D can potentially provide new theories beyond the standard model.

\textit{Experimental realization:} There have been numerous field-theoretical discussions on possible properties of topologically ordered phases in 3D ~\cite{BiYouXu14,JiangMesarosRan14,WangLevin14,JianQi14,WangWen15,WangLevin15,LinLevin15,ChenTiwariRyu15}. However, unlike the 2D case there are no materials that exhibit topological order (quasiparticle excitations) in 3D. In this work, we show that an interacting Weyl or Dirac semimetal is a good place to start for the following reasons. A symmetry preserving gap must result in topological order and fractionalization (c.f.~subsection~\ref{sec:intenable}). While it is entirely likely that interactions leads to a spontaneous symmetry breaking phase, we show that there is no obstruction to realizing an interacting phase that preserves symmetries. Such a gapping must support fractionalization, such as the $e/4$ charged Ising-like and $e/2$ charged semion-like quasiparticles in the bulk, as predicted by our work. These charged particles can in principle be measured using a shot noise experiment across a point contact. Moreover, the gapping procedure involves Pfaffian channels so there should be excitations that mirror those in a Pfaffian state. There would also be line-like excitations in 3D for which the experimental signature is not yet clear. Therefore, we believe an interaction-enabled, symmetry-preserving gapped Weyl/Dirac semimetal is a good candidate for realizing topologically ordered phases in 3D. As for the experimental verification of the anomalous interaction-enabled Dirac semimetal, the electronic energy spectrum of a single pair of momentum-separated Weyl nodes in the presence of time-reversal symmetry can be measured using \hyperlink{ARPES}{ARPES}, assuming spontaneous symmetry-breaking is absent. Although the proposed experimental signatures, if measured, will strongly point towards the existence of such states, we cannot claim that such signatures provide a smoking gun evidence yet. More work needs to be done for the complete characterization of the point-like and line-like topological order of these states and will be part of a future work.

Apart from the 3D topological order having a much richer structure than 2D topological order, the 3D case presented in this work is qualitatively different from the well-studied 2D case. In the 2D case, the massless Dirac surface state is anomalous and lives on the boundary of a higher dimensional bulk. This is qualitatively distinct from the 3D Dirac/Weyl (semi)metal, which does not require holographic projection from a 4D bulk. In fact, a single 3D Weyl fermion, which is supported on the boundary of a 4D topological insulator, cannot be gapped while preserving charge U(1) conservation even with many-body interaction due to chiral anomaly. This serves as a counter-example which distinguishes the gappability of 2D versus 3D boundary state. Thus, it is not a priori an expected result that a Dirac/Weyl (semi)metal can be gapped without breaking symmetries. Moreover, the topological origin of 3D Dirac/Weyl (semi)metals relies on the addition of non-local spatial symmetry, in the current 3D case, the $C_2$ screw rotation. This is distinct from the 2D Dirac surface case, where all symmetries are local.

%superconducting version and nodal systems
We conclude by discussing possible future directions. First, coupled wire constructions can also be applied in superconducting settings and more general nodal electronic systems. For example, a Dirac/Weyl metal can be turned into a topological superconductor~\cite{SchnyderRyuFurusakiLudwig08,Kitaevtable08,QiHughesRaghuZhang09} under appropriate intra-species (i.e.~intra-valley) $s$-wave pairing~\cite{QiWittenZhang13}. Pairing vortices host gapless chiral Majorana channels~\cite{QiWittenZhang13,GuQi15,LopesTeoRyu17}. An array of these chiral vortices can form the basis in modeling superconducting many-body topological phases in three dimensions. On the other hand, instead of considering superconductivity in the continuous bulk, inter-wire pairing can also be introduced in the coupled Dirac wire model and lead to new topological states~\cite{ParkTeoGilbertappearsoon}.

Dirac/Weyl (semi)metals are a specific type of nodal electronic matter. For example, nodal superconductors were studied in states with dx$^2$-y$^2$ pairing~\cite{RyuHatsugaiPRL02}, He$^3$ in its superfluid A-phase~\cite{Volovik3HeA,Volovikbook}, and non-centrosymmetric states~\cite{SchnyderRyuFlat,BrydonSchnyderTimmFlat}. Weyl and Dirac fermions were generalized in \TR and mirror symmetric systems to carry $\mathbb{Z}_2$ topological charge~\cite{morimotoFurusakiPRB14}. General classification and characterization of gapless nodal semimetals and superconductors were proposed~\cite{Sato_Crystalline_PRB14,ZhaoWangPRL13,ZhaoWangPRB14,ChiuSchnyder14,matsuuraNJP13,Volovikbook,RMP,HoravaPRL05}. It would be interesting to investigate the effect of strong many-body interactions in general nodal systems.

%coarse-graining implication in real space RG and interaction; vortex dynamics
Second, in section~\ref{sec:DiracSemimetal}, we described a coarse-graining procedure of the coupled wire model that resembles a real-space renormalization and allows one to integrate out high energy degrees of freedom. While this procedure was not required in the discussions that follow because the many-body interacting model we considered was exactly solvable, it may be useful in the analysis of generic interactions and disorder. The coarse-graining procedure relied on the formation of vortices, which were introduced extrinsically. Like superconducting vortices, it would be interesting as a theory and essential in application to study the mechanism where the vortices of Dirac mass can be generated dynamically. To this end, it may be helpful to explore the interplay between possible (anti)ferromagnetic orders and the spin-momentum locked Dirac fermion through antisymmetric exchange interactions like the Dzyaloshinskii-Moriya interaction~\cite{Dzyaloshinsky58,Moriya60}.

%topological order, threefold lattice and alternative fractionalization
Third, the symmetry-preserving many-body gapping interactions considered in section~\ref{sec:interaction} have a ground state that exhibits long-range entanglement. This entails degenerate ground states when the system is compactified on a closed three dimensional manifold, and fractional quasi-particle and quasi-string excitations or defects. These topological order properties were not elaborated in our current work but will be crucial in understanding the topological phase~\cite{SirotaRazaTeoappearsoon} as well as the future designs of detection and observation. It would also be interesting to explore possible relationships between the coupled wire construction and alternative exotic states in three dimensions, such as the Haah's code~\cite{Haah11,Haah13}.

Fourth, the many-body inter-wire backscatterings proposed in section~\ref{sec:interactionmodels} were based on a fractionalization scheme described in \ref{sec:gluing} that decomposes a chiral Dirac channel with $(c,\nu)=(1,1)$ into a decoupled pair of Pfaffian ones each with $(c,\nu)=(1/2,1/2)$. In theory, there are more exotic alternative partitions. For instance, if a Dirac channel can be split into three equal parts instead of two, an alternative coupled wire model that put Dirac channels on a honeycomb vortex lattice could be constructed by backscattering these fractionalized channels between adjacent pairs of wires. Such higher order decompositions may already be available as conformal embeddings in the \CFT context. For example, the affine $SU(2)$ Kac-Moody theory at level $k=16$ has the central charge $c=8/3$, and its variation may serve as the basis of a ``ternionic" model.

%Continuum model
%Material realization?
%Melting
%Defect modes
%Linear response theory

\begin{acknowledgments}%\section*{Acknowledgments}
This work is supported by the National Science Foundation under Grant No.~DMR-1653535. We thank Matthew Gilbert and Moon Jip Park for insightful discussions.
\end{acknowledgments}

\appendix

\section{Chiral modes along topological defects}\label{sec:chiralmodesapp}
In section~\ref{sec:DiracSemimetal}, we begin with the Dirac Hamiltonian \eqref{DiracHam} where the mass term winds around a vortex and as a consequence, it hosts a chiral Dirac channel along the vortex (also see figure~\ref{fig:Diracstring}). Here we will demonstrate an example of a simple vortex, and show that there is a chiral Dirac zero mode. In general, the correspondence between the number of protected chiral Dirac channels and the vortex winding is a special case of the Atiyah-Singer Index theorem~\cite{AtiyahSinger63} and falls in the physical classification of topological defects~\cite{TeoKane}.

First, say we start with the Hamiltonian from \eqref{DiracHam}. Then for simplicity we consider the particular Dirac mass $m({\bf r})=m_x({\bf r})+im_y({\bf r})=|m|e^{i\theta}$ that constitute a vortex along the $z$-axis, where $\theta$ is the polar angle on the $xy$-plane. By replacing $k_{x,y}\leftrightarrow-i\partial_{x,y}$, \eqref{DiracHam} becomes \begin{align}H({\bf r})=&\hbar v(-i\partial_xs_x-i\partial_ys_y+k_zs_z)\mu_z\nonumber\\&\;+|m|\cos\theta\mu_x+|m|\sin\theta\mu_y\label{DiracHamapp}\end{align} where $k_z$ is still a good quantum number because translation in $z$ is still preserved. The Hamiltonian can be transformed under a new basis into \begin{align}H'=UHU^{-1}=\left(\begin{smallmatrix}-\hbar vk_z&D\\D^\dagger&\hbar vk_z\end{smallmatrix}\right),\quad U =\left(\begin{smallmatrix}0&1&0&0\\0&0&1&0\\1&0&0&0\\0&0&0&1\end{smallmatrix}\right)\end{align} where the Dirac operator occupying the off-diagonal blocks is \begin{align}D^\dagger &=\left(\begin{smallmatrix}-2i\hbar v\partial_w&|m|e^{-i\theta}\\|m|e^{i\theta}&2i\hbar v \partial_{\bar{w}}\end{smallmatrix}\right)\nonumber\\&=e^{-i\theta\sigma_z}\left(\begin{smallmatrix}-i\hbar v(\partial_r-i \partial_\theta/r)&|m|\\|m|&i\hbar v(\partial_r+i\partial_\theta/r)\end{smallmatrix}\right)\end{align} where $w=x+iy=re^{i\theta}$ and $\sigma_z=\mathrm{diag}(1,-1)$. 

Now we separate the Hamiltonian \begin{align}H'(k_z)=\hbar vk_z\Gamma_5+\left(\begin{smallmatrix}0&D\\D^\dagger&0\end{smallmatrix}\right).\end{align} where $\Gamma_5=\mathrm{diag}(-\openone_2,\openone_2)$. We note that the zero momentum sector $H'(k_z=0)$ has a chiral symmetry since it anticommutes with with $\Gamma_5$, and it reduces to the Jackiw-Rossi vortex problem in two-dimensions~\cite{JackiwRossi81}. The Dirac operator $D^\dagger$ has only one normalizable zero mode $u_0(r)\propto e^{-|m|r/\hbar v}(e^{i\pi/4}, e^{-i\pi/4})^T$, while its conjugate $D$ has none. $H'(k_z=0)$ therefore has a zero eigenvector of $\psi_0(r)=(u_0(r),0)^T$, which is also an eigenvector of $\Gamma_5$. In the full Hamiltonian, the zero mode $\psi_0(r)$ has energy $-\hbar vk_z$ and corresponds a single mid-gap chiral Dirac channel.

\section{Symmetry transformations of Chern invariants}\label{sec:Chernapp}

In section~\ref{sec:anomaly}, we discussed the Chern numbers on two-dimensional momentum planes of the anomalous Dirac (semi)metal. It was claimed that the Chern numbers \eqref{1stChern} on the two planes at $k_x=\pm\pi/2$ (see figure~\ref{fig:Weylspectrum}) are of opposite signs because of the \AFTR and twofold $\mathcal{C}_2$ (screw) rotation symmetries. In this appendix we will derive the symmetry flipping operations on the Chern invariants.

We begin with a Bloch Hamiltonian $H({\bf k})$ that is symmetric under the operation $G({\bf k})$, \begin{align}H({\bf k})&=G(g{\bf k})H(g{\bf k})G(g{\bf k})^{-1}\end{align} if $G$ is unitary, or \begin{align}H({\bf k})&=G(g{\bf k})H(g{\bf k})^\ast G(g{\bf k})^{-1}\end{align} if it is antiunitary. Let $|u_m({\bf k})\rangle$ be the occupied states of $H({\bf k})$. We define $|u'_m({\bf k})\rangle=|Gu_m({\bf k})\rangle=G(g{\bf k})|u_m(g{\bf k})\rangle$ (or $|u'_m({\bf k})\rangle=|Gu_m({\bf k})\rangle=G(g{\bf k})|u_m(g{\bf k})^\ast\rangle$), which is also an occupied state of $H({\bf k})$, for unitary (resp.~antiunitary) symmetry.

The Chern number \eqref{1stChern} can equivalently be defined as \begin{align}\mathrm{Ch}_1(k_x)=\frac{i}{2\pi}\int_{\mathcal{N}_{k_x}}\mathrm{Tr}\left(\mathcal{F}_{\bf k}\right)\label{1stChernapp}\end{align} where $\mathrm{Tr}\left(\mathcal{F}_{\bf k}\right)=d\mathrm{Tr}\left(\mathcal{A}_k\right)$, $\mathcal{N}_{k_x}$ is the oriented $k_yk_z$-plane with fixed $k_x$, and $\mathcal{A}_k$ is the Berry connection of the occupied states $\mathcal{A}_{\bf k}^{mn}=\langle u_m({\bf k})|du_n({\bf k})\rangle$. The Berry connection transforms according to \begin{align}{\mathcal{A}'}_{\bf k}^{mn}&\equiv\langle u'_m({\bf k})|du'_n({\bf k})\rangle\\&=\langle u_m(g{\bf k})|G(g{\bf k})^\dagger d \left[G(g{\bf k})|u_n(g{\bf k})\rangle\right]\nonumber\\&=\mathcal{A}_{g{\bf k}}^{mn}+\langle u_m(g{\bf k})|\left[G(g {\bf k})^\dagger dG(g {\bf k})\right]|u_n(g{\bf k})\rangle\nonumber\end{align} for unitary $G$, or \begin{align}{\mathcal{A}'}_{\bf k}^{mn}&=\left(\mathcal{A}_{g{\bf k}}^{mn}\right)^\ast+\langle u_m(g{\bf k})^\ast|\left[G(g {\bf k})^\dagger dG(g {\bf k})\right]|u_n(g{\bf k})^\ast\rangle\nonumber\\&=-\mathcal{A}_{g{\bf k}}^{nm}+\langle u_m(g{\bf k})^\ast|\left[G(g {\bf k})^\dagger dG(g {\bf k})\right]|u_n(g{\bf k})^\ast\rangle\nonumber\end{align} if $G$ is antiunitary, because the connection is skew-hermitian $\mathcal{A}=-\mathcal{A}^\dagger$. Therefore \begin{align}%\mathrm{Tr}(\mathcal{A}'_{\bf k})&=\mathrm{Tr}(\mathcal{A}_{g{\bf k}})+\mathrm{Tr}\left\{P_{g \bf{k}}\wedge\left[G(g{\bf k})^\dagger dG(g {\bf k})\right]\right\},\nonumber\\
\mathcal{F}'_{\bf k}&=\mathcal{F}_{g{\bf k}}+d\mathrm{Tr}\left\{P_{g\bf{k}}\wedge\left(G(g{\bf k})^\dagger dG(g{\bf k})\right]\right\}\label{curvature}\end{align} for an unitary symmetry, or \begin{align}%\mathrm{Tr}(\mathcal{A}'_{\bf k})&=-\mathrm{Tr}(\mathcal{A}_{g{\bf k}})+\mathrm{Tr}\left\{P_{g \bf{k}}^\ast\wedge\left[G(g{\bf k})^\dagger dG(g {\bf k})\right]\right\},\nonumber\\
\mathcal{F}'_{\bf k}&=-\mathcal{F}_{g{\bf k}}+d\mathrm{Tr}\left\{P_{g\bf{k}}^\ast\wedge\left(G(g{\bf k})^\dagger dG(g{\bf k})\right]\right\}\label{curvature2}\end{align} for an antiunitary one. Here $P({\bf k})=\sum_n|u_n({\bf k})\rangle\langle u_n({\bf k})|$ is the projection operator on to the occupied energy states at momentum ${\bf k}$. Since the trace of Berry curvature $\mathrm{Tr}(\mathcal{F})$ does not depend on the gauge choice of occupied states, $\mathrm{Tr}(\mathcal{F}_{\bf k})=\mathrm{Tr}(\mathcal{F}'_{\bf k})$. We notice the final terms in both \eqref{curvature} and \eqref{curvature2} integrate to zero over the closed periodic momentum plane $\mathcal{N}_{k_x}$. This is because they are total derivatives, and unlike $\mathcal{A}_{\bf k}$, $P_{\bf k}$ and $G({\bf k})$ are defined non-singularly on the entire Brillouin zone (see \eqref{AFTRk} and \eqref{C2k}). 

Now we derive the relation between the Chern number \eqref{1stChernapp} between $k_x$ and $-k_x$ using the antiunitary \AFTR and the unitary $\mathcal{C}_2$ symmetries. The \AFTR symmetries flip all momentum axes $\mathcal{T}_{11},\mathcal{T}_{\bar{1}1}:(k_x,k_y,k_z)\mapsto(-k_x,-k_y,-k_z)$, while the $\mathcal{C}_2$ symmetry flips only two $\mathcal{C}_2:(k_x,k_y,k_z)\mapsto(-k_x,-k_y,k_z)$. Thus, $\mathcal{T}_{11},\mathcal{T}_{\bar{1}1}:\mathcal{N}_{k_x}\to\mathcal{N}_{-k_x}$ maps between opposite planes while preserving their orientations, but $\mathcal{C}_2:\mathcal{N}_{k_x}\to-\mathcal{N}_{-k_x}$ is orientation reversing. Lastly, we substitute \eqref{curvature} and \eqref{curvature2} into \eqref{1stChernapp}, and apply a change of integration variable ${\bf k}\leftrightarrow g{\bf k}$. The \AFTR and $\mathcal{C}_2$ requires the Chern number to flip under $k_x\leftrightarrow-k_x$ \begin{align}\mathrm{Ch}_1(k_x)=-\mathrm{Ch}_1(-k_x).\end{align}

%\bibliography{refFinal}



\end{document}